\documentclass[amsmath, amssymb,10pt,aps,prx,twocolumn,notitlepage,showpacs,superscriptaddress]{revtex4-1}
\usepackage{graphicx}
\usepackage{calc}
\usepackage{braket}
\usepackage{ulem}   
\usepackage{slashed}
\usepackage{wasysym}
\usepackage{dsfont}
\usepackage{amsthm,amsmath,amsfonts,amssymb,verbatim,color}
\usepackage{graphicx}
\usepackage{subfigure}
\usepackage{bm}
\usepackage{epsfig,slashed}
\usepackage[T1]{fontenc}
\usepackage[colorlinks=true,citecolor=blue,linkcolor=blue,urlcolor=blue]{hyperref}
\newcommand{\bD}{\textbf{D}}
\newcommand{\bk}{\textbf{k}}
\newcommand{\bj}{\textbf{j}}
\newcommand{\bE}{\textbf{E}}
\newcommand{\bv}{\textbf{v}}
\newcommand{\bA}{\textbf{A}}

\newcommand{\icm}{\ensuremath{~\textrm{cm}^{-1}}}

\begin{document}

\title{Optical signatures of multifold fermions in the chiral topological semimetal CoSi}

\author{B. Xu}
\affiliation{University of Fribourg, Department of Physics and Fribourg Center for Nanomaterials, Chemin du Mus\'{e}e 3, CH-1700 Fribourg, Switzerland}
\author{Z. Fang}
\affiliation{Department of Chemistry, University of Pennsylvania, Philadelphia, Pennsylvania 19104–6323, USA}
\author{M. A. S\'{a}nchez-Mart\'{i}nez}
\affiliation{Institut N\'{e}el, CNRS and Univ. Grenoble Alpes, 38042 Grenoble, France}
\author{J. W. F. Venderbos}
\author{Z. Ni}
\affiliation{Department of Physics and Astronomy, University of Pennsylvania, Philadelphia, Pennsylvania 19104, USA}
\author{T. Qiu}
\affiliation{Department of Chemistry, University of Pennsylvania, Philadelphia, Pennsylvania 19104–6323, USA}
\author{K. Manna}
\affiliation{Max-Planck-Institut fur Chemische Physik fester Stoffe, 01187 Dresden, Germany}
\author{K. Wang}
\affiliation{Maryland Quantum Materials Center, Department of Physics, University of Maryland, College Park, MD 20742, USA.}
\author{J. Paglione}
\affiliation{Maryland Quantum Materials Center, Department of Physics, University of Maryland, College Park, MD 20742, USA.}
\author{C. Bernhard}
\affiliation{University of Fribourg, Department of Physics and Fribourg Center for Nanomaterials,
Chemin du Mus\'{e}e 3, CH-1700 Fribourg, Switzerland}
\author{C. Felser}
\affiliation{Max-Planck-Institut fur Chemische Physik fester Stoffe, 01187 Dresden, Germany}
\author{E. J. Mele}
\affiliation{Department of Physics and Astronomy, University of Pennsylvania, Philadelphia, Pennsylvania 19104, USA}
\author{A. G. Grushin}
\affiliation{Institut N\'{e}el, CNRS and Univ. Grenoble Alpes, 38042 Grenoble, France}
\author{A. M. Rappe}
\affiliation{Department of Chemistry, University of Pennsylvania, Philadelphia, Pennsylvania 19104–6323, USA}
\author{Liang Wu}
\email{liangwu@sas.upenn.edu}
\affiliation{Department of Physics and Astronomy, University of Pennsylvania, Philadelphia, Pennsylvania 19104, USA}

\date{\today}

\begin{abstract}
We report the optical conductivity in high-quality crystals of the chiral topological semimetal CoSi, which hosts exotic quasiparticles known as multifold fermions. We find that the optical response is separated into several distinct regions as a function of frequency, each dominated by different types of quasiparticles. The low-frequency intraband response is captured by a narrow Drude peak from a high-mobility electron pocket of double Weyl quasi-particles, and the temperature dependence of the spectral weight is consistent with its Fermi velocity.  By subtracting the low-frequency sharp Drude and phonon peaks at low temperatures, we reveal two intermediate quasi-linear inter-band contributions separated by a kink at 0.2 eV. Using Wannier tight-binding models based on first-principle calculations, we link the optical conductivity above and below 0.2 eV to interband transitions near the double Weyl fermion  and a threefold fermion, respectively. We analyze and determine the chemical potential relative to the energy of the threefold fermion, revealing the importance of transitions between a linearly dispersing band and a flat band.
More strikingly, below 0.1 eV our data are best explained if spin-orbit coupling is included, suggesting that at these energies the optical response is governed by transitions between a previously unobserved four-fold spin-3/2 node and a Weyl node. Our comprehensive combined experimental and theoretical study provides a way to resolve different types of multifold fermions in CoSi at different energy. More broadly our results provide the necessary basis to interpret the burgeoning set of optical and transport experiments in chiral topological semimetals.
\end{abstract}

\maketitle

%
%

Topological semimetals are metals defined by topologically-protected degeneracies. In the solid state, their simplest realization features two bands that cross at a single node~\cite{MurakamiNJP07, WanPRB2011, BurkovPRL2011}, known as a Weyl node. Weyl nodes are degeneracies not protected by the crystalline symmetry, and the excitations around them behave as spin-$1/2$ quasiparticles, with a linear relationship between energy and momentum. They are found in a family of non-centrosymmetric transition monopnictides, including TaAs, TaP, NbAs and NbP~\cite{WengPRX2015, HuangNatComm2015, XuScience2015, LvPRX2015, lvNatPhys2015,xuNatphys2015, YangNatPhys2015,xuNatComm2016}, as well as in the magnetic compounds Co$_2$MnGa~\cite{belopolskiScience2019} and Co$_3$Sn$_2$S$_2$~\cite{liuScience2019, moraliScience2019}.

In general, crystal symmetries can protect band crossings with higher degeneracies at high-symmetry points, known as multifold crossings, around which the bands disperse linearly~\cite{wiederPRL2016, bradlynScience2016}. In particular, chiral crystals with non-symmorphic symmetries that lack inversion and mirror symmetry were predicted to realize a variety of multifold crossings: three-, four-, or sixfold crossings. The dispersion close to each multifold degeneracy is described by a higher-spin quasiparticle, such as a spin-1 fermion in the case of a three-fold crossing.  Analogous to the case of Weyl nodes, multifold nodes are monopoles of Berry curvature with integer charge. However, the topological charge (i.e., Chern number) of multifold crossings is higher than that of Weyl nodes, which have topological charge $\pm1$. Remarkably, multifold fermions have been shown to exist in the chiral semimetals CoSi, RhSi, PtAl, and PdGa, all of which crystallize in the chiral space group 198~\cite{changPRL2017, tangPRL2017, raoNature2019, sanchezPRL2019,takanePRL2019,schroterNatPhys2019, schroterArxiv2019}. CoSi and RhSi host a three-fold spin-$1$ fermion at the zone center (the $\Gamma$ point), which is expected to weakly split by the spin-orbit interaction into a fourfold spin-$3/2$ and a twofold spin-$1/2$ Weyl fermion. At the zone boundary (the $R$ point) they host a double spin-1/2 Weyl fermion, which is expected to split into a sixfold spin-1 fermion and a twofold degenerate Kramers pair \cite{changPRL2017, tangPRL2017}. However, the splitting of the spin 3/2 multifold and the Weyl node were not resolved in the previous photo-emission experiments due to the small energy scales that are involved~\cite{raoNature2019, sanchezPRL2019,takanePRL2019}. The three-fold nodes in these materials are fundamentally different from the triply degenerate point in achiral materials such as in MoP~\cite{lvNature2017}, which have zero Chern number and occur away from time-reversal invariant momenta~\cite{wengPRB2016, wengPRB22016,zhuPRX2016,changSciRep2017,lvNature2017}.

The multifold fermions in chiral topological semimetals are responsible for unusual and interesting optical responses such as gyrotropy~\cite{ma2015PRB, zhongPRL2016} and the quantized circular photogalvanic effect~\cite{dejuanNatComm2017,changPRL2017, flickerPRB2018, dejuanarxiv2019, changNatMat2018}.  The optical conductivity is a particularly useful tool to probe the multifold node at the zone center, since it is expected to dominate the response at low frequencies~\cite{sanchezPRB2019}.
An accurate measurement of the optical conductivity is also essential to precisely extract nonlinear optical responses such as second harmonic generation~\cite{wuNatPhys2017, patankarPRB2018} and photogalvanic effect precisely ~\cite{maNatPhys2017, osterhoudtNatMat2019, jiNatMat2019, maNatMat2019,  siricaPRL2019, reesarxiv2019, gaoNatComm2020}. Clarifying the nonlinear responses in these non-centrosymmetric topological semimetals acts to certify the presence and energy range where topological nodes are active~\cite{Ni2020arXivCPGE, Ni2020arXiv, reesarxiv2019}. Determining the carrier lifetime and energy range at which topological crossings are activated is the key to observing the quantized circularly photogalvanic effect, and to using them in the next generation of efficient topological optoelectronics \cite{dejuanNatComm2017}.

Linearly dispersing bands result in a characteristic low-frequency optical conductivity given by
$\sigma_1(\omega) \sim \omega^{ d-2}$, where $d$ is the spatial dimension~\cite{bacsiPRB2013}. Although the expected frequency-independent conductivity was observed in graphite \cite{kuzmenkoPRL2008} and graphene~\cite{nairScience2008}, and the observation of a linear $\omega$-dependence characteristic of three-dimensional linear bands~\cite{hosurPRL2012} is often challenging. The latter has been the subject of various experimental studies, for example in the Dirac semimetals Cd$_3$As$_2$~\cite{akrapPRL2016,neubauerPRB2016} and ZrTe$_5$~\cite{chenPRB2015,xuPRL2018, martinoPRL2019}, and the Weyl semimetal TaAs~\cite{xuPRB2016,kimuraPRB2017}. However, in Cd$_3$As$_2$ the band structure at low energy deviates from linearity, resulting in a sub-linear frequency dependence~\cite{akrapPRL2016}. In ZrTe$_5$, the band structure is quite sensitive to the growth method, and the bands can disperse quadratically along one direction, leading to a different frequency dependence of the optical conductivity at low temperature~\cite{chenPRB2015,xuPRL2018, martinoPRL2019}. In the case of TaAs, coexistence of trivial electron and hole pockets and the small energy difference between two kinds of Weyl nodes complicates the data analysis~\cite{xuPRB2016,kimuraPRB2017}. In RhSi, despite promising initial work~\cite{reesarxiv2019, maulanaPRR2020}, good agreement between theory and experiment on the linear scaling of the conductivity, which would  signal multifold fermions, has remained elusive. Furthermore, in many materials studied previously the Lifshitz energy, i.e. the energy below which topological quasi-particles are excited, is so low that it could not be resolved \cite{akrapPRL2016,neubauerPRB2016,chenPRB2015,xuPRL2018, martinoPRL2019,xuPRB2016,kimuraPRB2017}, a circumstance which makes these systems not ideal for studying linear optical conductivity, in particular the effect of spin-orbit coupling or correlation on the optical conductivity response.

The cubic chiral crystal CoSi is a promising material to reveal the signature of multifold fermions in an optical conductivity experiment. It has a low carrier density, a large Lifshiz energy of $\sim$ 0.6 eV \cite{tangPRL2017, takanePRL2019, raoNature2019, sanchez2019topological} and the multifold nodes located at the zone center and zone boundary are significantly split by an energy difference $\approx$ 0.2~eV, with no other bands expected at the Fermi level~\cite{raoNature2019, sanchezPRL2019,takanePRL2019}. Below 0.2 eV, inter-band excitations near the nodes at the zone boundary are expected to be Pauli-blocked, leaving only the linearly dispersing multifold fermions at the zone center~\cite{sanchezPRB2019, habePRB2019}. These properties motivate the choice of CoSi as an ideal candidate to display a clean linear relation between the conductivity and frequency, and also to reveal why and how deviations could occur. Optical conductivity on CoSi was first measured more than two decades ago, but an understanding of neither the Drude response nor the interband excitations was provided \cite{VANDERMAREL1998, MenaPRB2006}, most likely because it is a weakly correlated semimetal, as evidenced by a normal metallic Sommerfeld constant \cite{petrovaPRB2010}. Most importantly, the topological properties of CoSi and their implications for the optical response were not addressed, since the topology of CoSi was not known until very recently \cite{tangPRL2017, takanePRL2019, raoNature2019, sanchez2019topological}.

In the present work, we used Fourier-transform infrared (FTIR) spectroscopy to measure the conductivity over a broad range of 40 to 50\,000\icm, with temperature dependence from 10 K to 300 K. Based on improvements of the sample quality, we observed a Drude peak width of 2 meV and low onset energy of interband excitation of 20 meV, which are both around one order of magnitude lower than previous works \cite{VANDERMAREL1998, MenaPRB2006}.  Due to such improvement, we establish that the optical conductivity in CoSi is determined by the existence of multifold fermions at low frequencies. The temperature dependent spectral weight of the Drude response is consistent with the Fermi velocity of the  electron pocket consisting of the double Weyl quasi-particles. By subtracting a single sharp Drude peak and four narrow phonon peaks from the real part of the conductivity at 10 K, we reveal two approximately linear conductivity regimes separated by a kink at $\approx 0.2$ eV, culminating in a sharp peak at 0.56 eV. Using density functional theory (DFT) and tight-binding calculations, we link the two linear regimes to multifold excitations close to the $\Gamma$ ($\omega<0.2$ eV) and $R$ ($\omega> 0.2$ eV) points, and the peak to a saddle point in the band structure at the $M$ point. Crucially, below 0.2 eV our calculations and measurements reveal a slight deviation from a perfect linear conductivity, consistent with a chemical potential that crosses a nearly flat band of a threefold node at the zone center. This observation, combined with the band splitting due to spin-orbit coupling, suggests that the optical transitions below 0.2 eV involve the spin-3/2 and a spin-1/2 Weyl nodes at the zone center. Our work highlights that the location of the chemical potential with respect to the multifold fermion is critical to interpreting future optical and transport responses of multifold fermion materials, due to the presence of a threefold fermion with an approximately flat band at the zone center.

\begin{figure*}
\includegraphics[width=0.98\textwidth]{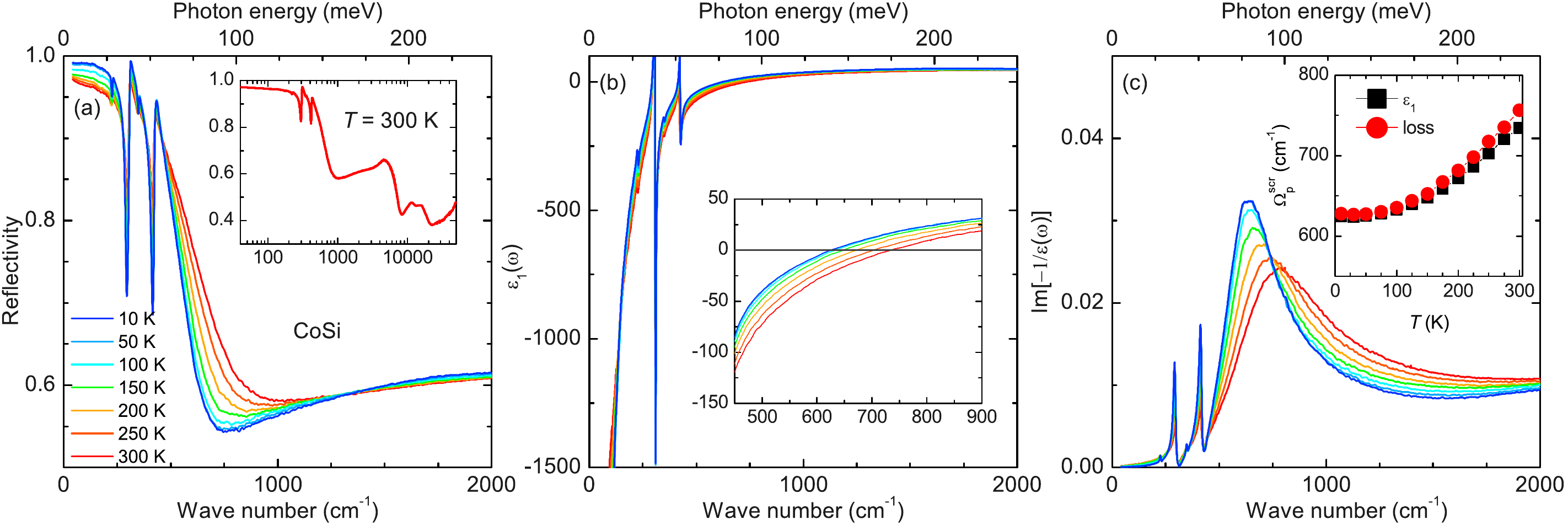}
\caption{(a) Temperature-dependent reflectivity spectra of a (001)-oriented CoSi crystal grown by CVT. The inset shows the  reflectivity from  40 to 50\,000\icm\ at room temperature.  (b) Temperature dependence of the real part of the dielectric function $\varepsilon_1(\omega)$. The inset shows an enlarged view to emphasize the zero crossings of $\varepsilon_1(\omega)$, which correspond to screened plasma frequencies at different temperatures. (c) Temperature-dependent loss function. The inset shows the screened plasma frequency of free carriers obtained from the zero crossings of $\varepsilon_1(\omega)$ and the peak in the loss function as a function of temperature.}
\label{Fig1}
\end{figure*}

\begin{figure}
\centering
\includegraphics[width=0.95\linewidth]{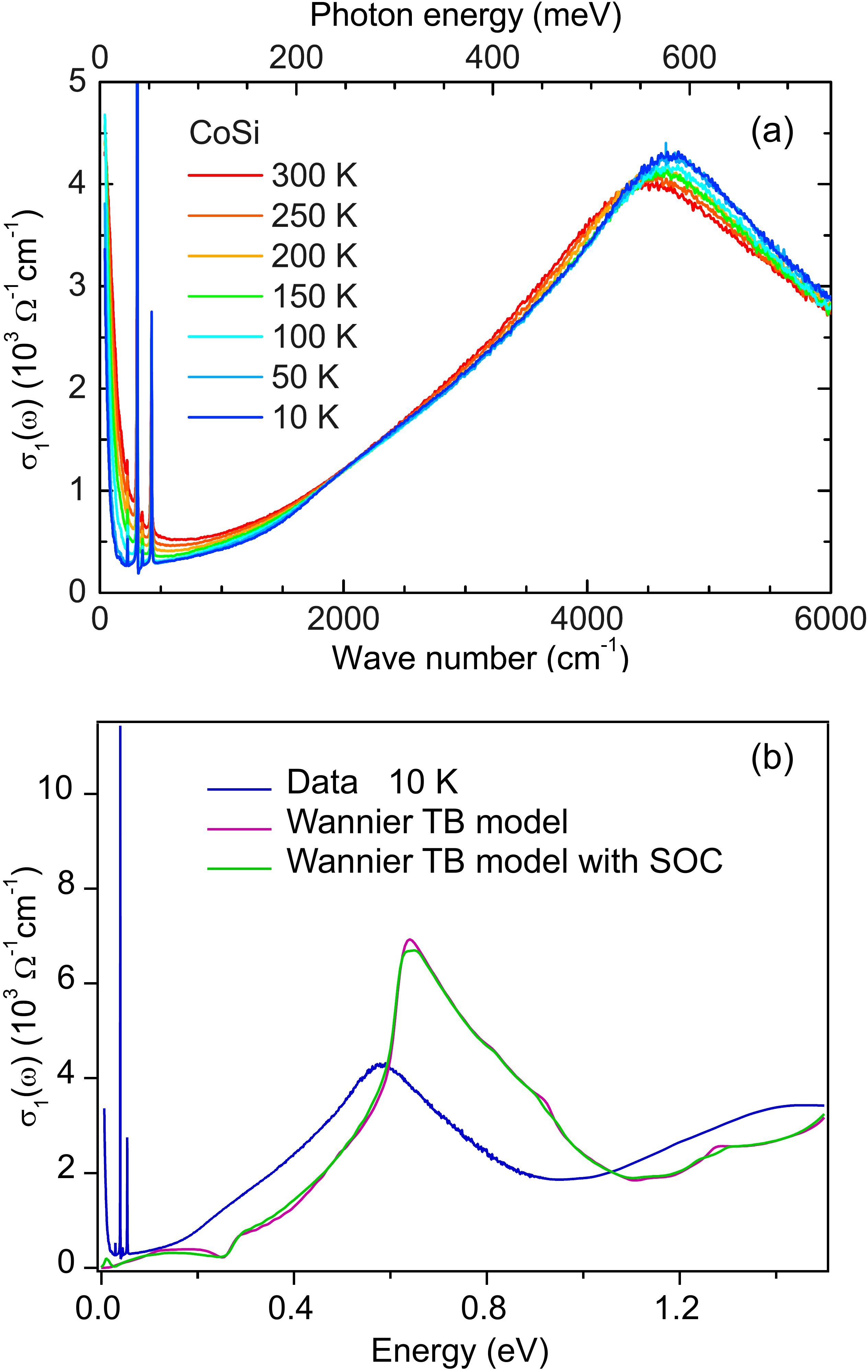}
\caption{ (a) Temperature-dependent optical conductivity spectra $\sigma_1(\omega)$ of a CVT-grown CoSi (001) crystal.   (b) Measured optical conductivity at 10 K along with the Wannier tight-binding (TB) calculation with and without spin-orbit coupling at 10 K.  }
\label{Fig2}
\end{figure}

%
%
%
High-quality single crystals of CoSi have been synthesized with a chemical vapor transport (CVT) method~\cite{sanchez2019topological} and a flux method~\cite{xuXPRB2019}. Since it was found that the Drude response is significantly sharper for unpolished samples~\cite{xuPRB2016, kimuraPRB2017}, all of the reflectivity measurements on CoSi were performed on as-grown flat shiny facets. The data in Fig.~\ref{Fig1}, Fig.~\ref{Fig2} and Fig.~\ref{Fig4} in the main text are all measured on the (001) facet of CoSi grown by CVT. Part of the data measured on the flux-grown sample is shown in Fig.~\ref{Fig5}. (A complete data set on the flux-grown sample is shown in supplementary information (SI) Appendix A.) The facet direction is confirmed by X-ray  Laue diffraction and second-harmonic generation measurement~\cite{wuNatPhys2017}.

%

Fig.~\ref{Fig1}(a) shows the measured temperature-dependent reflectivity spectra $R(\omega)$ of CoSi over a wide frequency range. In the low-frequency region ($< 1000$~\icm ), $R(\omega)$ shows a typical metallic response with a rather sharp plasma edge, below which it rapidly approaches unity. The plasma edge exhibits a rather strong temperature dependence. As the temperature decreases, it shifts continuously towards lower frequency and becomes steeper, indicating reductions in both the carrier density and scattering rate. The low value of the plasma edge ($\approx 700$~\icm) generally suggests a very small carrier density, consistent with the tiny Fermi surface in this material~\cite{takanePRL2019,raoNature2019,xuXPRB2019}. In addition, we identify four sharper phonon peaks that are located at about 225, 305, 345, and 420\icm and a stronger temperature dependence of the plasma edge than previous works \cite{VANDERMAREL1998, MenaPRB2006}.

Fig.~\ref{Fig1}(b) shows the real part of the dielectric function $\varepsilon_1(\omega)$. At low frequencies, $\varepsilon_1(\omega)$ is negative (a defining property of a metal) and can be described by the Drude model $\varepsilon(\omega) = \varepsilon_{\infty} - \omega_p^2/(\omega^2 + i\omega\gamma)$, where $\varepsilon_{\infty}$ is the high-frequency dielectric constant, $\omega_p$ is the Drude plasma frequency, and $\gamma$ is the electronic scattering rate. The inset shows an enlarged view to emphasize the zero crossing of $\varepsilon_1(\omega)$ at different temperatures. The zero crossing of $\varepsilon_1(\omega)$ corresponds to the screened plasma frequency $\omega_p^{\rm scr}$ of free carriers, which is related to the Drude plasma frequency through $\omega_p^{\rm scr} = \omega_p/\sqrt{\varepsilon_{\infty}}$. Fig.~\ref{Fig1}(c) shows the temperature-dependent loss function with the peak around  700~\icm\ being the screened plasma frequency. We observe that the temperature dependence of $\omega_p^{\rm scr}$ in the dielectric function and the loss function agree well with the minimum of the plasma edge of $R(\omega)$ $\approx 700$~\icm. As shown in the inset of Fig.~\ref{Fig1}(c), the screened plasma frequency decreases from about 740\icm\ at 300~K toward 625\icm\ at 10~K.

Fig.~\ref{Fig2}(a) displays the temperature dependence of the real part of the optical conductivity $\sigma_1(\omega)$ in the infrared range. In the low-frequency region, the free-carrier contribution to $\sigma_1(\omega)$ is seen as a Drude-like peak centered at zero frequency. Upon cooling, the Drude peak becomes narrower and loses spectral weight, implying that both the quasiparticle scattering rate and carrier density drop with decreasing temperature. This is consistent with the reflectivity analysis shown in Fig.~\ref{Fig1}. Along with the tail of the Drude peak, the inter-band optical conductivity  increases approximately linearly with $\omega$ (up to about 2\,000\icm\ at 300~K). More interestingly, at low temperatures an upturn kink emerges around 1\,600\icm\ (0.2~eV) in the spectrum of $\sigma_1(\omega)$, resulting in two regions of quasilinear behavior at higher and lower frequencies. The upturn of $\sigma_1(\omega)$ indicates that new inter-band excitations become allowed above $\approx$ 0.2~eV.  Fig.~\ref{Fig2}(b) shows the $\sigma_1(\omega)$ spectrum at 10 K on a larger energy scale up to 1.5~eV. It reveals a sharp peak at 0.56 eV  that is most likely arises from vertical transitions at the corresponding energy, between bands which disperse nearly parallel in a large region of momentum space.

\begin{figure}
\centering
\includegraphics[width=0.49\textwidth]{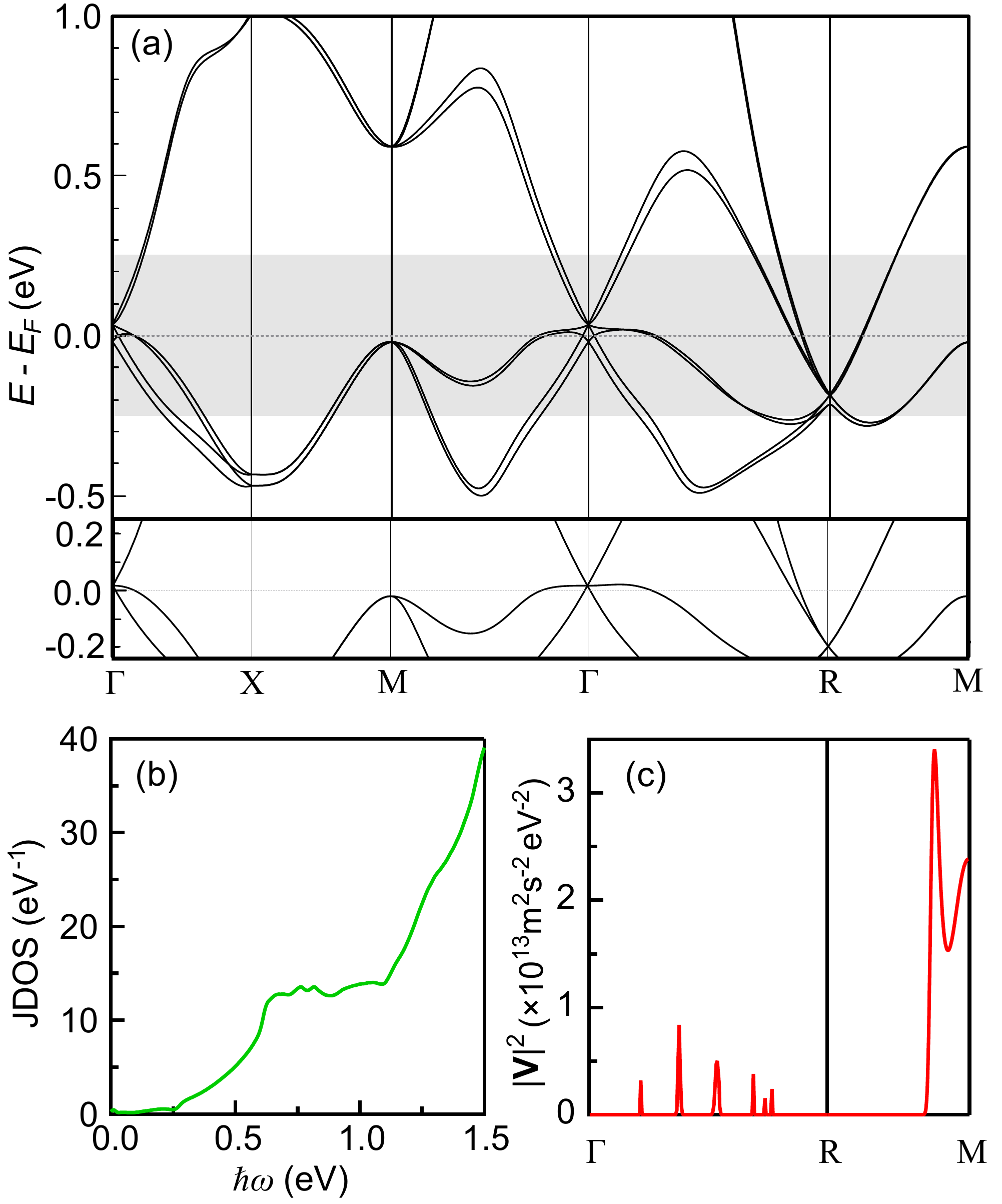}
\caption{ (a) Band structure of CoSi with spin-orbit coupling. The bottom inset shows the band structure without spin-orbit coupling in the same shaded energy window of (a). (b) Joint density of states as a function of energy.  (c) Momentum-resolved matrix element that contributes to the 0.62 eV inter-band transition along $\Gamma-R-M$ direction. }
\label{Fig3}
\end{figure}

\begin{figure}
\includegraphics[width=0.49\textwidth]{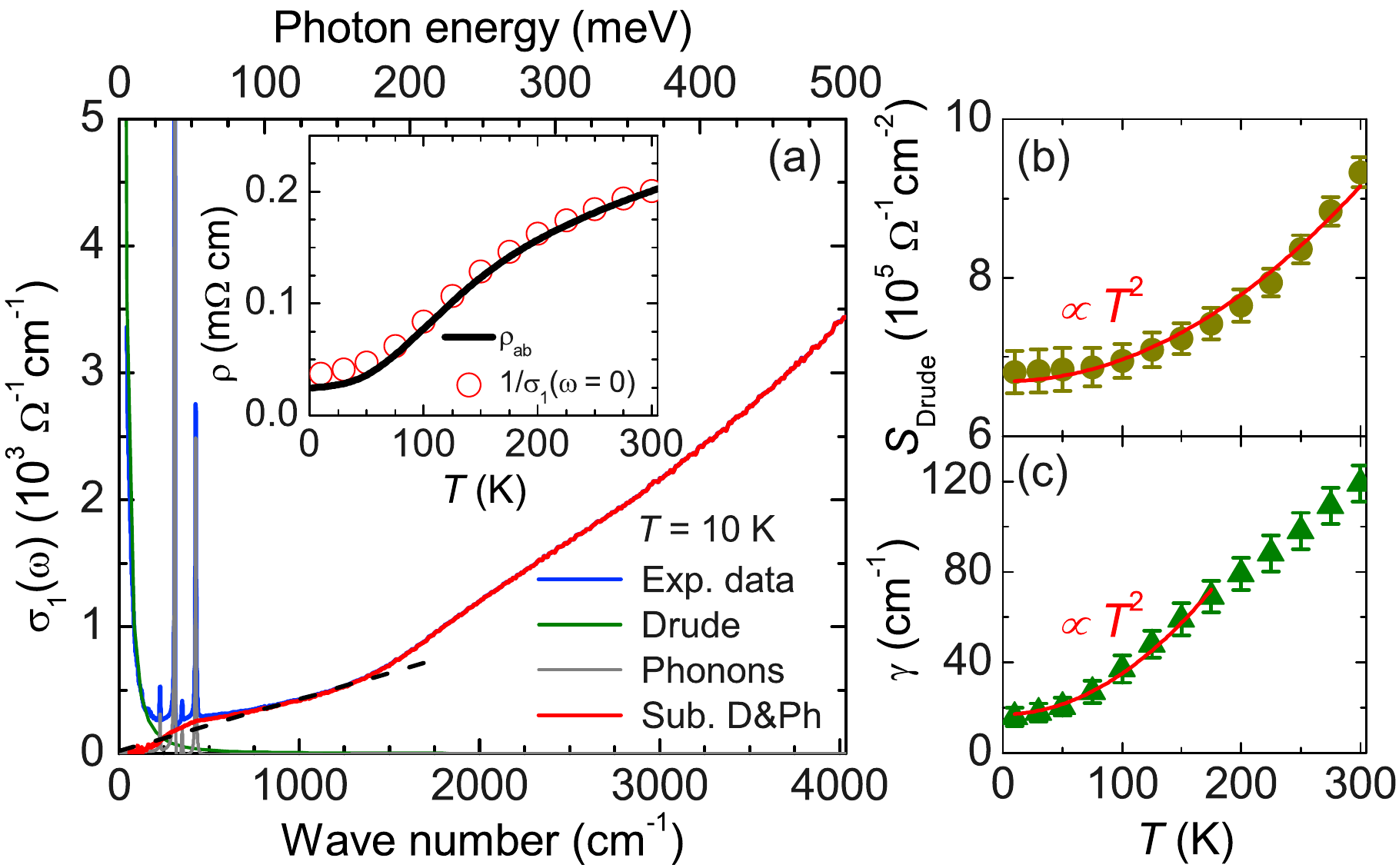}
\caption{(a) Optical conductivity of CoSi at 10~K (blue); thin solid lines represent fit contributions of the Drude term (green) and phonon modes (grey), as well as the remaining inter-band contribution (subtracting Drude and phonons) to $\sigma_1(\omega)$ (red). The black dashed line indicates the linear conductivity below 0.2 eV. Inset: Comparison of the dc resistivity, $\rho_{ab}$ (solid black line), with the zero-frequency values of the Drude fits to the conductivity data, $1/\sigma_1(\omega = 0)$ (red circles). Temperature dependences of (b) the Drude weight and (c) the electronic scattering rate.}
\label{Fig4}
\end{figure}

To determine the origin of the observed peak, we performed first-principles DFT band structure calculations (See Methods). We find the relaxed CoSi lattice constant to be $a=4.485$~\AA, which matches well with previous theoretical and experimental work~\cite{petrovaPRB2010, SeverinJPCM2018, takanePRL2019, habePRB2019}. The calculated electronic band structure is shown in Fig.~\ref{Fig3}~(a), also in agreement with previous reports~\cite{tangPRL2017, SeverinJPCM2018, habePRB2019}. The bottom inset of Fig.~\ref{Fig3}(a) shows the electronic bands calculated without  spin-orbit coupling. As previously reported~\cite{tangPRL2017}, in the absence of spin-orbit coupling, CoSi hosts a three-fold spin-1 fermion at $\Gamma$ and a double Weyl fermion at the zone-boundary $R$ point, which may be viewed as two degenerate spin-$1/2$ Weyl fermions with equal Chern number. The effect of spin-orbit coupling is to split the three-fold node at $\Gamma$ into a four-fold spin-$3/2$ fermion and a two-fold Weyl fermion, as expected from the addition rules for angular momenta\cite{tangPRL2017, changPRL2017}. At the $R$ point, the double Weyl node splits into a six-fold multifold fermion, which can be viewed as a double spin-$1$ multifold fermion, and a twofold degenerate Kramers pair. The energetic splitting of the multifold nodes is a measure of the strength of spin-orbit coupling when compared to the bandwidth; at $\Gamma$ we find a spin-orbit splitting of $\approx 18.1$ meV. Note that the band structure exhibits a saddle point at the $M$ point.

Furthermore, we calculated the maximally-localized Wannier functions that represent the DFT valence Bloch states using Wannier90~\cite{Mostofi2014}, and we constructed a tight-binding model in this Wannier function basis by fitting to the DFT band structure.  We use this model to calculate the inter-band contribution to the optical conductivity, defined as
\begin{equation}
        \sigma^{ab} (\hbar \omega) = \pi e^2 \hbar \sum_{m\neq n} \int \frac{d^3 \bk}{(2\pi)^3} \frac{f_{nm}}{E_{nm}} v_{nm}^a v_{mn}^b \delta (E_{mn} - \hbar    \omega).  \label{eq:conductivity_def}
\end{equation}
where $f_{nm}(\bk) = f_n(\bk) - f_m(\bk)$ is the occupation difference between Bloch states with band indices $n$ and $m$ at momentum $\bk$, $E_{nm}(\bk) = E_n(\bk) - E_m(\bk)$ is the energy difference between these two bands, and $v_{nm}^a (\bk)$ is the velocity matrix element along the $a$-direction. With these Wannier orbitals, this model can describe the bands up to $1.5$~eV above the Fermi level accurately. As a consequence of cubic symmetry, the conductivity tensor $ \sigma^{ab} (\hbar \omega) $ has only one independent component.

The optical conductivity, calculated for the band structure shown in Fig.~\ref{Fig3}(a), is shown in Fig.~\ref{Fig2}(b), in addition to the experimental result at 10 K. For comparison, we present the optical conductivity calculation with and without spin-orbit coupling with a Gaussian broadening of 5 meV, which is close to the  Drude peak width at low temperature. (See Appendices B and C for more details.) The difference between the two calculations is relatively small, as the spin-orbit coupling strength is weak ($\approx$ 20 meV) in this material. This suggests that a model without spin-orbit coupling is sufficient to describe the coarse features, broader than this energy scale, while spin-orbit coupling is needed to describe finer structure of the response. As mentioned above, the calculation is restricted to the inter-band contribution to the conductivity, whereas the experimental measurement also shows the (intra-band) Drude response, as well as the phonon contribution in the low-energy regime.

Let us first focus on the measured peak around 0.6~eV, which is only due to inter-band transitions. The calculated optical conductivity shows a peak at $\omega \approx 0.62$~eV, the position of which matches well with the experimentally-observed peak. Contributions to the conductivity peak come from all transitions at the peak energy range which are not Pauli-blocked and have nonzero velocity matrix elements.

The joint density of states (JDOS), shown in Fig.~\ref{Fig3}(b) for the case of vanishing spin-orbit coupling, is a measure of the number of transitions at a given energy, and is thus a significant indicator for the origins of the observed peak. As is clear from Fig.~\ref{Fig3}(b), the JDOS exhibits a shoulder-like feature at $\approx 0.62$ eV, characteristic of a saddle point, which suggests that the inter-band contributions primarily originate from the $M$ point. The JDOS is a momentum-integrated quantity; therefore, to unambiguously determine which inter-band transitions give rise to the peak, we plot the (momentum-resolved) integrand of Eq.~\eqref{eq:conductivity_def} in Fig.~\ref{Fig3}(c), and the momentum-resolved contribution in SI Appendix D. In particular, Fig.~\ref{Fig3}(c) we show the quantity $ |V(\bk)|^{2}=\sum_{n\neq m}  \frac{f_{nm}(\bk)}{E_{mn}} |v_{nm}^a (\bk) |^2 \delta (E_{mn} - \hbar    \omega) $ for fixed energy $\hbar \omega =0.62$ eV along $\Gamma-R$ and $R-M$, which is a measure of the matrix elements of all allowed transitions at the peak energy. Fig.~\ref{Fig3}(c) clearly confirms that the inter-band transitions giving rise to the peak originate mainly from the vicinity of the $M$ point, which was assigned to a different origin in previous works~\cite{reesarxiv2019, habePRB2019, maulanaPRR2020}. Note that the calculated peak in Fig. \ref{Fig2}(b) becomes lower and wider with increasing broadening  (See SI Appendix C).

A second feature of the calculated conductivity shown in Fig.~\ref{Fig2}(b) is a dip at around 0.25 eV. This is due to the curvature of the middle band in the three-fold node at the $\Gamma$ point. As shown in Fig.~\ref{Fig3} (a), when the energy of the incoming photon is very small, the allowed transitions should occur between the lower band and the middle band at momenta right near $\Gamma$. Away from $\Gamma$, the middle band curves downward in energy and becomes occupied, thus blocking transitions from the lower band to the middle band and providing the downward dip in the spectrum. As the energy of the incoming photon increases further, the transitions around the $R$ point become activated, providing the recovery from the dip and the continuation of the spectrum upward. However, this dip is not observed in experiments, probably due to a short lifetime of the hot carriers around 0.25 eV as decreasing the quasiparticle lifetime by broadening the Dirac delta function smears out this dip feature~\cite{habePRB2019}. (See SI Appendix C.)

Finally, we focus on the optical conductivity below 0.2~eV. In order to compare the calculation results with experiment below 0.2 eV accurately, we need to extract the inter-band contribution from the experimental data by subtracting the sharp Drude and phonon responses at low energy. In order to quantify the temperature dependence of the Drude response, in Fig.~\ref{Fig4}(a) we fit the low-energy part of $\sigma_1(\omega)$ using the Drude model. The inset of Fig.~\ref{Fig4}(a) shows the dc resistivity $\rho \equiv 1/\sigma_1(\omega = 0)$, derived from the fitted zero-frequency value (open red circles), which agrees well with the transport data (solid black curve), indicating that the subtraction of the Drude response is reliable. As shown in SI Appendix A, the linear Hall resistivity is dominated by a high-mobiliy electron channel, which indicates that the narrow Drude response comes from the electron pocket with Double Weyl quasi-particles at $R$. The corresponding free carrier weight, $S_{\text{Drude}} = \frac{\pi^2}{Z_0}\omega_p^2$, where $Z_0$ is the vacuum impedance, and the scattering rate, $\gamma$, are displayed in Fig.~\ref{Fig4}(b) and Fig.~\ref{Fig4}(c), respectively. The free carrier weight shows a $T^2$ temperature dependence as expected for massless Dirac electrons~\cite{Ashby2014}. The coefficient before the $T^2$ temperature dependence could be used to extract the Fermi velocity of the Dirac fermions (double Weyl fermions in this case) by the relation $\omega_p(T)^2=\omega_{p0}^2+\frac{2\pi}{9v^R_F}\frac{e^2}{\hbar^3}(k_BT)^2$~\cite{Ashby2014,Wang2020NP}. We estimated that $v^R_F$ for the double Weyl fermions is around 1.4 $\times$ 10$^5$ m/s, which agrees with the DFT calculation in Fig. \ref{Fig3}(a) bottom within 4 $\%$. The scattering rate decreases as the temperature is lowered, following a $T^{2}$ temperature dependence at low temperatures, as shown by the red solid curve through the data points. Moreover, the value of $\gamma$ becomes extremely small at low temperatures, which is at least one order of magnitude narrower than the value reported in RhSi~\cite{reesarxiv2019, maulanaPRR2020,Ni2020arXiv} and in previous studies of CoSi~\cite{VANDERMAREL1998, MenaPRB2006}.

The most striking feature of the optical conductivity is its approximately linear behavior up to $\approx 0.2$ eV, obtained after subtracting the single sharp Drude peak and four narrow phonons, as indicated by the black dashed line in Fig.~\ref{Fig4}(a). After Drude subtraction, the optical conductivity derives from vertical inter-band transitions, and in the case of CoSi we expect that, at low frequencies, these transitions occur between bands associated with the multi-fold fermions. Below $\approx$0.2 eV, vertical transitions at the $R$ point are Pauli blocked, as is clear from Fig.~\ref{Fig3}(a), except for the tiny 11 meV peak shown in Fig.~\ref{Fig2}(b), which is associated with the inter-band transitions between spin-orbit split bands along the $R-M$ line (see SI Appendix E for more discussion). The contribution from these transitions is much smaller, however, than the vertical transitions from the multi-fold fermions at the $\Gamma$ point~\cite{sanchezPRB2019}. Therefore, the linear conductivity below 0.2 eV is mainly attributed to these inter-band transitions near the  $\Gamma$ point.

For multi-fold fermions with linear band dispersion, the optical conductivity at low frequencies (where the linear approximation holds) was shown to have a linear frequency dependence~\cite{sanchezPRB2019}, which can be understood qualitatively from dimensional analysis, $\sigma_1(\omega) \sim \omega^{ d-2}$. In the isotropic limit of the dispersion, i.e., when full rotation symmetry is present, the slope of the conductivity of a single multi-fold fermion only depends on the Fermi velocity $v^{\Gamma}_F$ and is inversely proportional to it~\cite{sanchezPRB2019}.

To quantitatively understand the linear slope of the observed conductivity, and to determine whether it originates from the low-energy multi-fold fermions, we now analyze the low-frequency regime of the optical conductivity in detail. As a first step, we consider a $k\cdot p$ model for the three-fold spin-1 fermion node shown in the lower panel of Fig.~\ref{Fig3}(a) in the absence of spin-orbit coupling. Inter-band transitions are allowed from the partially occupied lower linear band to the central flat band, but are forbidden between the two linearly dispersing bands due to angular momentum selection rules~\cite{sanchezPRB2019}. For this model, the analytic formula for the optical conductivity at $T \rightarrow 0$~K is  $ \frac{e^2 \omega}{3 \pi \hbar v^{\Gamma}_F}\theta(\hbar \omega - E_0)$\cite{sanchezPRB2019}, where $E_0$ is the energy of the three-fold node measured from the Fermi energy.  Using a Fermi velocity $v^{\Gamma}_F$ = 1.9 $\times$ 10$^5$ m/s  from a fit to the linear band in Fig.~\ref{Fig3}(a) (details of the fit are included in SI Appendix F), we plot the analytical conductivity  in Fig. S9 (orange curve) in SI Appendix F, together with the experimental result. The linear model falls lower than the experimental curves for both samples, failing to capture the shoulder-like features around 50 meV in the data, mainly due to the absence of quadratic corrections to the flat band.

\begin{figure}
\centering
\includegraphics[width=0.49\textwidth]{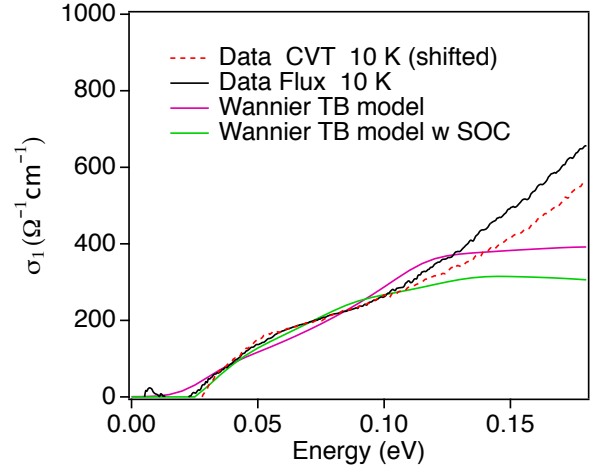}
\caption{Inter-band optical conductivity in CoSi from measurement in both CVT- and flux- grown samples (red and black curves) along with Wannier tight-binding calculations  without and with spin-orbit coupling (SOC). }
\label{Fig5}
\end{figure}

To better understand the origin of the discrepancies, we go beyond the linear model and compare the optical conductivity data to that obtained by the Wannier tight-binding model based on DFT introduced above.  The lower-energy regime of optical conductivity in Fig. \ref{Fig2}(b) of the  Wannier tight-bind models is shown in Fig. \ref{Fig5}.   Overall, this analysis establishes that the chemical potential lies below the threefold node at $\Gamma$ and highlights that deviations from a linear band structure near multifold fermions have an important qualitative impact on the optical conductivity. A simpler four-band tight-binding model for CoSi~\cite{changPRL2017,flickerPRB2018,sanchezPRB2019}, that captures the symmetries of space group 198 with three parameters that we fit the to first-principles band structure in Fig.\ref{Fig3}(a) reaches the same conclusions (see SI Appendix G).

Next, we assess the role of spin-orbit coupling at low frequencies.  The DFT-based Wannier tight-binding model result with spin-orbit coupling (green curve of Fig.~\ref{Fig5}), from which we have subtracted the small peak feature at 11 meV shown in Fig.~\ref{Fig2}(b), since it originates from the inter-band excitations between spin-orbit-split bands along the $R-M$ line (see SI Appendix D). This calculation displays close agreement with the measured conductivity in the flux-grown sample below 0.1 eV. This agreement supports that the low-frequency shoulder below $0.1$ eV observed in experiments arises from transitions between the spin-orbit split linear and flat threefold fermion bands, which are composed by a spin-3/2 fourfold node and a spin-1/2 Weyl node.

Finally, we discuss the distinctions between the two crystals studied. Our calculated conductivity agrees well with the inter-band conductivity data from the flux-grown CoSi sample, while the conductivity data from the CVT-grown sample are  higher at low frequencies (by $\approx$ 80 $\Omega^{-1}$ cm$^{-1}$ below 0.1~eV). It turns out that this difference is not due to the observation that the carrier density in the CVT-grown sample is $\approx$ 10 $\%$ higher than the flux-grown sample by comparing the Drude spectral weight.
As shown in Appendices G and H, assuming a higher chemical potential does not reach as good  agreement as the flux-grown sample. This is an indication that either the additional contribution in the CVT-grown sample is not coming from interband transition~\cite{maulanaPRR2020,Ni2020arXiv} or it is related with surface Fermi arcs as (001) (CVT) sample has surface states but (111) (flux) sample does not~\cite{changPRL2020}.  Nevertheless, the average slopes of both samples and the calculations are comparable after shifting down the experimental result on the CVT sample by $\approx$ 80 $\Omega^{-1}$ cm$^{-1}$ (See Fig.~\ref{Fig5}), suggesting that there is agreement concerning the Fermi velocity (see SI Appendix H for more discussions).

In summary, we revealed the topological origin of the optical conductivity in CoSi based on the development of high-quality crystals. Our analysis shows that locating the chemical potential accurately with respect to the the central flat band is crucial to understanding the optical response. We have shown that both intraband and interband optical conductivity in different frequency regimes are dominated by the existence of different topological multifold  fermions, as well as a band structure saddle point. Most notably, we provide the evidence of the existence of the fourfold spin-3/2  quasi-particles for the first time. Our results not only shed light on the interpretation of circular photogalvanic measurement in this family of materials \cite{dejuanNatComm2017,changPRL2017,Ni2020arXiv,Ni2020arXivCPGE,reesarxiv2019}, but also pave the way to study optical signatures in other chiral topological semimetals in the future \cite{bradlynScience2016,changNatMat2018}. \\

\emph{Materials and Methods.} Millimeter-sized single crystals of CoSi have been synthesized with both a chemical vapor transport method~\cite{sanchez2019topological} and a flux method~\cite{xuXPRB2019}. Temperature-dependent infrared reflectance measurements were performed on  as-grown flat shiny facets using a Bruker VERTEX 70v FTIR spectrometer with an \emph{in situ} gold overfilling technique~\cite{Homes1993}, see SI Appendix, section A for details.
We performed first-principles density functional theory (DFT) band structure calculations with the Quantum Espresso code~\cite{RamerPRB99, RappePRB1990, GiannozziJPCM2009}. We used norm-conserving pseudopotentials generated with the OPIUM code~\cite{Opium}, the  Perdew-Burke-Ernzerhof generalized gradient approximation functional~\cite{PerdewPRL1996}, and spin-orbit coupling to relax the structure and to calculate the band structures.

\emph{Acknowledgments.} We thank  F. de Juan, C.L. Kane and Y. Zhang for helpful discussion. L. W. and N. Z. are supported by Army Research Office under Grant W911NF1910342 (nonlinear optical measurement and data analysis). J.V. and E. J. M. are also supported by a seed grant from Materials Research Science and Engineering Center (MRSEC) at Penn under the Grant DMR-1720530 (theory calculation). A. M. R. was supported by the Department of Energy, Office of Basic Energy Sciences, under Grant No. DE-FG02-07ER46431. Computational support was provided by the National Energy Research Scientific Computing Center (NERSC). B.X and C.B. are supported by the Schweizerische Nationalfonds (SNF) by Grant No. 200020-172611. M. A. S. M acknowledges support from the European Union's Horizon 2020 research and innovation programme under the Marie-Sklodowska-Curie grant agreement No. 754303 and the GreQuE Cofund programme.  A. G. G. is supported by the ANR under the grant ANR-18-CE30-0001-01 (TOPODRIVE) and the European Union Horizon 2020 research and innovation programme under grant agreement No. 829044 (SCHINES). K. M. and C. F. acknowledge financial support from the European Research Council (ERC) Advanced Grant No.742068 ``TOP-MAT'' and Deutsche Forschungsgemeinschaft (Project-ID 258499086 and FE 633\/30-1) (CVT growth). Research (flux growth) at the University of Maryland was supported by the Gordon and Betty Moore Foundation's EPiQS Initiative through Grant No. GBMF9071, and the Maryland Quantum Materials Center. This research was supported in part by the National Science Foundation under Grant No. NSF PHY11-25915. \\

B. X., Z. F. and M. A. S. M contributed equally to this work.

\clearpage
\appendix

\renewcommand{\thefigure}{S\arabic{figure}}

\setcounter{figure}{0}

\section {Reflectivity measurement, Kramers-Kronig analysis, Optical conductivity data on flux-grown sample and Hall resistivity data}

\begin{figure*}
\includegraphics[width=\textwidth]{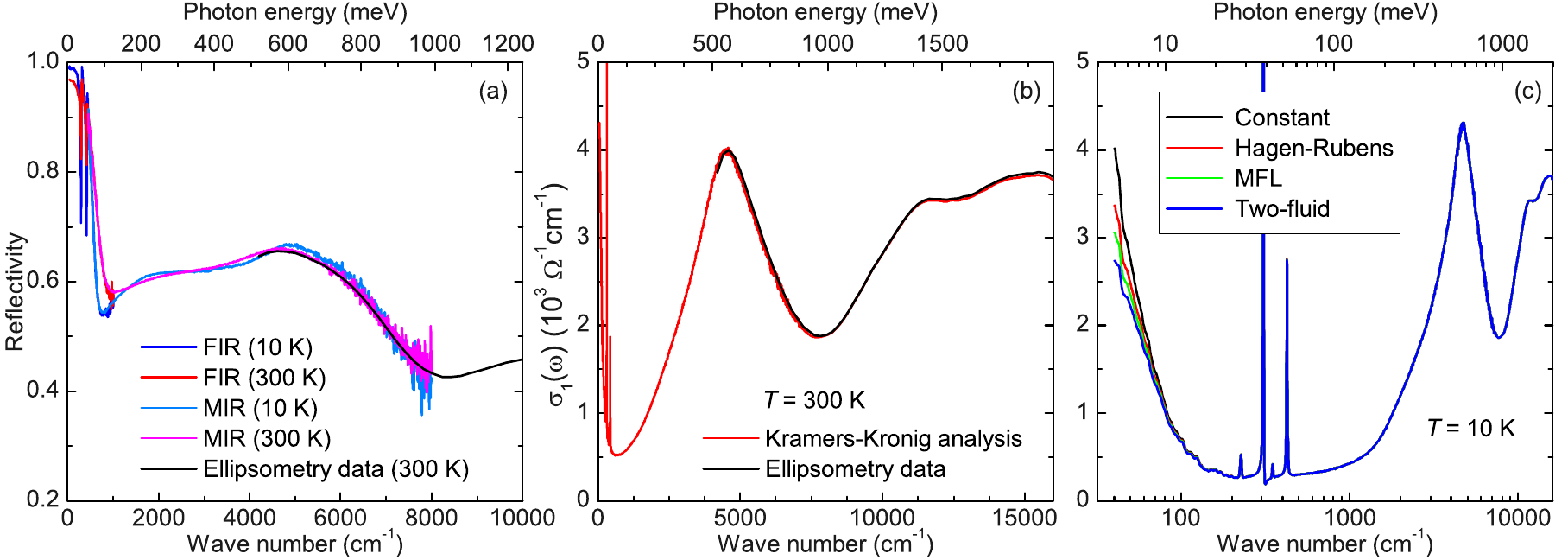}
\caption{ (color online) (a) Reflectivity spectra $R(\omega)$ measured for different frequency ranges with a Fourier transform infrared reflectometer and with a grating-based spectroscopic ellipsometer. (b) Comparison of the optical conductivity spectrum at 300 K as obtained from the ellipsometry data and a Kramers-Kronig analysis of the reflectivity data. (c) The optical conductivity at 10 K obtained by the Kramers-Kronig analysis of $R(\omega)$ with different low-frequency extrapolations, such as Constant ($R(\omega) =$ constant), Hagen-Rubens ($R(\omega) = 1 - A\sqrt{\omega}$), Marginal Fermi Liquid ($R(\omega) = 1 - A\omega$), and Two Fluid ($R(\omega) = 1 - A\omega^2$).}
\label{Fig:KK}
\end{figure*}
The reflectivity $R(\omega)$ of CoSi was measured at a near-normal angle of incidence using a Bruker VERTEX 70v Fourier transform infrared spectrometer. An \emph{in situ} gold overfilling technique~\cite{Homes1993} was used to obtain the absolute reflectivity. As shown in Figure~\ref{Fig:KK}(a), the reflectivity spectra have been measured over a very broad frequency range from 40 to 8\,000\icm\ on a shiny as-grown sample surface by using a series of combinations of sources, beamsplitters and detectors. The reflectivity spectra at different temperatures from 300 to 10~K were collected with an ARS-Helitran cryostat. Fig.~\ref{Fig:KK}(a) also shows the raw spectra measured for different frequency ranges at 10 and 300~K. It is evident that all spectra exhibit the same temperature dependence and are overlapping very well in the region where the spectra have been connected to perform the Kramers-Kronig analysis. This confirms the accuracy and reproducibility of the measured reflectivity spectra.

The optical conductivity $\sigma_1(\omega)$ was obtained from a Kramers-Kronig analysis of $R(\omega)$~\cite{Dressel2002}. For the low-frequency extrapolation we used a Hagen-Rubens function
$R = 1 - 2\sqrt{\frac{2\epsilon_0\omega}{\sigma_{DC}}}$, where $\sigma_{DC}$ is the dc conductivity determined by the temperature-dependent DC four-terminal resistivity data. On the high-frequency side, the ellipsometry data were used, for which the spectrum in the near-infrared to ultraviolet range (4\,000 -- 50\,000\icm) was measured at room temperature with a commercial ellipsometer (Woollam VASE), and the Kramers-Kronig analysis was anchored by the room temperature ellipsometry data, as shown by the spectra in Fig.~\ref{Fig:KK}(b).

To verify the reliability of Kramers-Kronig analysis at low frequencies, in Fig.~\ref{Fig:KK}(c) we show how different low-frequency extrapolations during the Kramers-Kronig analysis of reflectivity influence the obtained spectra of optical conductivity. It shows that the uncertainty is very small, especially in the energy range above 20 meV where we studied the low-energy interband transitions associated with multifold fermions, since our reflectivity data have been measured with high accuracy down to very low energies of about 5 meV.

\begin{figure}
\includegraphics[width=0.8 \linewidth]{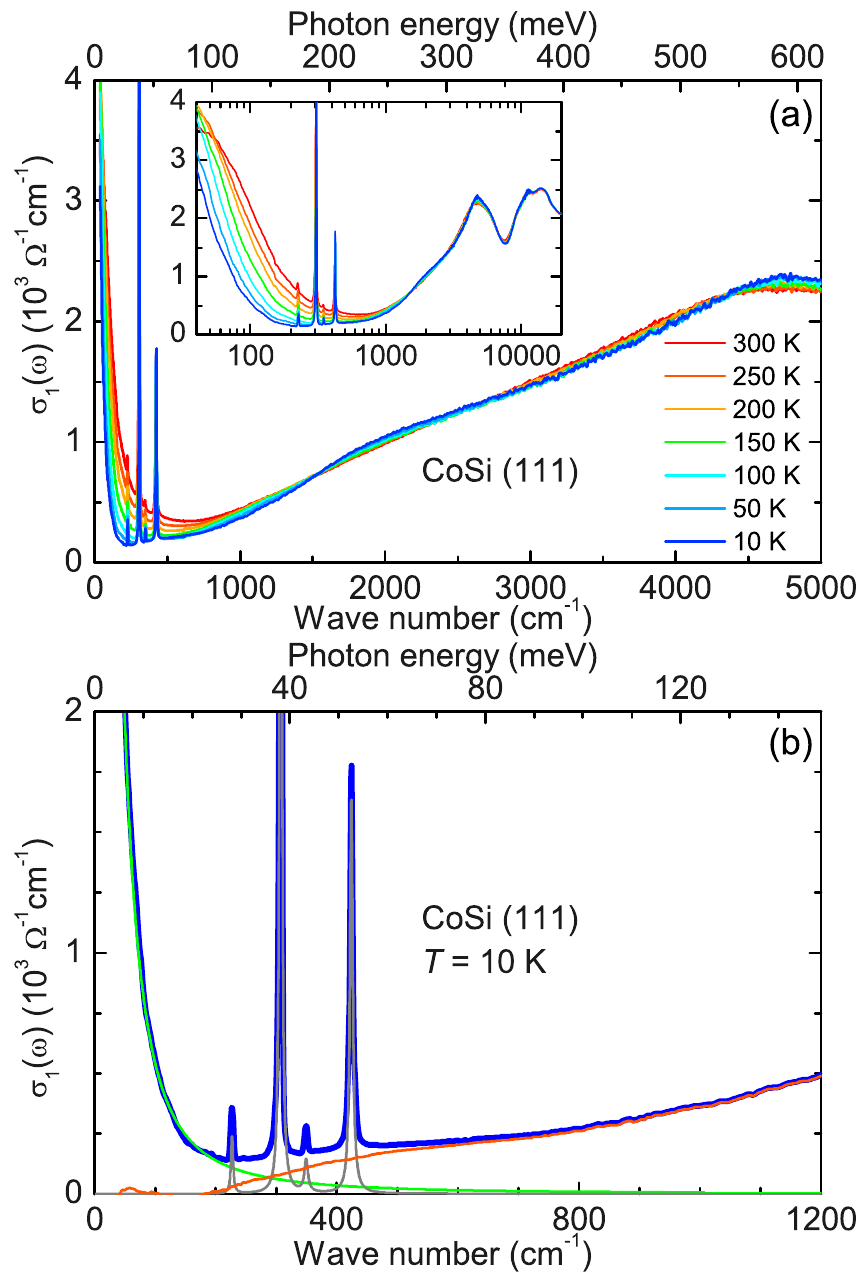}
\caption{(a) Temperature-dependent optical conductivity spectra $\sigma_1(\omega)$ of a flux-grown CoSi (111) sample.  (b) Optical conductivity of CoSi at 10~K (blue); thin solid lines represent fit contributions of the Drude term (green) and phonon modes (grey), as well as the rest of the interband contributions to $\sigma_1(\omega)$ (red).
\label{fig:S1} }
\end{figure}

In Fig.~\ref{fig:S1}, we show the optical conductivity data on the flux-grown CoSi (111) sample, measured under the same conditions as the CVT sample reported in the main text. Because of the cubic symmetry, the conductivity is the same on (001) and (111) facets. The temperature dependent real part of the conductivity is shown in Fig.~\ref{fig:S1}(a). The results are very similar to the CVT sample.  The 10~K curve also shows a kink around 0.2~eV. It also has a peak around 0.6 eV and sharp Drude and phonon peaks at low temperature. In  Fig.~\ref{fig:S1} (b), we show the 10~K conductivity (blue) and the remainder after subtracting single  sharp Drude (green) and four sharp phonon (gray) contributions. The remaining conductivity spectrum is shown in red and is also shown in Fig.~5 in the main text (black).

\begin{figure}
\includegraphics[width=0.95 \linewidth]{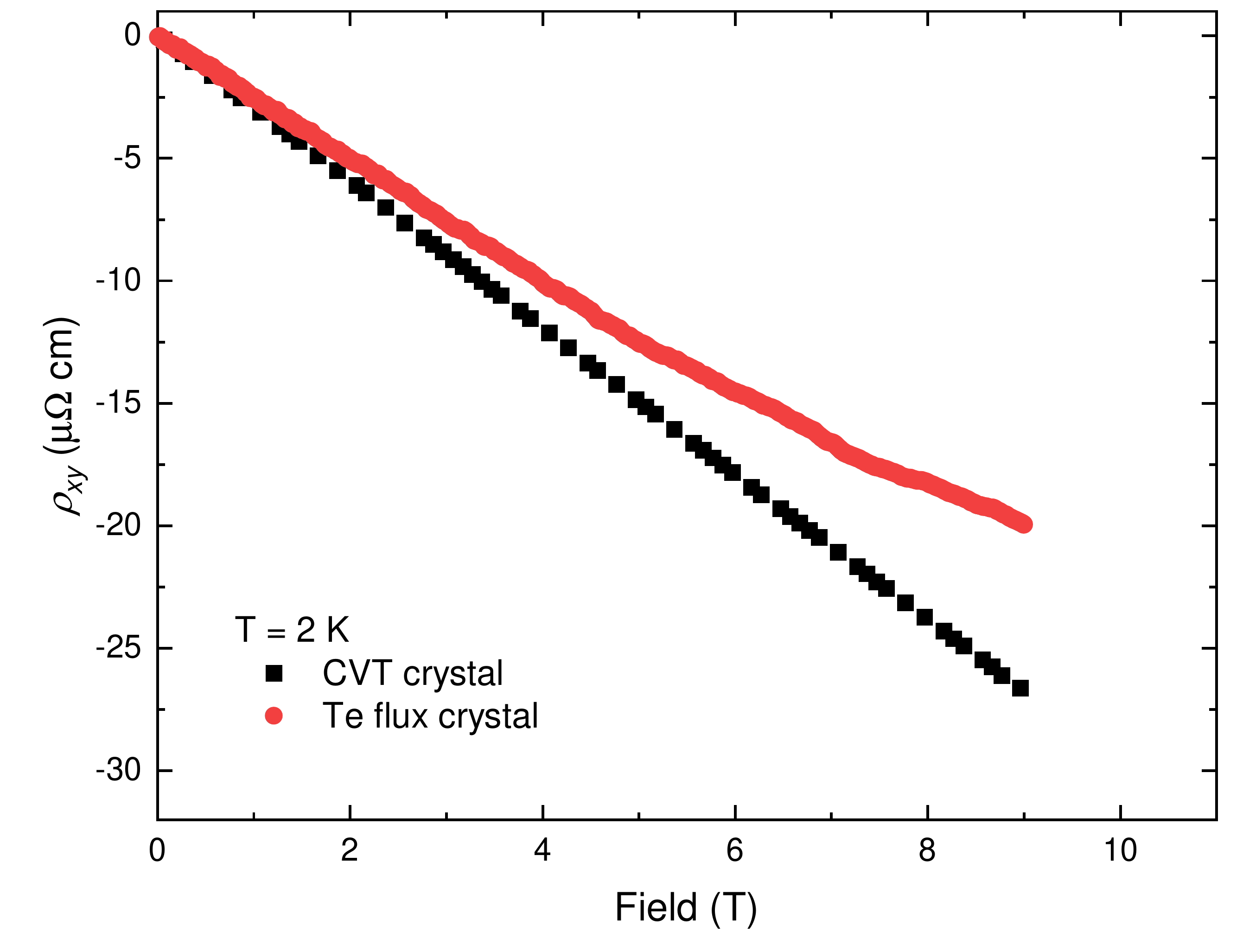}
\caption{ Hall resistivity data on a CVT-grown and a flux-grown CoSi at 2 K.
\label{fig:S2} }
\end{figure}

In Fig.~\ref{fig:S2}, we show the typical Hall resistivity data on a CVT-grown and a flux-grown CoSi at 2 K.  The linear negative Hall resistivity shows the dominating high-mobility electrons from the pocket at the $R$ point. The carrier density  is 2.1 $\times$ 10 $^{20}$ cm$^{-3}$ and 2.8 $\times$ 10 $^{20}$ cm$^{-3}$ for the CVT-grown and flux-grown samples, respectively. The mobility is $\approx$ 2,000 cm$^2$/Vs for the CVT sample while it is $\approx$ 6,700 cm$^2$/Vs for the flux-grown sample as one could see some quantum oscillations at high field in the latter sample. The residual resistivity ratio (RRR) for these two samples are 10 and 34 respectively. Note that these two samples are different from those measured by optical conductivity.

As shown in the inset of Fig. 4(a) of the main text, the RRR in the two samples measured by optical conductivity is $\sim$ 10, while it was 1.5 in previous studies \cite{VANDERMAREL1998, MenaPRB2006}. While the room temperature resistivity is similar, the low-temperature value in our sample is usually one order of magnitude smaller than previous ones \cite{VANDERMAREL1998, MenaPRB2006}.  We also observe much sharper phonon peaks in our crystal, which most likely indicates much better crystal quality in terms of lower carrier density and higher mobility.  The onset of interband excitations in our samples is around 20 meV while it was 125 meV in previous studies \cite{VANDERMAREL1998, MenaPRB2006}, which is consistent  with lower chemical potential (Pauli-blocking energy) in our samples.

\section {Optical conductivity formula}

The complex optical conductivity $\sigma(\omega)$ is defined through
\begin{equation}
j^a (\omega) = \sigma^{ab}(\omega) E^b (\omega),
\end{equation}
where the superscripts $a$ and $b$ represent Cartesian directions, $\bj (\omega)$ is the Fourier component of $\bj (t) = \sum_{\omega} \bj (\omega) e^{-i \omega t}$, and $\bE (\omega)$ is the Fourier component of $\bE (t) = \sum_{\omega} \bE (\omega) e^{-i \omega t}$. Using linear response theory, the complex optical conductivity is expressed as
\begin{equation}
\sigma^{ab} (\omega) = i e^2 \hbar \sum_{m \neq n} \int \frac{d^3 \bk}{8 \pi^3} \frac{f_{nm}}{E_{nm}} \frac{v_{nm}^a v_{mn}^b}{E_{nm} - \hbar \omega + i\eta},
\end{equation}
where $f_{nm} (\bk) = f_n (\bk) - f_m (\bk)$ is the Fermi-Dirac occupation difference between the $n^\text{th}$ and $m^\text{th}$ band at $\bk$, $E_{nm} (\bk) = E_n (\bk) - E_m (\bk)$ is the band energy difference between these two bands, $v_{nm}^a = <n\bk | v^a | m\bk>$ is the velocity matrix element, and $\eta$ accounts phenomenologically for disorder-related broadening.

The real part of the complex linear conductivity, which we refer to as the linear conductivity in this paper, is
\begin{equation}
\sigma^{ab} (\omega) = -\pi e^2 \hbar \sum_{m \neq n} \int \frac{d^3 \bk}{8 \pi^3} \frac{f_{nm}}{E_{nm}} v_{nm}^a v_{mn}^b \delta (E_{nm} - \hbar \omega). \label{eq:conductivity}
\end{equation}
This expression is symmetric under exchange $m \leftrightarrow n$ if $a = b$. Given any positive $\hbar \omega$, the delta function will be nonzero only when $E_{nm} > 0$, or $n > m$, so the summation over bands where $m \neq n$ could be safely replaced by the summation where $n \in \text{unocc}$ and $m \in \text{occ}$, where \text{unocc} and \text{occ} refer to unoccupied and occupied states respectively. The conductivity is also symmetric under the exchange $\omega \leftrightarrow -\omega$. Note that, when computing the optical conductivity from non-SOC bands, the conductivity should be multiplied by a factor of 2 to account for the spin degeneracy of the bands.

\section {Computational details of first-principles calculations}

\begin{figure*}
\includegraphics[width=0.7\linewidth]{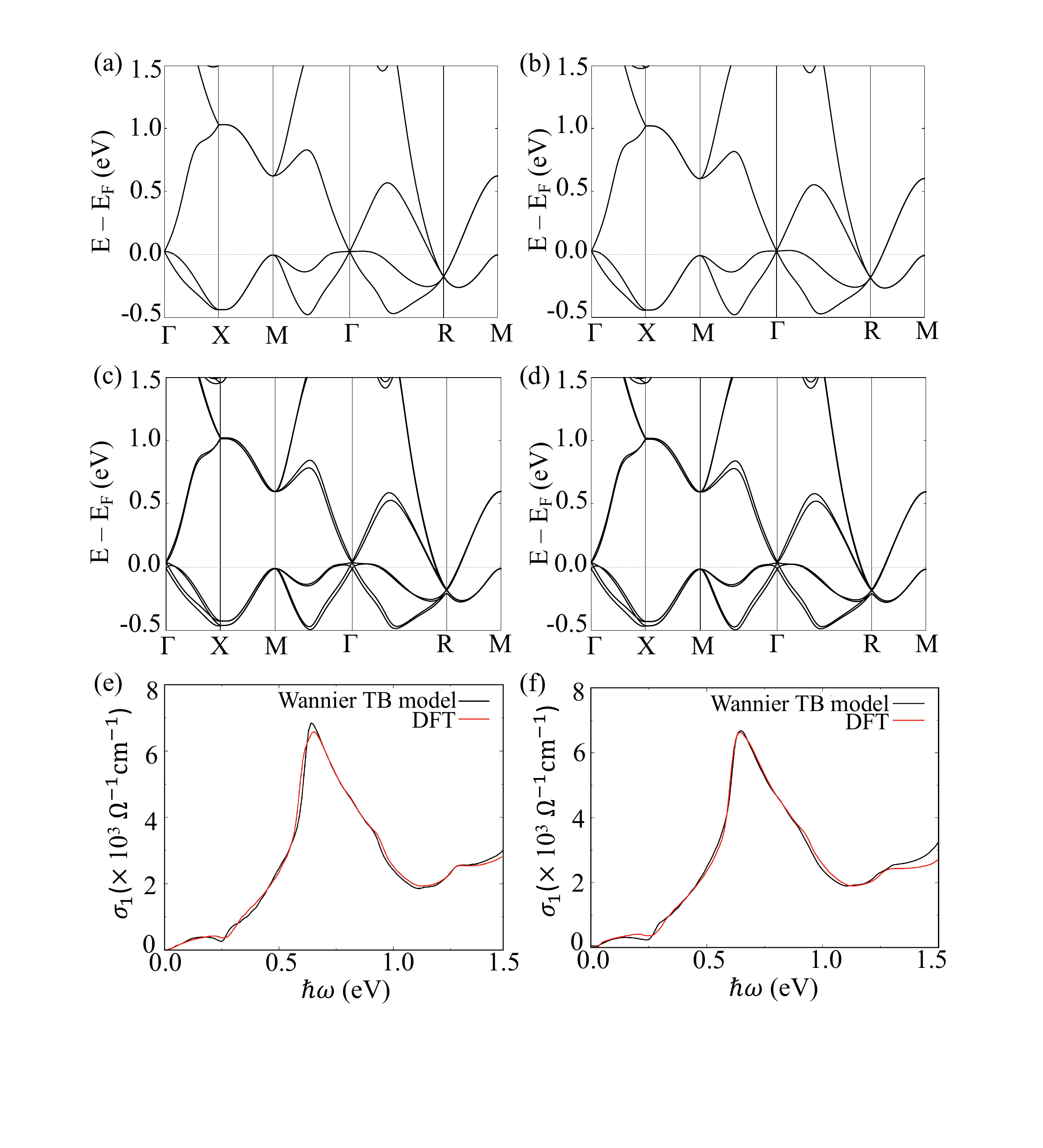}
\caption{(a) The band structure calculated from first-principles calculations without SOC; (b) The band structure calculated from the Wannier tight-binding model without SOC. (c) The band structure calculated from first-principles calculations with SOC. (d) The band structure calculated from Wannier tight-binding model with SOC. (e) Comparison between the conductivity calculated from first-principles calculations with non-SOC bands and that from Wannier tight-binding model; (f) Comparison between the conductivity calculated from first-principles calculations with SOC bands and that from the Wannier tight-binding model. \label{fig:S5} }
\end{figure*}

We performed first-principles density functional theory (DFT) calculations in the software package of Quantum Espresso, with norm-conserving pseudopotentials generated from the OPIUM package. We chose the kinetic energy cutoff for th wavefunctions to be 90 Ry, and we used an $8 \times 8 \times 8$ k-point sampling grid for the crystal structure and band structure calculations. The relaxed lattice constants of CoSi are $a = 4.485$~\AA.

\begin{figure}[tb]
\centering
\includegraphics[width=\linewidth]{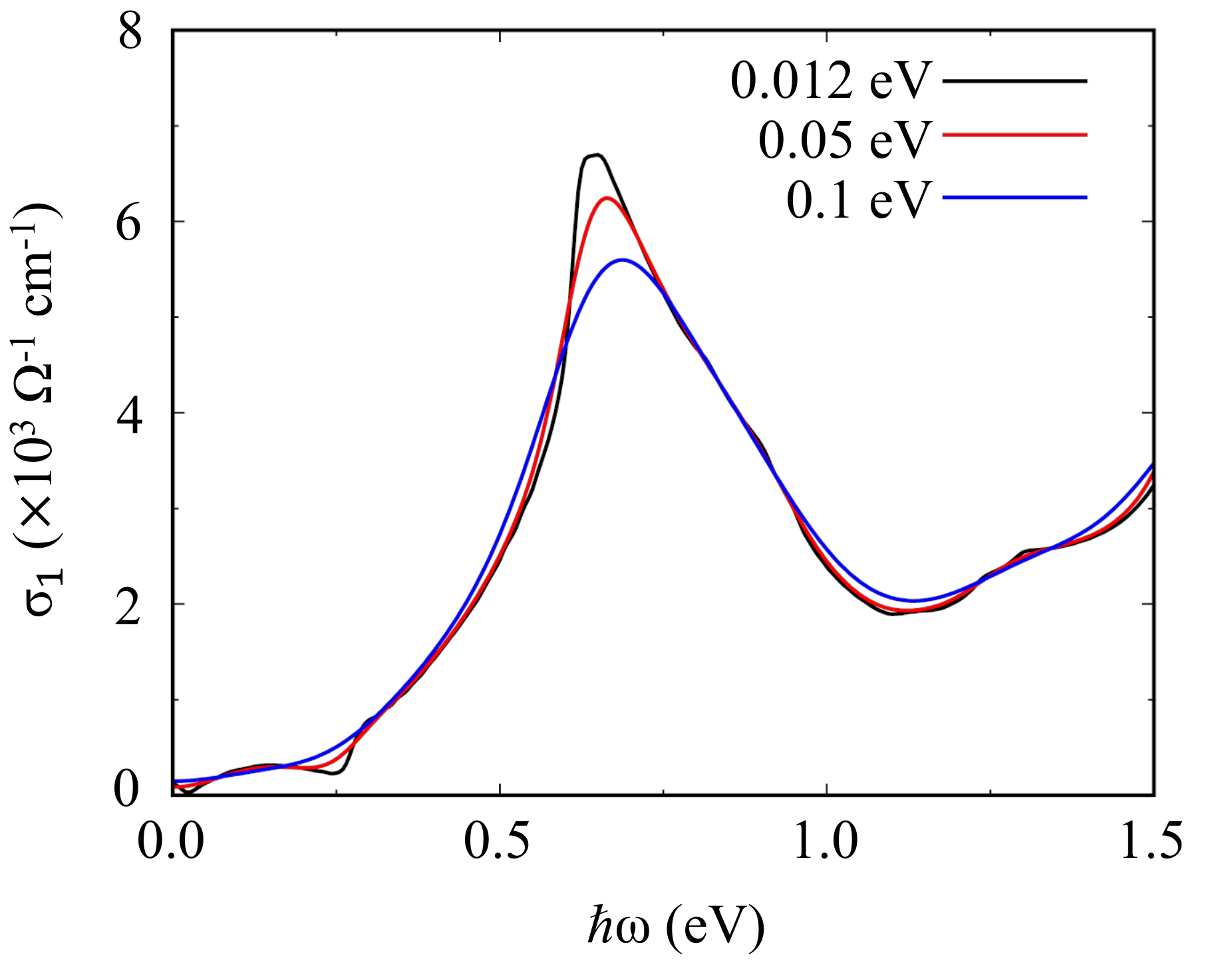}
\caption{The optical conductivity calculated from the Wannier TB model (w SOC) with different smearing width and the same number of $500 \times 500 \times 500$ k-points  \label{smearing} }
\end{figure}

Next, we used the software package Wannier90 to construct a tight-binding model in the basis of maximally-localized Wannier functions, fitted to the first-principles calculation results. To construct this model, we chose the initial projection functions to be $s$- and $d$-orbitals of Co, and $s$- and $p$-orbitals of Si; the converged maximally-localized Wannier functions that span the Bloch state valence manifold are well-localized in real space. The model can describe the electronic bands up to 1.5~eV above the Fermi level accurately, allowing us to calculate the conductivity accurately up to that energy scale. The comparison between the bands obtained from first-principles calculations and from this tight-binding model is shown in Fig.~\ref{fig:S5} (a-d). Here, we also calculated the conductivity with different smearing width shown in Fig. \ref{smearing}. As the smearing width increases, the dip around 0.25 eV get smeared out and the peak around 0.6 eV gets broadened.

We calculate the conductivity from DFT directly and from the solutions of the tight-binding model, and these are compared in Fig.~\ref{fig:S5} (e-f). In both calculations, to achieve faster convergence, the delta function $\delta(E_{nm} - \hbar \omega)$ in Eq.~\ref{eq:conductivity} is replaced by a broadened delta function $\Tilde{\delta} (E_{nm} - \hbar \omega)$, expressed in the form of Gaussian function
\begin{equation}
G(E_{mn}, \hbar \omega, \sigma) = \frac{1}{\sqrt{\pi} \sigma}  e^{-(\frac{E_{mn} - \hbar \omega}{\sigma})^2}
\end{equation}
where $\sigma$, the width of the function, represents the carrier lifetime in materials. Since one of our primary goals is to focus on the slope of the linear conductivity at low frequency, which is controlled by the slope of the dispersing band of the spin-1 three-fold degenerate node at $\Gamma$, we choose carefully the size of the k-point sampling and the smearing width $\sigma$ such that the Gaussian function can capture the energy change between adjacent k-points. In other words, we choose $\frac{dE}{dk} \approx \frac{\Delta E}{\Delta k}$, where $\frac{dE}{dk}$ is the slope of the dispersing bands around $\Gamma$, which, in our case, is 1.231~eV$\cdot$\AA, and $\Delta k$ is the distance between adjacent k-points in a chosen k-point grid. For example, if we use a $150 \times 150 \times 150$ k-point grid to calculate the optical conductivity, the smearing width is chosen to be $\sigma = 12$~meV, which are used in Fig.~\ref{fig:S5} (e-f) for the DFT and Wannier tight-binding calculation. In the main text, we used $500 \times 500 \times 500$ k-point grid and $\sigma = 5$~meV for the Wannier tight-binding calculation.  We also perform a calculation with a  Gaussian broadening of 1.2 meV and 3 meV, and the latter is the experimental Drude peak width at 10 K, and the results with four different broadening (1.2 meV, 3 meV, 5 meV and 12 meV) overlaps with each other. Note that 3 meV is the low-temperature Drude peak width in both samples, which is a reasonable estimate of the broadening.

The calculations of the velocity matrix elements in Quantum Espresso is implemented through $v_{nm}^a = p_{nm}^a / m_0$, where $m_0$ is the electron mass and $p_{nm}^a$ is the momentum matrix element. For the tight-binding model from Wannier90, the velocity matrix elements are calculated from $<n\bk | \bv | m\bk> = -\frac{i}{\hbar} (E_{m\bk} - E_{n\bk}) \bA_{nm} (\bk)$, where the Berry connection is $\bA_{nm} (\bk) = <n \bk | i \nabla_\bk | m \bk>$.

\section {Velocity matrix elements and momentum-resolved contribution}

\begin{figure}
\includegraphics[width= \linewidth]{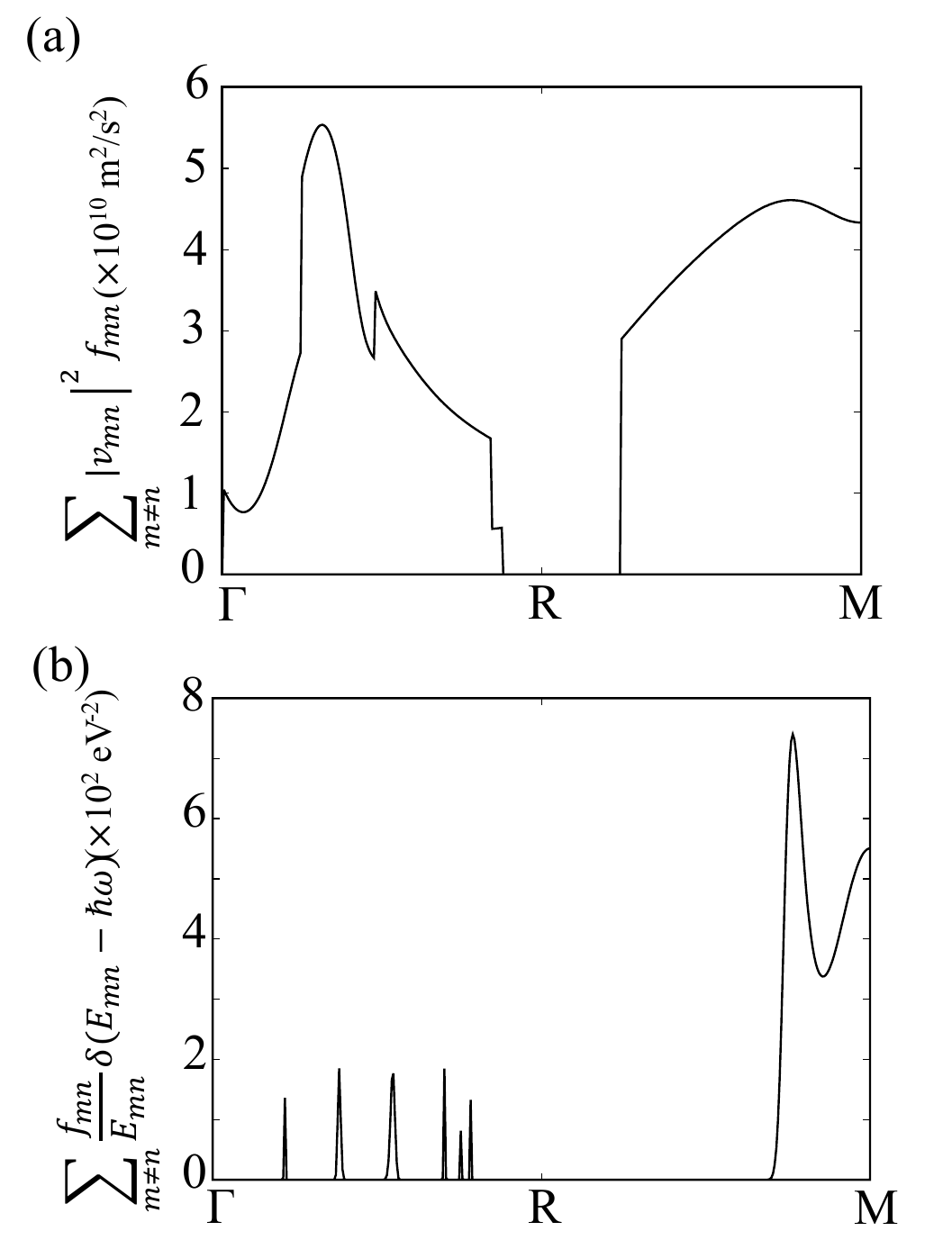}
\caption{(a) The sum of velocity matrix elements (including Fermi-Dirac occupation) $\sum_{nm} v_{nm}^a (\bk) v_{mn}^a (\bk) f_{nm}(\bk)$, and (b) the sum of delta function (including Fermi-Dirac occupation) $\sum_{nm} \delta (E_{mn} (\bk) - \hbar  \omega) \frac{f_{nm}(\bk)}{E_{nm} (\bk)}$.}
\label{momentum}
\end{figure}

Along the $\Gamma-R-M$ lines, the contributions to the peak at 0.62~eV in the optical conductivity originate mainly from the $R-M$ line, as can be seen from the main text. This can be further analyzed by focusing on the velocity matrix elements (Fig.~\ref{momentum}(a)) and momentum-resolved JDOS (Fig.~\ref{momentum}(b)) individually. As can be seen from Fig.~\ref{momentum}, the magnitude of the velocity matrix elements are comparable on both lines, but the JDOS is much larger on the $R-M$ line than on the $\Gamma-R$ line. This suggests that the energy difference between the bands along the $R-M$ line are much closer to 0.62~eV, which makes them contribute more to the peak at 0.62~eV in the conductivity.

\section {Contributions of spin-orbit-split bands along the $R-M$ line}

\begin{figure}
\includegraphics[width=\linewidth]{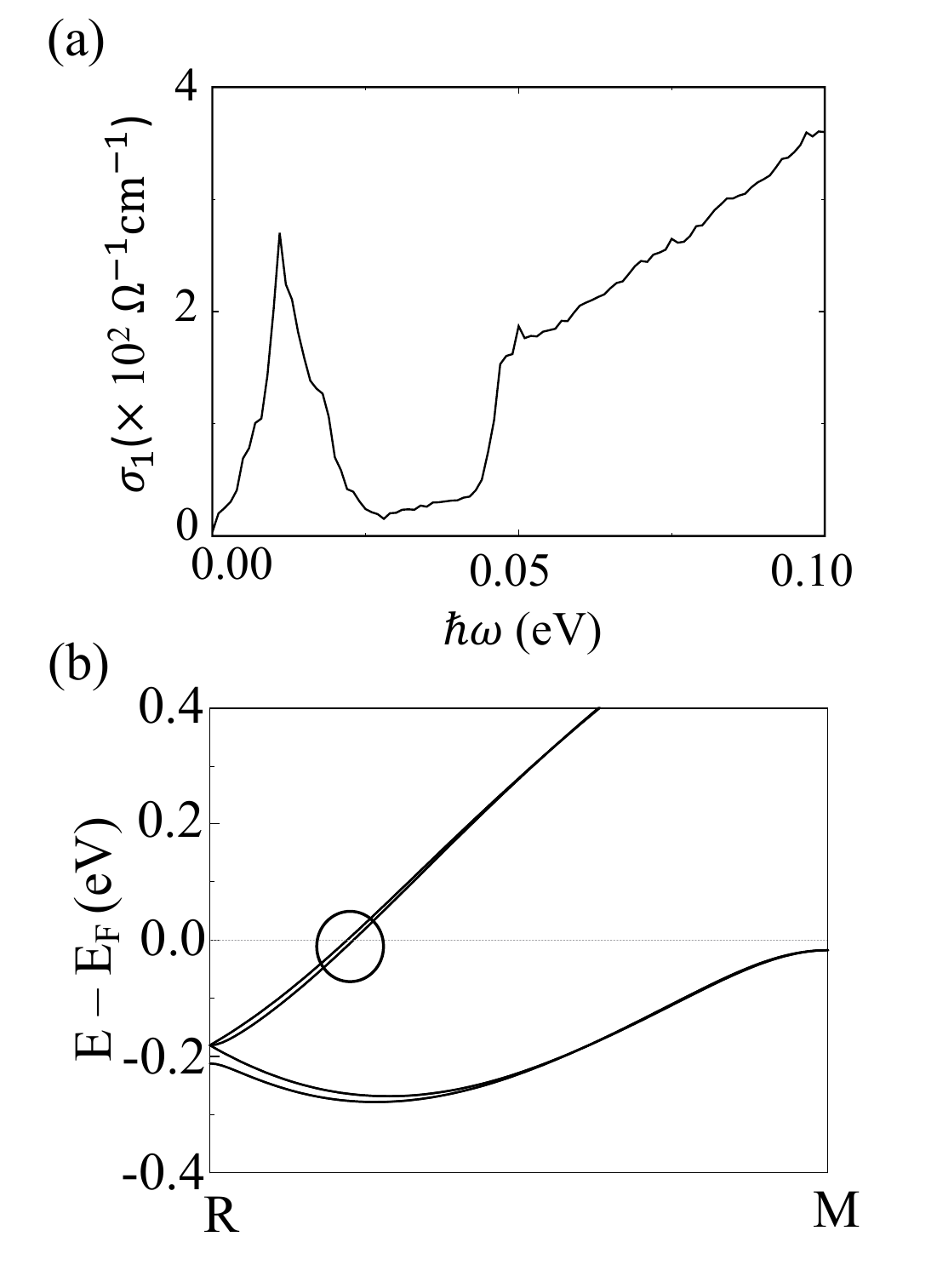}
\caption{(a) The calculated optical conductivity from SOC-split bands. (b) The calculated band structure (including SOC) along the $R-M$ high-symmetry line of CoSi.}
\label{peak}
\end{figure}

When turning on SOC, a peak at around 0.011~eV appears in the calculated optical conductivity, as shown in Fig.~\ref{peak}(a). This peak originates from the spin-orbit-split bands along the $R-M$ high-symmetry line, as can be seen in Fig.~\ref{peak}(b). Without SOC, these bands are degenerate in energy; since the pair of bands along the $R-M$ line crosses the Fermi level, transitions are enabled at k-points where one band is below the Fermi level and the other is above $E_F$. The transition energy, or the position of the peak, indicates the SOC strength of CoSi.

Note that when calculating the peak, the smearing width $\sigma$ has to be much smaller than the photon energy step; otherwise the peak will lead to nonzero conductivity at $\hbar \omega = 0$ eV. In the main text and this section, we used the smearing width $\sigma = 0.0012$ eV and the energy step $\Delta (\hbar \omega) = 0.005$ eV, and correspondingly a very dense kpoint sampling scheme, in order to show the position of the peak. In Appendix G, we used the smearing width $\sigma = 0.012$ eV.

\section{Linear Models}
\begin{figure}
\includegraphics[width=0.95 \linewidth]{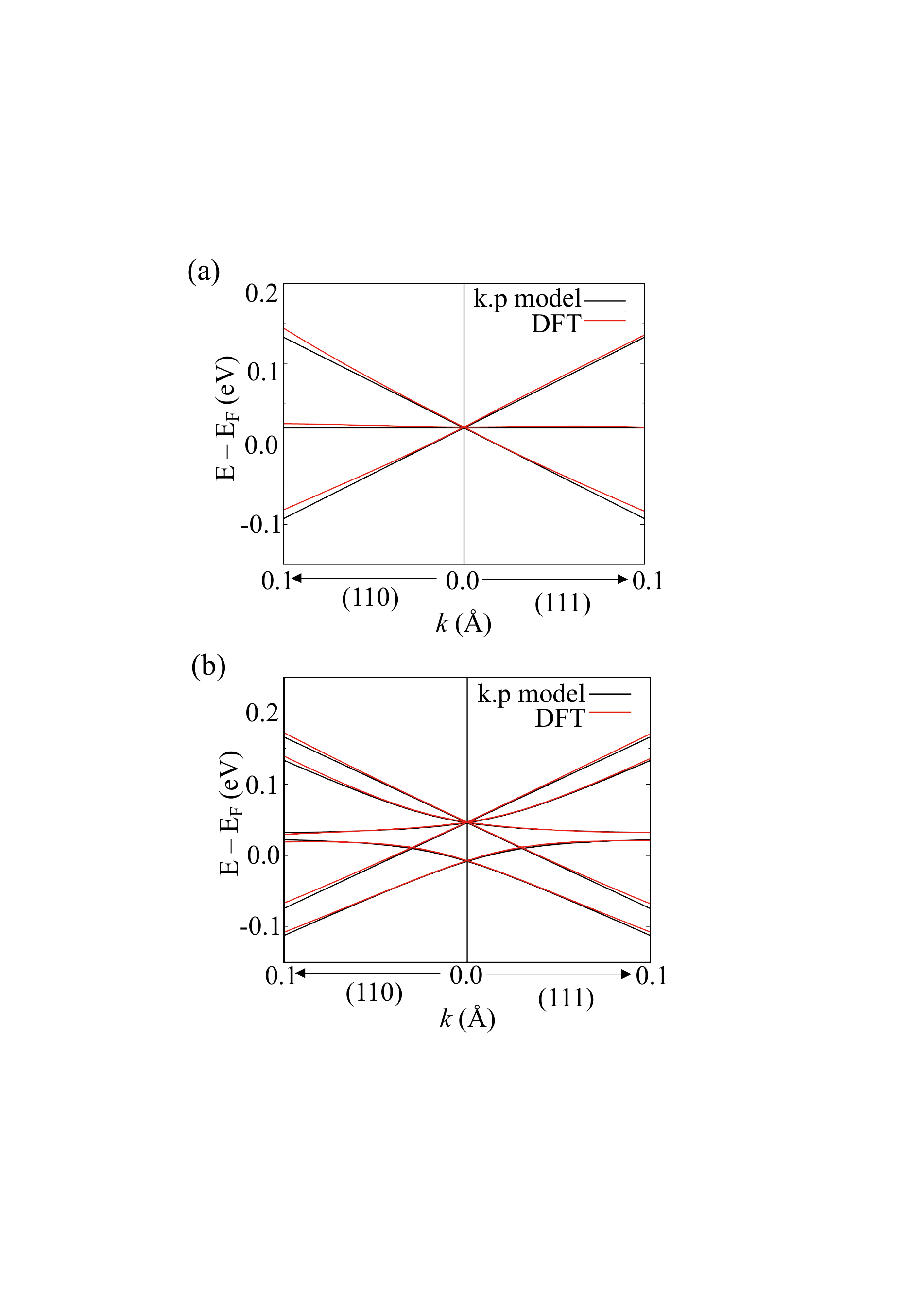}
\caption{The comparison between the CoSi band structures near $\Gamma$ obtained from DFT (black) and from the $k\cdot p$ model (red) without SOC (a) and with SOC (b).}
\label{k.p_fitting}
\end{figure}

The low-energy Hamiltonian without SOC, describing the spin-1 fermion node at $\Gamma$ is
\begin{equation}
H = E_0 \hat{I}_{3\times3} + \hbar v_F
\begin{pmatrix}
0 & i k_x & -i k_y \\
-i k_x & 0 & i k_z \\
i
k_y & -i k_z & 0
\end{pmatrix},
\end{equation}
where the parameter $E_0 = E_{\text{node}} - E_f = 0.0222$ eV is the energy difference between the node and Fermi energy, and $v_F = 1.231$ eV$\cdot$\AA\, is the Fermi velocity. The fitted band structures along $M-\Gamma-R$ high-symmetry lines are shown in Fig.~\ref{k.p_fitting} (a).

\begin{figure}
\includegraphics[width=0.95 \linewidth]{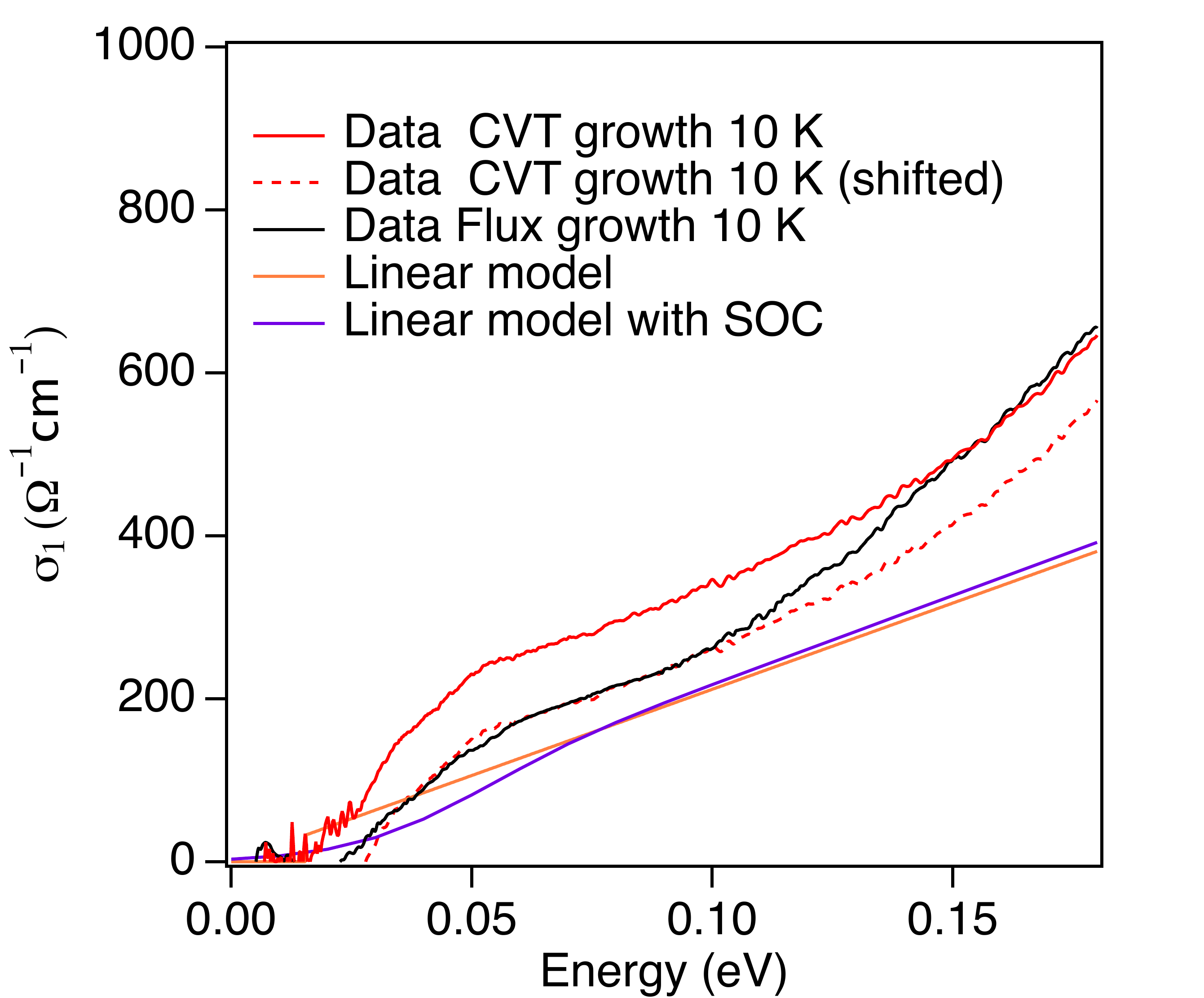}
\caption{Inter-band optical conductivity in CoSi from measurement in both CVT- and flux- grown samples (red and black curves) along with the linear model without and  with spin-orbit coupling. The dashed red line has a down-shift of 80 $\Omega^{-1}cm^{-1}$ from  the solid red curve.}
\label{linearmodel}
\end{figure}

The low-energy Hamiltonian with SOC, describing the spin-$\frac{3}{2}$ and spin-$\frac{1}{2}$ node at $\Gamma$, is

\begin{equation}
    H = E_0 \hat{I}_{6\times6} + \epsilon \hat{\Delta} + \lambda_1 \bD_1 \cdot \bk + \lambda_2 \bD_2 \cdot \bk + \lambda_3 \bD_3 \cdot \bk + \lambda_4 \bD_4 \cdot \bk
\end{equation}

where $\bD_i \cdot \bk = D_{ix} k_x + D_{iy} k_y + D_{iz} k_z$ (i = 1, 2, 3, 4), and

\begin{gather}
    \hat{\Delta} = \begin{pmatrix}
    1 & 0 & 0 & 0 & 0 & 0 \\
    0 & 1 & 0 & 0 & 0 & 0 \\
    0 & 0 & 1 & 0 & 0 & 0 \\
    0 & 0 & 0 & 1 & 0 & 0 \\
    0 & 0 & 0 & 0 & -2 & 0 \\
    0 & 0 & 0 & 0 & 0 & -2
    \end{pmatrix}, \nonumber  \allowdisplaybreaks \\
    D_{1x} = \begin{pmatrix}
    0 & \frac{\sqrt{3}}{2} & 0 & 0 & -\frac{\sqrt{3}}{2\sqrt{2}} & 0 \\
    \frac{\sqrt{3}}{2} & 0 & 1 & 0 & 0 & -\frac{1}{2\sqrt{2}} \\
    0 & 1 & 0 & \frac{\sqrt{3}}{2} & \frac{1}{2\sqrt{2}} & 0 \\
    0 & 0 & \frac{\sqrt{3}}{2} & 0 & 0 & \frac{\sqrt{3}}{2\sqrt{2}} \\
    -\frac{\sqrt{3}}{2\sqrt{2}} & 0 & \frac{1}{2\sqrt{2}} & 0 & 0 & 1 \\
    0 & -\frac{1}{2\sqrt{2}} & 0 & \frac{\sqrt{3}}{2\sqrt{2}} & 1 & 0
    \end{pmatrix}, \nonumber  \allowdisplaybreaks \\
    D_{1y} = \begin{pmatrix}
    0 & -\frac{\sqrt{3}i}{2} & 0 & 0 & \frac{\sqrt{3}i}{2\sqrt{2}} & 0 \\
    \frac{\sqrt{3}i}{2} & 0 & -i & 0 & 0 & \frac{i}{2\sqrt{2}} \\
    0 & i & 0 & -\frac{\sqrt{3}i}{2} & \frac{i}{2\sqrt{2}} & 0\\
    0 & 0 & \frac{\sqrt{3}i}{2} & 0 & 0 & \frac{\sqrt{3}i}{2\sqrt{2}} \\
    -\frac{\sqrt{3}i}{2\sqrt{2}} & 0 & -\frac{i}{2\sqrt{2}} & 0 & 0 & -i \\
    0 & -\frac{i}{2\sqrt{2}} & 0 & -\frac{\sqrt{3}i}{2\sqrt{2}} & i & 0
    \end{pmatrix}, \nonumber  \allowdisplaybreaks \\
    D_{1z} = \begin{pmatrix}
    \frac{3}{2} & 0 & 0 & 0 & 0 & 0 \\
    0 & \frac{1}{2} & 0 & 0 & \frac{1}{\sqrt{2}} & 0\\
    0 & 0 & -\frac{1}{2} & 0 & 0 & \frac{1}{\sqrt{2}} \\
    0 & 0 & 0 & -\frac{3}{2} & 0 & 0 \\
    0 & \frac{1}{\sqrt{2}} & 0 & 0 & 1 & 0 \\
    0 & 0 & \frac{1}{\sqrt{2}} & 0 & 0 & -1
    \end{pmatrix}, \nonumber \allowdisplaybreaks \\
    D_{2x} = \begin{pmatrix}
    0 & 0 & 0 & -\frac{3}{2} & 0 & 0 \\
    0 & 0 & -\frac{3}{2} & 0 & 0 & 0 \\
    -\frac{3}{2} & 0 & 0 & 0 & 0 & 0\\
    -\frac{3}{2} & 0 & 0 & 0 & 0 & 0 \\
    0 & 0 & 0 & 0 & 0 & \frac{3}{2} \\
    0 & 0 & 0 & 0 & \frac{3}{2} & 0
    \end{pmatrix}, \quad   \nonumber  \allowdisplaybreaks
	\\
    D_{2y} = \begin{pmatrix}
    0 & 0 & 0 & -\frac{3i}{2} & 0 & 0 \\
    0 & 0 & \frac{3i}{2} & 0 & 0 & 0 \\
    0 & -\frac{3i}{2} & 0 & 0 & 0 & 0 \\
    \frac{3i}{2} & 0 & 0 & 0 & 0 & 0 \\
    0 & 0 & 0 & 0 & 0 & -\frac{3i}{2} \\
    0 & 0 & 0 & 0 & \frac{3i}{2} & 0
    \end{pmatrix}  ,\quad  \nonumber \allowdisplaybreaks
	\\
    D_{2z} = \begin{pmatrix}
    -\frac{3}{2} & 0 & 0 & 0 & 0 & 0 \\
    0 & \frac{3}{2} & 0 & 0 & 0 & 0 \\
    0 & 0 & -\frac{3}{2} & 0 & 0 & 0 \\
    0 & 0 & 0 & \frac{3}{2} & 0 & 0 \\
    0 & 0 & 0 & 0 & \frac{3}{2} & 0 \\
    0 & 0 & 0 & 0 & 0 & -\frac{3}{2}
    \end{pmatrix}, \nonumber  \allowdisplaybreaks \\
    D_{3x} = \begin{pmatrix}
    0 & -\frac{\sqrt{3}}{2} & 0 & 0 & -\frac{\sqrt{3}}{\sqrt{2}} & 0 \\
    -\frac{\sqrt{3}}{2} & 0 & 1 & 0 & 0 & \frac{1}{\sqrt{2}} \\
    0 & 1 & 0 & -\frac{\sqrt{3}}{2} & -\frac{1}{\sqrt{2}} & 0 \\
    0 & 0 & -\frac{\sqrt{3}}{2} & 0 & 0 & \frac{\sqrt{3}}{\sqrt{2}} \\
    -\frac{\sqrt{3}}{\sqrt{2}} & 0 & -\frac{1}{\sqrt{2}} & 0 & 0 & -\frac{1}{2} \\
    0 & \frac{1}{\sqrt{2}} & 0 & \frac{\sqrt{3}}{\sqrt{2}} & -\frac{1}{2} & 0
    \end{pmatrix}, \nonumber \allowdisplaybreaks \\
    D_{3y} = \begin{pmatrix}
    0 & 0 & 0 & \frac{3 i}{2} & 0 & 0 \\
    0 & 0 & \frac{i}{2} & 0 & 0 & i \sqrt{2} \\
    0 & -\frac{i}{2} & 0 & 0 & i \sqrt{2} & 0 \\
    -\frac{3 i}{2} & 0 & 0 & 0 & 0 & 0 \\
    0 & 0 & -i \sqrt{2} & 0 & 0 & \frac{i}{2} \\
    0 & -i \sqrt{2} & 0 & 0 & -\frac{i}{2} & 0
    \end{pmatrix}, \nonumber  \allowdisplaybreaks \\
    D_{3z} = \begin{pmatrix}
    0 & 0 & \frac{\sqrt{3}}{2} & 0 & 0 & \frac{\sqrt{3}}{\sqrt{2}} \\
    0 & -1 & 0 & -\frac{\sqrt{3}}{2} & \frac{1}{\sqrt{2}} & 0 \\
    \frac{\sqrt{3}}{2} & 0 & 1 & 0 & 0 & \frac{1}{\sqrt{2}} \\
    0 & -\frac{\sqrt{3}}{2} & 0 & 0 & -\frac{\sqrt{3}}{\sqrt{2}} & 0 \\
    0 & \frac{1}{\sqrt{2}} & 0 & -\frac{\sqrt{3}}{\sqrt{2}} & -\frac{1}{2} & 0 \\
    -\frac{\sqrt{3}}{\sqrt{2}} & 0 & \frac{1}{\sqrt{2}} & 0 & 0 & \frac{1}{2}
    \end{pmatrix}, \nonumber \allowdisplaybreaks \\
    D_{4x} = \begin{pmatrix}
    0 & 0 & 0 & -\frac{3}{2} & 0 & 0 \\
    0 & 0 & \frac{1}{2} & 0 & 0 & -\frac{1}{\sqrt{2}} \\
    0 & \frac{1}{2} & 0 & 0 & \frac{1}{\sqrt{2}} & 0 \\
    -\frac{3}{2} & 0 & 0 & 0 & 0 & 0 \\
    0 & 0 & \frac{1}{\sqrt{2}} & 0 & 0 & -1 \\
    0 & -\frac{1}{\sqrt{2}} & 0 & 0 & -1 & 0 \\
    \end{pmatrix}, \nonumber  \allowdisplaybreaks \\
    D_{4y} = \begin{pmatrix}
    0 & -\frac{\sqrt{3}i}{2} & 0 & 0 & \frac{\sqrt{3}i}{2\sqrt{2}} & 0 \\
    \frac{\sqrt{3}i}{2} & 0 & i & 0 & 0 & -\frac{i}{2 \sqrt{2}} \\
    0 & -i & 0 & -\frac{\sqrt{3}i}{2} & -\frac{i}{2 \sqrt{2}} & 0 \\
    0 & 0 & \frac{\sqrt{3}i}{2} & 0 & 0 & \frac{\sqrt{3}i}{2\sqrt{2}} \\
    -\frac{\sqrt{3}i}{2\sqrt{2}} & 0 & \frac{i}{2 \sqrt{2}} & 0 & 0 & i \\
    0 & \frac{i}{2 \sqrt{2}} & 0 & -\frac{\sqrt{3}i}{2\sqrt{2}} & -i & 0
    \end{pmatrix}, \nonumber \allowdisplaybreaks \\
    D_{4z} = \begin{pmatrix}
    0 & 0 & \frac{\sqrt{3}}{2} & 0 & 0 & \frac{\sqrt{3}}{2\sqrt{2}} \\
    0 & 1 & 0 & -\frac{\sqrt{3}}{2} & \frac{1}{2 \sqrt{2}} & 0 \\
    \frac{\sqrt{3}}{2} & 0 & -1 & 0 & 0 & \frac{1}{2 \sqrt{2}} \\
    0 & -\frac{\sqrt{3}}{2} & 0 & 0 & \frac{\sqrt{3}}{2\sqrt{2}} & 0 \\
    0 & \frac{1}{2 \sqrt{2}} & 0 & \frac{\sqrt{3}}{2\sqrt{2}} & -1 & 0 \\
    \frac{\sqrt{3}}{2\sqrt{2}} & 0 & \frac{1}{2 \sqrt{2}} & 0 & 0 & 1
    \end{pmatrix}.
\end{gather}

The fitted parameters are shown in Table~\ref{tab:parameters_SOC}, and the fitted band structures along $M-\Gamma-R$ high-symmetry lines are shown in Fig.~\ref{k.p_fitting} (b).

By using the band structure in Fig.~\ref{k.p_fitting} (a) and (b), the calculated conductivity with a 12 meV broadening is shown in Fig.~\ref{linearmodel}.

\begin{table}[htb]
\centering
\begin{tabular}{cccccc}
\hline $E_0$ (meV) & $\epsilon$ (meV) & $\lambda_1$ (eV \AA) & $\lambda_2$ (eV \AA) & $\lambda_3$ (eV \AA) & $\lambda_4$ (eV \AA) \\ \hline
43.7 & 18.4&  0.7904 & -0.0052 & -0.0222 & 0.0219 \\ \hline
\end{tabular}
\caption{The fitted parameters of the tight-binding model with SOC}
\label{tab:parameters_SOC}
\end{table}

\section{Four-band tight-binding model}

\begin{figure}
\includegraphics[width=0.95\linewidth]{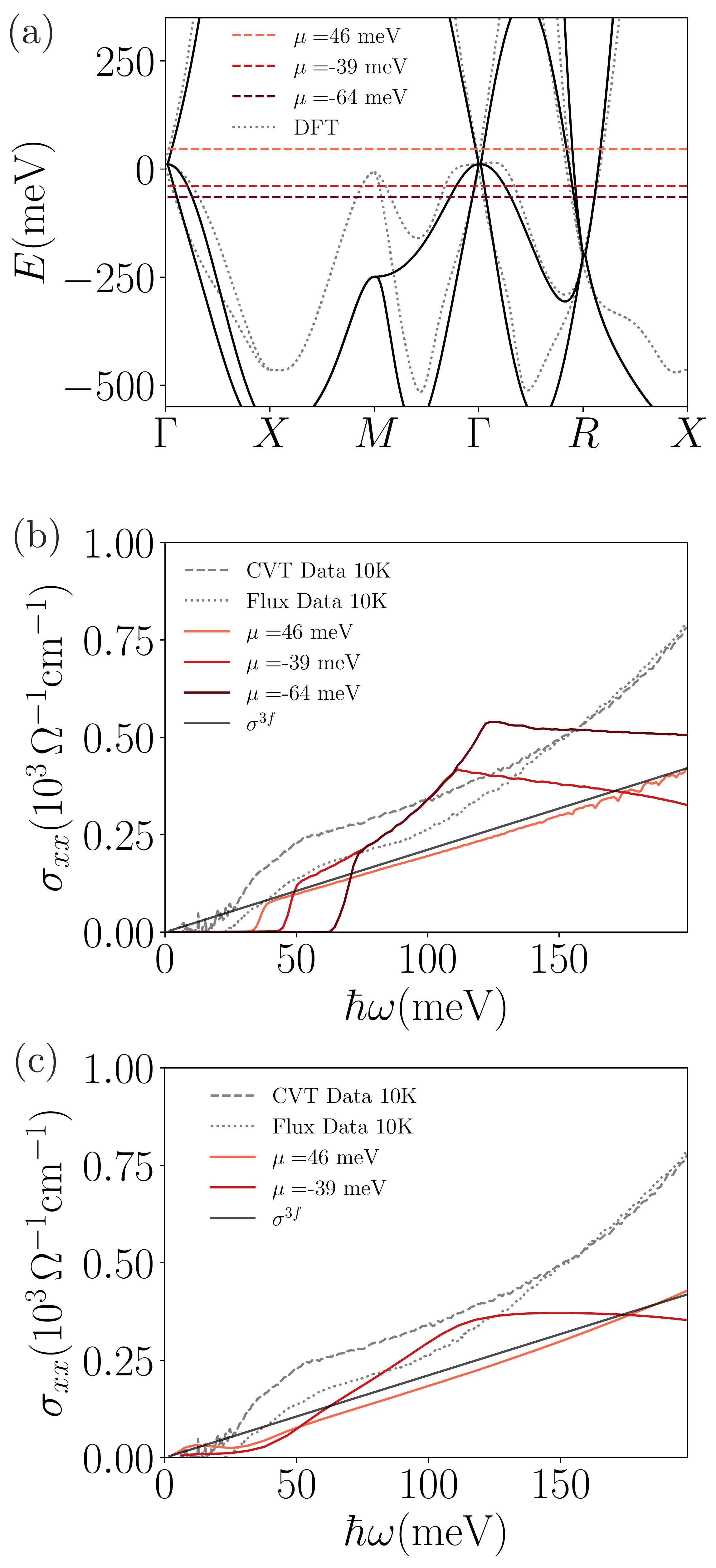}
\caption{\label{fig:TBvp055} (a) Band structure of CoSi obtained using a four-band tight-binding model with $v_1=1.29$, $v_2=0.25$ and $v_p=0.55$ compared to DFT bands (dotted grey). Fermi levels for different chemical potentials are marked as horizontal dashed lines. (b) Optical conductivity for the chemical potentials shown in (a) with 0 meV broadening.  For reference we show as a solid line the optical conductivity of a threefold fermion $\sigma^{\mathrm{3f}}  = \frac{\omega}{3 \pi v_F}$ with $v_F=v_p/2$. (c) Optical conductivity at two fixed chemical potentials, one above and one below the $\Gamma$ node, calculated with a finite broadening of $12$~meV. The curve labelled $\mu=-39$ meV is shown in Fig. 5(a) of the main text as a blue line.}
\end{figure}

\begin{figure}
\includegraphics[width=0.95\linewidth]{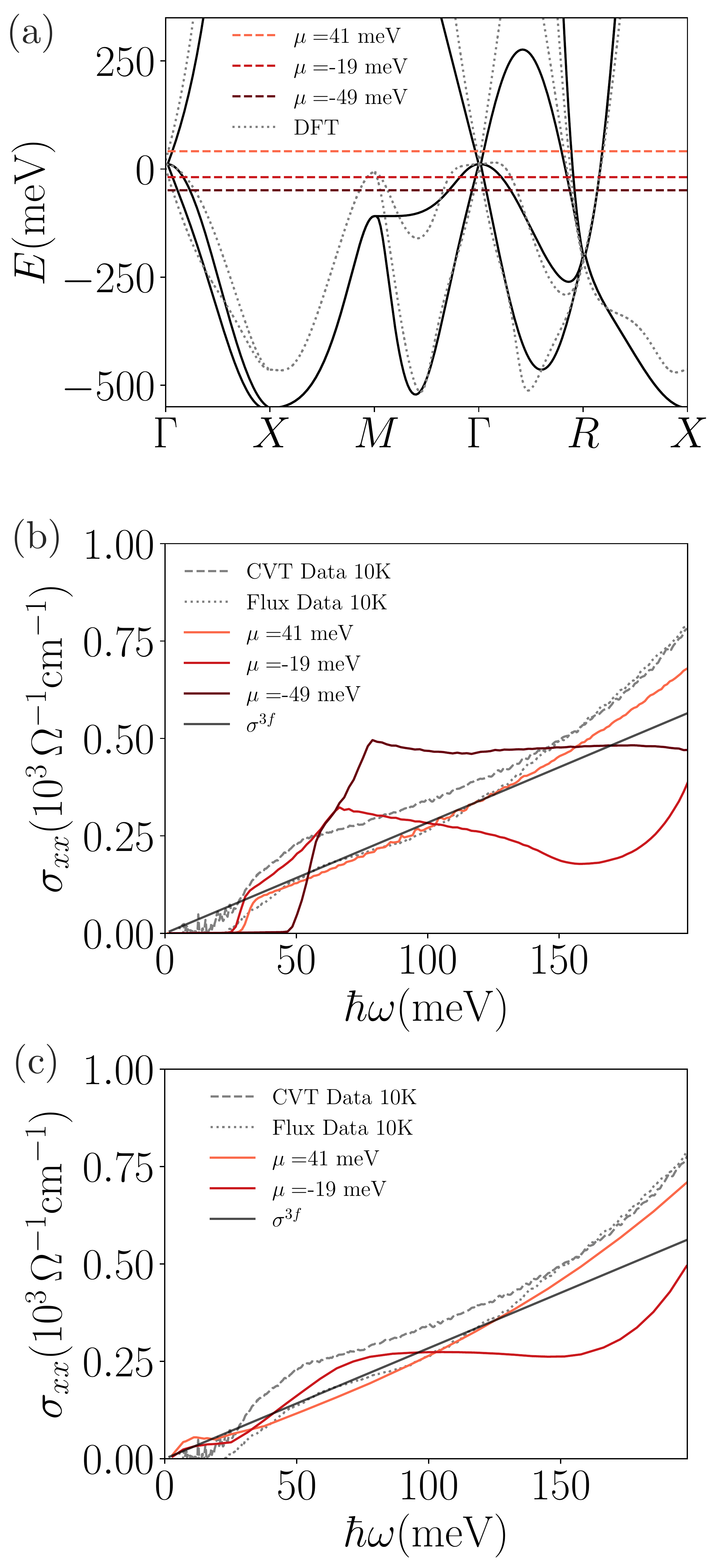}
\caption{\label{fig:TBvp041} Same as Fig. \ref{fig:TBvp055} with $v_p=0.41$. (a) Band structure of CoSi obtained using a four-band tight-binding model with $v_1=1.29$, $v_2=0.25$ and $v_p=0.41$ compared to DFT bands (dotted grey). Fermi levels for different chemical potentials are marked as horizontal dashed lines. (b) Optical conductivity at the chemical potentials in (a). For reference, we show as a solid line the optical conductivity of a threefold fermion $\sigma^{\mathrm{3f}}  = \frac{\omega}{3 \pi v_F}$ with $v_F=v_p/2$. (c) Optical conductivity at two fixed chemical potentials, one above and one below the $\Gamma$ node calculated with a finite broadening of $12$~meV. }
\end{figure}

Following Ref.~\cite{chang_unconventional_2017}, we have constructed a four-band tight-binding model for CoSi that is consistent with its crystal symmetry to describe the closest four energy bands to the Fermi level of this material. In general, this model can describe any material in space group 198, and without spin-orbit coupling it is defined through three material-dependent parameters: $v_1$, $v_p$, and $v_2$. In addition to these paramaters~\cite{chang_unconventional_2017} we incorporate the orbital embedding as described in Appendix F of Ref.~\cite{sanchezPRB2019}. The orbital embedding $x$ describes the position (or embedding) of orbitals in real space. It enters as a momentum-dependent unitary transformation~\cite{flicker_chiral_2018}, leaving the band structure unaltered, but modifying its eigenfunctions and, as a result, the predicted optical conductivity.

To describe CoSi specifically, we fix $x_{\mathrm{CoSi}}=0.3865$, as determined by crystallographic databases, and find $v_1$, $v_p$, and $v_2$ by fitting the tight-binding spectrum to the DFT bands. To capture the separation in energy between the multi-fold nodes at $\Gamma$ and $R$, we fix it to match that found by DFT, which equals $210$~meV. By expanding the Hamiltonian to linear order in momentum around $\Gamma$ and $R$, we see that the parameter $v_p$ sets the Fermi velocity of the three-fold and double Weyl fermion at the $\Gamma$ and $R$ points to $v^{\Gamma}_{F}=v_p/2$ and $v^{\Gamma}_{R}=v_p/(2\sqrt{3})$, respectively  \cite{manes_existence_2012}. To set the right energy scale in the tight-binding model we add a constant energy shift of $E_0=0.551$~eV, $H=H_{198}(v_1,v_p,v_2;x)+(E_0-\mu)\mathbb{I}_{4\times 4}$. All chemical potentials are measured with respect to this energy shift.

It is illustrative then to compare two different fits, one that matches well the Fermi velocity near the $\Gamma$ point, and a second one to match the Fermi velocity at the $R$ point. For $v_p=0.55$, we fit the Fermi velocity of the bands near the $\Gamma$ point (see Fig.~\ref{fig:TBvp055} (a)). Upon closer inspection, we find that it provides a better description of the upper band of the three-fold crossing at the $\Gamma$ point than that of the middle and lower bands. For $v_p=0.41$, we fit to the Fermi velocity of the bands near the $R$ point (see Fig.~\ref{fig:TBvp041} (a)). Despite being a good fit for the $R$ point, we observe this to be a fair description also of the lower and middle bands at the $\Gamma$ point away from the $\Gamma$ node. For both fits, we obtain $v_1=1.29$ and $v_2=0.25$ for the remaining tight-binding parameters.

In Fig.~\ref{fig:TBvp055} and Fig.~\ref{fig:TBvp041} (b) we compare the optical conductivities obtained for different chemical potentials crossing the node with 0 meV broadening. The (c) panels show the optical conductivity for selected chemical potentials and with a finite phenomenological Lorentzian disorder broadening of $12$~meV. For both values of $v_p$, we find that a chemical potential that crosses the bands below the $\Gamma$ node results in a peak and dip structure. In the case of $v_p=0.55$, which amounts to $v_\mathrm{F}=1.23$ eV$\cdot$\AA, the conductivity has an increasing trend when lowering the chemical potential, which agrees well with the trend observed in Fig.~\ref{chemical_potential} in Appendix H. The curve with $v_p=0.55$, $\mu=-39$ meV shown in Fig.~\ref{fig:TBvp055}(c) falls closest to the Wannier tight-binding calculation, and it is shown in Fig. 5 (a) of the main text.

It is interesting to note that although a good agreement between the four-band tight-binding model and the flux sample data is found for $v_p=0.41$ with the chemical potential above the node, and a 12~meV broadening (Fig.~\ref{fig:TBvp041} (c)), this value of the chemical potential does not agree with the rest of our observations. As discussed at length in this appendix and in the main text, from transport results to ab-initio calculations we find that CoSi is a compensated semimetal with a hole pocket at the $\Gamma$, implying that the chemical potential in the sample is below the node. In this case, the curve with $\mu=-39$ meV $v_p=0.55$ in Fig.~\ref{fig:TBvp055} (c) results in the best fit to the data and first-principles calculations.

\begin{figure*}
\includegraphics[width=0.7\linewidth]{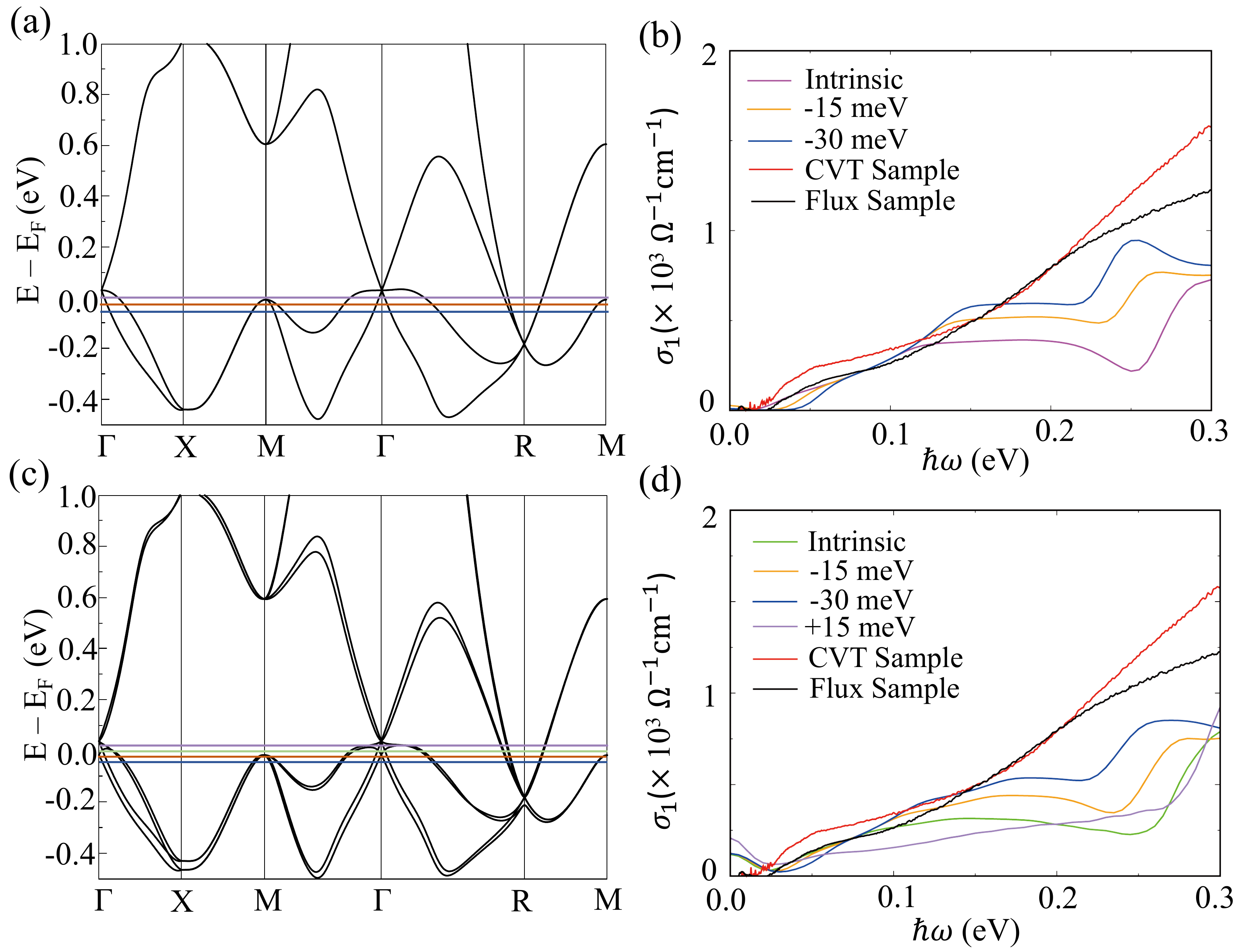}
\caption{(a) The position of several chosen chemical potentials superimposed on the calculated band structure without SOC. (b) The optical conductivity calculated with each of the chosen chemical potentials. (c, d) Same as (a, b), but incorporating SOC. }
\label{chemical_potential}
\end{figure*}

\section{The effect of chemical potential on the calculated conductivity in the Wannier tight-binding models}

The chemical potential in real materials could be different from that obtained in DFT, due to extrinsic effects such as the presence of impurities. Therefore, we have calculated the optical conductivity with different chemical potentials. The conductivity calculated without SOC coupling is shown in Fig.~\ref{chemical_potential}. Under the condition that the chosen chemical potential is lower than the flat middle band near the $\Gamma$ point, transitions will always occur between the lower band and the middle band when the energy of the incoming photon is small ($< 0.1$~eV), and occur near the $R$ point when the energy is relatively large ($> 0.2$~eV). Therefore, changing the chemical potential will influence the position and the shape of the dip structure ($\sigma_1(\omega\approx$~0.25~eV)) in two ways. Firstly, as the chemical potential decreases, more transitions could occur between the lower band and the middle band, and this would lead to an increase in conductivity at the downward part of the dip structure. Secondly, more transitions could also occur near the $R$ point due to the decrease in chemical potential, so it is also expected that the upward region of the dip structure will increase. Combining these two results, decreasing the Fermi level will tend to flatten (and perhaps eventually remove) the dip structure.

When turning on SOC, the conductivity calculated with different chemical potentials are shown in Fig.~\ref{chemical_potential}. The conclusions drawn from the non-SOC case can also be applied here; namely, decreasing the chemical potential will smooth the dip structure. Furthermore, we studied the case where the chemical potential is placed between the middle band pair (the purple curve in Fig.~\ref{chemical_potential} (c)). In this case, the number of transitions between the lower band pair and the middle band pair is reduced, which leads to a decrease in the lower-frequency region. However, the transitions which occur near the $R$ point are almost unaffected, which justifies the fact that the slope of the upward part of the dip structure is similar to that of the intrinsic Fermi level.

To summarize, assuming a lower chemical potential increases the conductivity below $0.2$ eV, but still the agreement with the CVT-grown sample is not as good as the agreement achieved for the flux-grown sample. This is an indication that either the additional contribution in the CVT-grown sample is not coming from inter-band transition or the origin of defects in the CVT sample is slightly different. Therefore, in terms of the discrepancy between the Wannier tight-binding model with SOC and the CVT sample, a more likely scenario is that the lower chemical potential in the CVT-grown sample leads to a bigger flat hole pocket at the  $\Gamma$ point than the flux-grown sample. This could result in a second flatter Drude peak (with a small spectral weight) corresponding to the relatively heavy holes~\cite{xuXPRB2019}. This seems to be the case in RhSi which displays a second, flat Drude peak due to its even larger hole pocket~~\cite{maulanaPRR2020}.

%
%


\begin{thebibliography}{79}%
\makeatletter
\providecommand \@ifxundefined [1]{%
 \@ifx{#1\undefined}
}%
\providecommand \@ifnum [1]{%
 \ifnum #1\expandafter \@firstoftwo
 \else \expandafter \@secondoftwo
 \fi
}%
\providecommand \@ifx [1]{%
 \ifx #1\expandafter \@firstoftwo
 \else \expandafter \@secondoftwo
 \fi
}%
\providecommand \natexlab [1]{#1}%
\providecommand \enquote  [1]{``#1''}%
\providecommand \bibnamefont  [1]{#1}%
\providecommand \bibfnamefont [1]{#1}%
\providecommand \citenamefont [1]{#1}%
\providecommand \href@noop [0]{\@secondoftwo}%
\providecommand \href [0]{\begingroup \@sanitize@url \@href}%
\providecommand \@href[1]{\@@startlink{#1}\@@href}%
\providecommand \@@href[1]{\endgroup#1\@@endlink}%
\providecommand \@sanitize@url [0]{\catcode `\\12\catcode `\$12\catcode
  `\&12\catcode `\#12\catcode `\^12\catcode `\_12\catcode `\%12\relax}%
\providecommand \@@startlink[1]{}%
\providecommand \@@endlink[0]{}%
\providecommand \url  [0]{\begingroup\@sanitize@url \@url }%
\providecommand \@url [1]{\endgroup\@href {#1}{\urlprefix }}%
\providecommand \urlprefix  [0]{URL }%
\providecommand \Eprint [0]{\href }%
\providecommand \doibase [0]{http://dx.doi.org/}%
\providecommand \selectlanguage [0]{\@gobble}%
\providecommand \bibinfo  [0]{\@secondoftwo}%
\providecommand \bibfield  [0]{\@secondoftwo}%
\providecommand \translation [1]{[#1]}%
\providecommand \BibitemOpen [0]{}%
\providecommand \bibitemStop [0]{}%
\providecommand \bibitemNoStop [0]{.\EOS\space}%
\providecommand \EOS [0]{\spacefactor3000\relax}%
\providecommand \BibitemShut  [1]{\csname bibitem#1\endcsname}%
\let\auto@bib@innerbib\@empty
\bibitem [{\citenamefont {Murakami}(2007)}]{MurakamiNJP07}%
  \BibitemOpen
  \bibfield  {author} {\bibinfo {author} {\bibfnamefont {S.}~\bibnamefont
  {Murakami}},\ }\href@noop {} {\bibfield  {journal} {\bibinfo  {journal} {New
  Journal of Physics}\ }\textbf {\bibinfo {volume} {9}},\ \bibinfo {pages}
  {356} (\bibinfo {year} {2007})}\BibitemShut {NoStop}%
\bibitem [{\citenamefont {Wan}\ \emph {et~al.}(2011)\citenamefont {Wan},
  \citenamefont {Turner}, \citenamefont {Vishwanath},\ and\ \citenamefont
  {Savrasov}}]{WanPRB2011}%
  \BibitemOpen
  \bibfield  {author} {\bibinfo {author} {\bibfnamefont {X.}~\bibnamefont
  {Wan}}, \bibinfo {author} {\bibfnamefont {A.~M.}\ \bibnamefont {Turner}},
  \bibinfo {author} {\bibfnamefont {A.}~\bibnamefont {Vishwanath}}, \ and\
  \bibinfo {author} {\bibfnamefont {S.~Y.}\ \bibnamefont {Savrasov}},\
  }\href@noop {} {\bibfield  {journal} {\bibinfo  {journal} {Physical Review
  B}\ }\textbf {\bibinfo {volume} {83}},\ \bibinfo {pages} {205101} (\bibinfo
  {year} {2011})}\BibitemShut {NoStop}%
\bibitem [{\citenamefont {Burkov}\ and\ \citenamefont
  {Balents}(2011)}]{BurkovPRL2011}%
  \BibitemOpen
  \bibfield  {author} {\bibinfo {author} {\bibfnamefont {A.}~\bibnamefont
  {Burkov}}\ and\ \bibinfo {author} {\bibfnamefont {L.}~\bibnamefont
  {Balents}},\ }\href@noop {} {\bibfield  {journal} {\bibinfo  {journal}
  {Physical review letters}\ }\textbf {\bibinfo {volume} {107}},\ \bibinfo
  {pages} {127205} (\bibinfo {year} {2011})}\BibitemShut {NoStop}%
\bibitem [{\citenamefont {Weng}\ \emph {et~al.}(2015)\citenamefont {Weng},
  \citenamefont {Fang}, \citenamefont {Fang}, \citenamefont {Bernevig},\ and\
  \citenamefont {Dai}}]{WengPRX2015}%
  \BibitemOpen
  \bibfield  {author} {\bibinfo {author} {\bibfnamefont {H.}~\bibnamefont
  {Weng}}, \bibinfo {author} {\bibfnamefont {C.}~\bibnamefont {Fang}}, \bibinfo
  {author} {\bibfnamefont {Z.}~\bibnamefont {Fang}}, \bibinfo {author}
  {\bibfnamefont {B.~A.}\ \bibnamefont {Bernevig}}, \ and\ \bibinfo {author}
  {\bibfnamefont {X.}~\bibnamefont {Dai}},\ }\href@noop {} {\bibfield
  {journal} {\bibinfo  {journal} {Physical Review X}\ }\textbf {\bibinfo
  {volume} {5}},\ \bibinfo {pages} {011029} (\bibinfo {year}
  {2015})}\BibitemShut {NoStop}%
\bibitem [{\citenamefont {Huang}\ \emph {et~al.}(2015)\citenamefont {Huang},
  \citenamefont {Xu}, \citenamefont {Belopolski}, \citenamefont {Lee},
  \citenamefont {Chang}, \citenamefont {Wang}, \citenamefont {Alidoust},
  \citenamefont {Bian}, \citenamefont {Neupane}, \citenamefont {Zhang} \emph
  {et~al.}}]{HuangNatComm2015}%
  \BibitemOpen
  \bibfield  {author} {\bibinfo {author} {\bibfnamefont {S.-M.}\ \bibnamefont
  {Huang}}, \bibinfo {author} {\bibfnamefont {S.-Y.}\ \bibnamefont {Xu}},
  \bibinfo {author} {\bibfnamefont {I.}~\bibnamefont {Belopolski}}, \bibinfo
  {author} {\bibfnamefont {C.-C.}\ \bibnamefont {Lee}}, \bibinfo {author}
  {\bibfnamefont {G.}~\bibnamefont {Chang}}, \bibinfo {author} {\bibfnamefont
  {B.}~\bibnamefont {Wang}}, \bibinfo {author} {\bibfnamefont {N.}~\bibnamefont
  {Alidoust}}, \bibinfo {author} {\bibfnamefont {G.}~\bibnamefont {Bian}},
  \bibinfo {author} {\bibfnamefont {M.}~\bibnamefont {Neupane}}, \bibinfo
  {author} {\bibfnamefont {C.}~\bibnamefont {Zhang}},  \emph {et~al.},\
  }\href@noop {} {\bibfield  {journal} {\bibinfo  {journal} {Nature
  Communications}\ }\textbf {\bibinfo {volume} {6}} (\bibinfo {year}
  {2015})}\BibitemShut {NoStop}%
\bibitem [{\citenamefont {Xu}\ \emph {et~al.}(2015{\natexlab{a}})\citenamefont
  {Xu}, \citenamefont {Belopolski}, \citenamefont {Alidoust}, \citenamefont
  {Neupane}, \citenamefont {Bian}, \citenamefont {Zhang}, \citenamefont
  {Sankar}, \citenamefont {Chang}, \citenamefont {Yuan}, \citenamefont {Lee}
  \emph {et~al.}}]{XuScience2015}%
  \BibitemOpen
  \bibfield  {author} {\bibinfo {author} {\bibfnamefont {S.-Y.}\ \bibnamefont
  {Xu}}, \bibinfo {author} {\bibfnamefont {I.}~\bibnamefont {Belopolski}},
  \bibinfo {author} {\bibfnamefont {N.}~\bibnamefont {Alidoust}}, \bibinfo
  {author} {\bibfnamefont {M.}~\bibnamefont {Neupane}}, \bibinfo {author}
  {\bibfnamefont {G.}~\bibnamefont {Bian}}, \bibinfo {author} {\bibfnamefont
  {C.}~\bibnamefont {Zhang}}, \bibinfo {author} {\bibfnamefont
  {R.}~\bibnamefont {Sankar}}, \bibinfo {author} {\bibfnamefont
  {G.}~\bibnamefont {Chang}}, \bibinfo {author} {\bibfnamefont
  {Z.}~\bibnamefont {Yuan}}, \bibinfo {author} {\bibfnamefont {C.-C.}\
  \bibnamefont {Lee}},  \emph {et~al.},\ }\href@noop {} {\bibfield  {journal}
  {\bibinfo  {journal} {Science}\ }\textbf {\bibinfo {volume} {349}},\ \bibinfo
  {pages} {613} (\bibinfo {year} {2015}{\natexlab{a}})}\BibitemShut {NoStop}%
\bibitem [{\citenamefont {Lv}\ \emph {et~al.}(2015{\natexlab{a}})\citenamefont
  {Lv}, \citenamefont {Weng}, \citenamefont {Fu}, \citenamefont {Wang},
  \citenamefont {Miao}, \citenamefont {Ma}, \citenamefont {Richard},
  \citenamefont {Huang}, \citenamefont {Zhao}, \citenamefont {Chen} \emph
  {et~al.}}]{LvPRX2015}%
  \BibitemOpen
  \bibfield  {author} {\bibinfo {author} {\bibfnamefont {B.}~\bibnamefont
  {Lv}}, \bibinfo {author} {\bibfnamefont {H.}~\bibnamefont {Weng}}, \bibinfo
  {author} {\bibfnamefont {B.}~\bibnamefont {Fu}}, \bibinfo {author}
  {\bibfnamefont {X.}~\bibnamefont {Wang}}, \bibinfo {author} {\bibfnamefont
  {H.}~\bibnamefont {Miao}}, \bibinfo {author} {\bibfnamefont {J.}~\bibnamefont
  {Ma}}, \bibinfo {author} {\bibfnamefont {P.}~\bibnamefont {Richard}},
  \bibinfo {author} {\bibfnamefont {X.}~\bibnamefont {Huang}}, \bibinfo
  {author} {\bibfnamefont {L.}~\bibnamefont {Zhao}}, \bibinfo {author}
  {\bibfnamefont {G.}~\bibnamefont {Chen}},  \emph {et~al.},\ }\href@noop {}
  {\bibfield  {journal} {\bibinfo  {journal} {Physical Review X}\ }\textbf
  {\bibinfo {volume} {5}},\ \bibinfo {pages} {031013} (\bibinfo {year}
  {2015}{\natexlab{a}})}\BibitemShut {NoStop}%
\bibitem [{\citenamefont {Lv}\ \emph {et~al.}(2015{\natexlab{b}})\citenamefont
  {Lv}, \citenamefont {Xu}, \citenamefont {Weng}, \citenamefont {Ma},
  \citenamefont {Richard}, \citenamefont {Huang}, \citenamefont {Zhao},
  \citenamefont {Chen}, \citenamefont {Matt}, \citenamefont {Bisti} \emph
  {et~al.}}]{lvNatPhys2015}%
  \BibitemOpen
  \bibfield  {author} {\bibinfo {author} {\bibfnamefont {B.}~\bibnamefont
  {Lv}}, \bibinfo {author} {\bibfnamefont {N.}~\bibnamefont {Xu}}, \bibinfo
  {author} {\bibfnamefont {H.}~\bibnamefont {Weng}}, \bibinfo {author}
  {\bibfnamefont {J.}~\bibnamefont {Ma}}, \bibinfo {author} {\bibfnamefont
  {P.}~\bibnamefont {Richard}}, \bibinfo {author} {\bibfnamefont
  {X.}~\bibnamefont {Huang}}, \bibinfo {author} {\bibfnamefont
  {L.}~\bibnamefont {Zhao}}, \bibinfo {author} {\bibfnamefont {G.}~\bibnamefont
  {Chen}}, \bibinfo {author} {\bibfnamefont {C.}~\bibnamefont {Matt}}, \bibinfo
  {author} {\bibfnamefont {F.}~\bibnamefont {Bisti}},  \emph {et~al.},\
  }\href@noop {} {\bibfield  {journal} {\bibinfo  {journal} {Nature Physics}\
  }\textbf {\bibinfo {volume} {11}},\ \bibinfo {pages} {724} (\bibinfo {year}
  {2015}{\natexlab{b}})}\BibitemShut {NoStop}%
\bibitem [{\citenamefont {Xu}\ \emph {et~al.}(2015{\natexlab{b}})\citenamefont
  {Xu}, \citenamefont {Alidoust}, \citenamefont {Belopolski}, \citenamefont
  {Yuan}, \citenamefont {Bian}, \citenamefont {Chang}, \citenamefont {Zheng},
  \citenamefont {Strocov}, \citenamefont {Sanchez}, \citenamefont {Chang} \emph
  {et~al.}}]{xuNatphys2015}%
  \BibitemOpen
  \bibfield  {author} {\bibinfo {author} {\bibfnamefont {S.-Y.}\ \bibnamefont
  {Xu}}, \bibinfo {author} {\bibfnamefont {N.}~\bibnamefont {Alidoust}},
  \bibinfo {author} {\bibfnamefont {I.}~\bibnamefont {Belopolski}}, \bibinfo
  {author} {\bibfnamefont {Z.}~\bibnamefont {Yuan}}, \bibinfo {author}
  {\bibfnamefont {G.}~\bibnamefont {Bian}}, \bibinfo {author} {\bibfnamefont
  {T.-R.}\ \bibnamefont {Chang}}, \bibinfo {author} {\bibfnamefont
  {H.}~\bibnamefont {Zheng}}, \bibinfo {author} {\bibfnamefont {V.~N.}\
  \bibnamefont {Strocov}}, \bibinfo {author} {\bibfnamefont {D.~S.}\
  \bibnamefont {Sanchez}}, \bibinfo {author} {\bibfnamefont {G.}~\bibnamefont
  {Chang}},  \emph {et~al.},\ }\href@noop {} {\bibfield  {journal} {\bibinfo
  {journal} {Nature Physics}\ }\textbf {\bibinfo {volume} {11}},\ \bibinfo
  {pages} {748} (\bibinfo {year} {2015}{\natexlab{b}})}\BibitemShut {NoStop}%
\bibitem [{\citenamefont {Yang}\ \emph {et~al.}(2015)\citenamefont {Yang},
  \citenamefont {Liu}, \citenamefont {Sun}, \citenamefont {Peng}, \citenamefont
  {Yang}, \citenamefont {Zhang}, \citenamefont {Zhou}, \citenamefont {Zhang},
  \citenamefont {Guo}, \citenamefont {Rahn} \emph {et~al.}}]{YangNatPhys2015}%
  \BibitemOpen
  \bibfield  {author} {\bibinfo {author} {\bibfnamefont {L.}~\bibnamefont
  {Yang}}, \bibinfo {author} {\bibfnamefont {Z.}~\bibnamefont {Liu}}, \bibinfo
  {author} {\bibfnamefont {Y.}~\bibnamefont {Sun}}, \bibinfo {author}
  {\bibfnamefont {H.}~\bibnamefont {Peng}}, \bibinfo {author} {\bibfnamefont
  {H.}~\bibnamefont {Yang}}, \bibinfo {author} {\bibfnamefont {T.}~\bibnamefont
  {Zhang}}, \bibinfo {author} {\bibfnamefont {B.}~\bibnamefont {Zhou}},
  \bibinfo {author} {\bibfnamefont {Y.}~\bibnamefont {Zhang}}, \bibinfo
  {author} {\bibfnamefont {Y.}~\bibnamefont {Guo}}, \bibinfo {author}
  {\bibfnamefont {M.}~\bibnamefont {Rahn}},  \emph {et~al.},\ }\href@noop {}
  {\bibfield  {journal} {\bibinfo  {journal} {Nature Physics}\ }\textbf
  {\bibinfo {volume} {11}},\ \bibinfo {pages} {728} (\bibinfo {year}
  {2015})}\BibitemShut {NoStop}%
\bibitem [{\citenamefont {Xu}\ \emph {et~al.}(2016{\natexlab{a}})\citenamefont
  {Xu}, \citenamefont {Weng}, \citenamefont {Lv}, \citenamefont {Matt},
  \citenamefont {Park}, \citenamefont {Bisti}, \citenamefont {Strocov},
  \citenamefont {Gawryluk}, \citenamefont {Pomjakushina}, \citenamefont
  {Conder} \emph {et~al.}}]{xuNatComm2016}%
  \BibitemOpen
  \bibfield  {author} {\bibinfo {author} {\bibfnamefont {N.}~\bibnamefont
  {Xu}}, \bibinfo {author} {\bibfnamefont {H.}~\bibnamefont {Weng}}, \bibinfo
  {author} {\bibfnamefont {B.}~\bibnamefont {Lv}}, \bibinfo {author}
  {\bibfnamefont {C.~E.}\ \bibnamefont {Matt}}, \bibinfo {author}
  {\bibfnamefont {J.}~\bibnamefont {Park}}, \bibinfo {author} {\bibfnamefont
  {F.}~\bibnamefont {Bisti}}, \bibinfo {author} {\bibfnamefont {V.~N.}\
  \bibnamefont {Strocov}}, \bibinfo {author} {\bibfnamefont {D.}~\bibnamefont
  {Gawryluk}}, \bibinfo {author} {\bibfnamefont {E.}~\bibnamefont
  {Pomjakushina}}, \bibinfo {author} {\bibfnamefont {K.}~\bibnamefont
  {Conder}},  \emph {et~al.},\ }\href@noop {} {\bibfield  {journal} {\bibinfo
  {journal} {Nature communications}\ }\textbf {\bibinfo {volume} {7}},\
  \bibinfo {pages} {11006} (\bibinfo {year} {2016}{\natexlab{a}})}\BibitemShut
  {NoStop}%
\bibitem [{\citenamefont {Belopolski}\ \emph {et~al.}(2019)\citenamefont
  {Belopolski}, \citenamefont {Manna}, \citenamefont {Sanchez}, \citenamefont
  {Chang}, \citenamefont {Ernst}, \citenamefont {Yin}, \citenamefont {Zhang},
  \citenamefont {Cochran}, \citenamefont {Shumiya}, \citenamefont {Zheng} \emph
  {et~al.}}]{belopolskiScience2019}%
  \BibitemOpen
  \bibfield  {author} {\bibinfo {author} {\bibfnamefont {I.}~\bibnamefont
  {Belopolski}}, \bibinfo {author} {\bibfnamefont {K.}~\bibnamefont {Manna}},
  \bibinfo {author} {\bibfnamefont {D.~S.}\ \bibnamefont {Sanchez}}, \bibinfo
  {author} {\bibfnamefont {G.}~\bibnamefont {Chang}}, \bibinfo {author}
  {\bibfnamefont {B.}~\bibnamefont {Ernst}}, \bibinfo {author} {\bibfnamefont
  {J.}~\bibnamefont {Yin}}, \bibinfo {author} {\bibfnamefont {S.~S.}\
  \bibnamefont {Zhang}}, \bibinfo {author} {\bibfnamefont {T.}~\bibnamefont
  {Cochran}}, \bibinfo {author} {\bibfnamefont {N.}~\bibnamefont {Shumiya}},
  \bibinfo {author} {\bibfnamefont {H.}~\bibnamefont {Zheng}},  \emph
  {et~al.},\ }\href@noop {} {\bibfield  {journal} {\bibinfo  {journal}
  {Science}\ }\textbf {\bibinfo {volume} {365}},\ \bibinfo {pages} {1278}
  (\bibinfo {year} {2019})}\BibitemShut {NoStop}%
\bibitem [{\citenamefont {Liu}\ \emph {et~al.}(2019)\citenamefont {Liu},
  \citenamefont {Liang}, \citenamefont {Liu}, \citenamefont {Xu}, \citenamefont
  {Li}, \citenamefont {Chen}, \citenamefont {Pei}, \citenamefont {Shi},
  \citenamefont {Mo}, \citenamefont {Dudin} \emph {et~al.}}]{liuScience2019}%
  \BibitemOpen
  \bibfield  {author} {\bibinfo {author} {\bibfnamefont {D.}~\bibnamefont
  {Liu}}, \bibinfo {author} {\bibfnamefont {A.}~\bibnamefont {Liang}}, \bibinfo
  {author} {\bibfnamefont {E.}~\bibnamefont {Liu}}, \bibinfo {author}
  {\bibfnamefont {Q.}~\bibnamefont {Xu}}, \bibinfo {author} {\bibfnamefont
  {Y.}~\bibnamefont {Li}}, \bibinfo {author} {\bibfnamefont {C.}~\bibnamefont
  {Chen}}, \bibinfo {author} {\bibfnamefont {D.}~\bibnamefont {Pei}}, \bibinfo
  {author} {\bibfnamefont {W.}~\bibnamefont {Shi}}, \bibinfo {author}
  {\bibfnamefont {S.}~\bibnamefont {Mo}}, \bibinfo {author} {\bibfnamefont
  {P.}~\bibnamefont {Dudin}},  \emph {et~al.},\ }\href@noop {} {\bibfield
  {journal} {\bibinfo  {journal} {Science}\ }\textbf {\bibinfo {volume}
  {365}},\ \bibinfo {pages} {1282} (\bibinfo {year} {2019})}\BibitemShut
  {NoStop}%
\bibitem [{\citenamefont {Morali}\ \emph {et~al.}(2019)\citenamefont {Morali},
  \citenamefont {Batabyal}, \citenamefont {Nag}, \citenamefont {Liu},
  \citenamefont {Xu}, \citenamefont {Sun}, \citenamefont {Yan}, \citenamefont
  {Felser}, \citenamefont {Avraham},\ and\ \citenamefont
  {Beidenkopf}}]{moraliScience2019}%
  \BibitemOpen
  \bibfield  {author} {\bibinfo {author} {\bibfnamefont {N.}~\bibnamefont
  {Morali}}, \bibinfo {author} {\bibfnamefont {R.}~\bibnamefont {Batabyal}},
  \bibinfo {author} {\bibfnamefont {P.~K.}\ \bibnamefont {Nag}}, \bibinfo
  {author} {\bibfnamefont {E.}~\bibnamefont {Liu}}, \bibinfo {author}
  {\bibfnamefont {Q.}~\bibnamefont {Xu}}, \bibinfo {author} {\bibfnamefont
  {Y.}~\bibnamefont {Sun}}, \bibinfo {author} {\bibfnamefont {B.}~\bibnamefont
  {Yan}}, \bibinfo {author} {\bibfnamefont {C.}~\bibnamefont {Felser}},
  \bibinfo {author} {\bibfnamefont {N.}~\bibnamefont {Avraham}}, \ and\
  \bibinfo {author} {\bibfnamefont {H.}~\bibnamefont {Beidenkopf}},\
  }\href@noop {} {\bibfield  {journal} {\bibinfo  {journal} {arXiv preprint
  arXiv:1903.00509}\ } (\bibinfo {year} {2019})}\BibitemShut {NoStop}%
\bibitem [{\citenamefont {Wieder}\ \emph {et~al.}(2016)\citenamefont {Wieder},
  \citenamefont {Kim}, \citenamefont {Rappe},\ and\ \citenamefont
  {Kane}}]{wiederPRL2016}%
  \BibitemOpen
  \bibfield  {author} {\bibinfo {author} {\bibfnamefont {B.~J.}\ \bibnamefont
  {Wieder}}, \bibinfo {author} {\bibfnamefont {Y.}~\bibnamefont {Kim}},
  \bibinfo {author} {\bibfnamefont {A.}~\bibnamefont {Rappe}}, \ and\ \bibinfo
  {author} {\bibfnamefont {C.}~\bibnamefont {Kane}},\ }\href@noop {} {\bibfield
   {journal} {\bibinfo  {journal} {Physical review letters}\ }\textbf {\bibinfo
  {volume} {116}},\ \bibinfo {pages} {186402} (\bibinfo {year}
  {2016})}\BibitemShut {NoStop}%
\bibitem [{\citenamefont {Bradlyn}\ \emph {et~al.}(2016)\citenamefont
  {Bradlyn}, \citenamefont {Cano}, \citenamefont {Wang}, \citenamefont
  {Vergniory}, \citenamefont {Felser}, \citenamefont {Cava},\ and\
  \citenamefont {Bernevig}}]{bradlynScience2016}%
  \BibitemOpen
  \bibfield  {author} {\bibinfo {author} {\bibfnamefont {B.}~\bibnamefont
  {Bradlyn}}, \bibinfo {author} {\bibfnamefont {J.}~\bibnamefont {Cano}},
  \bibinfo {author} {\bibfnamefont {Z.}~\bibnamefont {Wang}}, \bibinfo {author}
  {\bibfnamefont {M.}~\bibnamefont {Vergniory}}, \bibinfo {author}
  {\bibfnamefont {C.}~\bibnamefont {Felser}}, \bibinfo {author} {\bibfnamefont
  {R.~J.}\ \bibnamefont {Cava}}, \ and\ \bibinfo {author} {\bibfnamefont
  {B.~A.}\ \bibnamefont {Bernevig}},\ }\href@noop {} {\bibfield  {journal}
  {\bibinfo  {journal} {Science}\ }\textbf {\bibinfo {volume} {353}},\ \bibinfo
  {pages} {aaf5037} (\bibinfo {year} {2016})}\BibitemShut {NoStop}%
\bibitem [{\citenamefont {Chang}\ \emph
  {et~al.}(2017{\natexlab{a}})\citenamefont {Chang}, \citenamefont {Xu},
  \citenamefont {Wieder}, \citenamefont {Sanchez}, \citenamefont {Huang},
  \citenamefont {Belopolski}, \citenamefont {Chang}, \citenamefont {Zhang},
  \citenamefont {Bansil}, \citenamefont {Lin} \emph {et~al.}}]{changPRL2017}%
  \BibitemOpen
  \bibfield  {author} {\bibinfo {author} {\bibfnamefont {G.}~\bibnamefont
  {Chang}}, \bibinfo {author} {\bibfnamefont {S.-Y.}\ \bibnamefont {Xu}},
  \bibinfo {author} {\bibfnamefont {B.~J.}\ \bibnamefont {Wieder}}, \bibinfo
  {author} {\bibfnamefont {D.~S.}\ \bibnamefont {Sanchez}}, \bibinfo {author}
  {\bibfnamefont {S.-M.}\ \bibnamefont {Huang}}, \bibinfo {author}
  {\bibfnamefont {I.}~\bibnamefont {Belopolski}}, \bibinfo {author}
  {\bibfnamefont {T.-R.}\ \bibnamefont {Chang}}, \bibinfo {author}
  {\bibfnamefont {S.}~\bibnamefont {Zhang}}, \bibinfo {author} {\bibfnamefont
  {A.}~\bibnamefont {Bansil}}, \bibinfo {author} {\bibfnamefont
  {H.}~\bibnamefont {Lin}},  \emph {et~al.},\ }\href@noop {} {\bibfield
  {journal} {\bibinfo  {journal} {Physical review letters}\ }\textbf {\bibinfo
  {volume} {119}},\ \bibinfo {pages} {206401} (\bibinfo {year}
  {2017}{\natexlab{a}})}\BibitemShut {NoStop}%
\bibitem [{\citenamefont {Tang}\ \emph {et~al.}(2017)\citenamefont {Tang},
  \citenamefont {Zhou},\ and\ \citenamefont {Zhang}}]{tangPRL2017}%
  \BibitemOpen
  \bibfield  {author} {\bibinfo {author} {\bibfnamefont {P.}~\bibnamefont
  {Tang}}, \bibinfo {author} {\bibfnamefont {Q.}~\bibnamefont {Zhou}}, \ and\
  \bibinfo {author} {\bibfnamefont {S.-C.}\ \bibnamefont {Zhang}},\ }\href@noop
  {} {\bibfield  {journal} {\bibinfo  {journal} {Physical review letters}\
  }\textbf {\bibinfo {volume} {119}},\ \bibinfo {pages} {206402} (\bibinfo
  {year} {2017})}\BibitemShut {NoStop}%
\bibitem [{\citenamefont {Rao}\ \emph {et~al.}(2019)\citenamefont {Rao},
  \citenamefont {Li}, \citenamefont {Zhang}, \citenamefont {Tian},
  \citenamefont {Li}, \citenamefont {Fu}, \citenamefont {Tang}, \citenamefont
  {Wang}, \citenamefont {Li}, \citenamefont {Fan} \emph
  {et~al.}}]{raoNature2019}%
  \BibitemOpen
  \bibfield  {author} {\bibinfo {author} {\bibfnamefont {Z.}~\bibnamefont
  {Rao}}, \bibinfo {author} {\bibfnamefont {H.}~\bibnamefont {Li}}, \bibinfo
  {author} {\bibfnamefont {T.}~\bibnamefont {Zhang}}, \bibinfo {author}
  {\bibfnamefont {S.}~\bibnamefont {Tian}}, \bibinfo {author} {\bibfnamefont
  {C.}~\bibnamefont {Li}}, \bibinfo {author} {\bibfnamefont {B.}~\bibnamefont
  {Fu}}, \bibinfo {author} {\bibfnamefont {C.}~\bibnamefont {Tang}}, \bibinfo
  {author} {\bibfnamefont {L.}~\bibnamefont {Wang}}, \bibinfo {author}
  {\bibfnamefont {Z.}~\bibnamefont {Li}}, \bibinfo {author} {\bibfnamefont
  {W.}~\bibnamefont {Fan}},  \emph {et~al.},\ }\href@noop {} {\bibfield
  {journal} {\bibinfo  {journal} {Nature}\ }\textbf {\bibinfo {volume} {567}},\
  \bibinfo {pages} {496} (\bibinfo {year} {2019})}\BibitemShut {NoStop}%
\bibitem [{\citenamefont {Sanchez}\ \emph
  {et~al.}(2019{\natexlab{a}})\citenamefont {Sanchez}, \citenamefont
  {Belopolski}, \citenamefont {Cochran}, \citenamefont {Xu}, \citenamefont
  {Yin}, \citenamefont {Chang}, \citenamefont {Xie}, \citenamefont {Manna},
  \citenamefont {S{\"u}{\ss}}, \citenamefont {Huang} \emph
  {et~al.}}]{sanchezPRL2019}%
  \BibitemOpen
  \bibfield  {author} {\bibinfo {author} {\bibfnamefont {D.~S.}\ \bibnamefont
  {Sanchez}}, \bibinfo {author} {\bibfnamefont {I.}~\bibnamefont {Belopolski}},
  \bibinfo {author} {\bibfnamefont {T.~A.}\ \bibnamefont {Cochran}}, \bibinfo
  {author} {\bibfnamefont {X.}~\bibnamefont {Xu}}, \bibinfo {author}
  {\bibfnamefont {J.-X.}\ \bibnamefont {Yin}}, \bibinfo {author} {\bibfnamefont
  {G.}~\bibnamefont {Chang}}, \bibinfo {author} {\bibfnamefont
  {W.}~\bibnamefont {Xie}}, \bibinfo {author} {\bibfnamefont {K.}~\bibnamefont
  {Manna}}, \bibinfo {author} {\bibfnamefont {V.}~\bibnamefont {S{\"u}{\ss}}},
  \bibinfo {author} {\bibfnamefont {C.-Y.}\ \bibnamefont {Huang}},  \emph
  {et~al.},\ }\href@noop {} {\bibfield  {journal} {\bibinfo  {journal}
  {Nature}\ }\textbf {\bibinfo {volume} {567}},\ \bibinfo {pages} {500}
  (\bibinfo {year} {2019}{\natexlab{a}})}\BibitemShut {NoStop}%
\bibitem [{\citenamefont {Takane}\ \emph {et~al.}(2019)\citenamefont {Takane},
  \citenamefont {Wang}, \citenamefont {Souma}, \citenamefont {Nakayama},
  \citenamefont {Nakamura}, \citenamefont {Oinuma}, \citenamefont {Nakata},
  \citenamefont {Iwasawa}, \citenamefont {Cacho}, \citenamefont {Kim} \emph
  {et~al.}}]{takanePRL2019}%
  \BibitemOpen
  \bibfield  {author} {\bibinfo {author} {\bibfnamefont {D.}~\bibnamefont
  {Takane}}, \bibinfo {author} {\bibfnamefont {Z.}~\bibnamefont {Wang}},
  \bibinfo {author} {\bibfnamefont {S.}~\bibnamefont {Souma}}, \bibinfo
  {author} {\bibfnamefont {K.}~\bibnamefont {Nakayama}}, \bibinfo {author}
  {\bibfnamefont {T.}~\bibnamefont {Nakamura}}, \bibinfo {author}
  {\bibfnamefont {H.}~\bibnamefont {Oinuma}}, \bibinfo {author} {\bibfnamefont
  {Y.}~\bibnamefont {Nakata}}, \bibinfo {author} {\bibfnamefont
  {H.}~\bibnamefont {Iwasawa}}, \bibinfo {author} {\bibfnamefont
  {C.}~\bibnamefont {Cacho}}, \bibinfo {author} {\bibfnamefont
  {T.}~\bibnamefont {Kim}},  \emph {et~al.},\ }\href@noop {} {\bibfield
  {journal} {\bibinfo  {journal} {Physical review letters}\ }\textbf {\bibinfo
  {volume} {122}},\ \bibinfo {pages} {076402} (\bibinfo {year}
  {2019})}\BibitemShut {NoStop}%
\bibitem [{\citenamefont {Schr{\"o}ter}\ \emph
  {et~al.}(2019{\natexlab{a}})\citenamefont {Schr{\"o}ter}, \citenamefont
  {Pei}, \citenamefont {Vergniory}, \citenamefont {Sun}, \citenamefont {Manna},
  \citenamefont {de~Juan}, \citenamefont {Krieger}, \citenamefont {S{\"u}ss},
  \citenamefont {Schmidt}, \citenamefont {Dudin} \emph
  {et~al.}}]{schroterNatPhys2019}%
  \BibitemOpen
  \bibfield  {author} {\bibinfo {author} {\bibfnamefont {N.~B.}\ \bibnamefont
  {Schr{\"o}ter}}, \bibinfo {author} {\bibfnamefont {D.}~\bibnamefont {Pei}},
  \bibinfo {author} {\bibfnamefont {M.~G.}\ \bibnamefont {Vergniory}}, \bibinfo
  {author} {\bibfnamefont {Y.}~\bibnamefont {Sun}}, \bibinfo {author}
  {\bibfnamefont {K.}~\bibnamefont {Manna}}, \bibinfo {author} {\bibfnamefont
  {F.}~\bibnamefont {de~Juan}}, \bibinfo {author} {\bibfnamefont {J.~A.}\
  \bibnamefont {Krieger}}, \bibinfo {author} {\bibfnamefont {V.}~\bibnamefont
  {S{\"u}ss}}, \bibinfo {author} {\bibfnamefont {M.}~\bibnamefont {Schmidt}},
  \bibinfo {author} {\bibfnamefont {P.}~\bibnamefont {Dudin}},  \emph
  {et~al.},\ }\href@noop {} {\bibfield  {journal} {\bibinfo  {journal} {Nature
  Physics}\ ,\ \bibinfo {pages} {1}} (\bibinfo {year}
  {2019}{\natexlab{a}})}\BibitemShut {NoStop}%
\bibitem [{\citenamefont {Schr{\"o}ter}\ \emph
  {et~al.}(2019{\natexlab{b}})\citenamefont {Schr{\"o}ter}, \citenamefont
  {Stolz}, \citenamefont {Manna}, \citenamefont {de~Juan}, \citenamefont
  {Vergniory}, \citenamefont {Krieger}, \citenamefont {Pei}, \citenamefont
  {Dudin}, \citenamefont {Kim}, \citenamefont {Cacho} \emph
  {et~al.}}]{schroterArxiv2019}%
  \BibitemOpen
  \bibfield  {author} {\bibinfo {author} {\bibfnamefont {N.}~\bibnamefont
  {Schr{\"o}ter}}, \bibinfo {author} {\bibfnamefont {S.}~\bibnamefont {Stolz}},
  \bibinfo {author} {\bibfnamefont {K.}~\bibnamefont {Manna}}, \bibinfo
  {author} {\bibfnamefont {F.}~\bibnamefont {de~Juan}}, \bibinfo {author}
  {\bibfnamefont {M.~G.}\ \bibnamefont {Vergniory}}, \bibinfo {author}
  {\bibfnamefont {J.~A.}\ \bibnamefont {Krieger}}, \bibinfo {author}
  {\bibfnamefont {D.}~\bibnamefont {Pei}}, \bibinfo {author} {\bibfnamefont
  {P.}~\bibnamefont {Dudin}}, \bibinfo {author} {\bibfnamefont {T.~K.}\
  \bibnamefont {Kim}}, \bibinfo {author} {\bibfnamefont {C.}~\bibnamefont
  {Cacho}},  \emph {et~al.},\ }\href@noop {} {\bibfield  {journal} {\bibinfo
  {journal} {arXiv preprint arXiv:1907.08723}\ } (\bibinfo {year}
  {2019}{\natexlab{b}})}\BibitemShut {NoStop}%
\bibitem [{\citenamefont {Lv}\ \emph {et~al.}(2017)\citenamefont {Lv},
  \citenamefont {Feng}, \citenamefont {Xu}, \citenamefont {Gao}, \citenamefont
  {Ma}, \citenamefont {Kong}, \citenamefont {Richard}, \citenamefont {Huang},
  \citenamefont {Strocov}, \citenamefont {Fang} \emph {et~al.}}]{lvNature2017}%
  \BibitemOpen
  \bibfield  {author} {\bibinfo {author} {\bibfnamefont {B.}~\bibnamefont
  {Lv}}, \bibinfo {author} {\bibfnamefont {Z.-L.}\ \bibnamefont {Feng}},
  \bibinfo {author} {\bibfnamefont {Q.-N.}\ \bibnamefont {Xu}}, \bibinfo
  {author} {\bibfnamefont {X.}~\bibnamefont {Gao}}, \bibinfo {author}
  {\bibfnamefont {J.-Z.}\ \bibnamefont {Ma}}, \bibinfo {author} {\bibfnamefont
  {L.-Y.}\ \bibnamefont {Kong}}, \bibinfo {author} {\bibfnamefont
  {P.}~\bibnamefont {Richard}}, \bibinfo {author} {\bibfnamefont {Y.-B.}\
  \bibnamefont {Huang}}, \bibinfo {author} {\bibfnamefont {V.}~\bibnamefont
  {Strocov}}, \bibinfo {author} {\bibfnamefont {C.}~\bibnamefont {Fang}},
  \emph {et~al.},\ }\href@noop {} {\bibfield  {journal} {\bibinfo  {journal}
  {Nature}\ }\textbf {\bibinfo {volume} {546}},\ \bibinfo {pages} {627}
  (\bibinfo {year} {2017})}\BibitemShut {NoStop}%
\bibitem [{\citenamefont {Weng}\ \emph
  {et~al.}(2016{\natexlab{a}})\citenamefont {Weng}, \citenamefont {Fang},
  \citenamefont {Fang},\ and\ \citenamefont {Dai}}]{wengPRB2016}%
  \BibitemOpen
  \bibfield  {author} {\bibinfo {author} {\bibfnamefont {H.}~\bibnamefont
  {Weng}}, \bibinfo {author} {\bibfnamefont {C.}~\bibnamefont {Fang}}, \bibinfo
  {author} {\bibfnamefont {Z.}~\bibnamefont {Fang}}, \ and\ \bibinfo {author}
  {\bibfnamefont {X.}~\bibnamefont {Dai}},\ }\href@noop {} {\bibfield
  {journal} {\bibinfo  {journal} {Physical Review B}\ }\textbf {\bibinfo
  {volume} {93}},\ \bibinfo {pages} {241202} (\bibinfo {year}
  {2016}{\natexlab{a}})}\BibitemShut {NoStop}%
\bibitem [{\citenamefont {Weng}\ \emph
  {et~al.}(2016{\natexlab{b}})\citenamefont {Weng}, \citenamefont {Fang},
  \citenamefont {Fang},\ and\ \citenamefont {Dai}}]{wengPRB22016}%
  \BibitemOpen
  \bibfield  {author} {\bibinfo {author} {\bibfnamefont {H.}~\bibnamefont
  {Weng}}, \bibinfo {author} {\bibfnamefont {C.}~\bibnamefont {Fang}}, \bibinfo
  {author} {\bibfnamefont {Z.}~\bibnamefont {Fang}}, \ and\ \bibinfo {author}
  {\bibfnamefont {X.}~\bibnamefont {Dai}},\ }\href@noop {} {\bibfield
  {journal} {\bibinfo  {journal} {Physical Review B}\ }\textbf {\bibinfo
  {volume} {94}},\ \bibinfo {pages} {165201} (\bibinfo {year}
  {2016}{\natexlab{b}})}\BibitemShut {NoStop}%
\bibitem [{\citenamefont {Zhu}\ \emph {et~al.}(2016)\citenamefont {Zhu},
  \citenamefont {Winkler}, \citenamefont {Wu}, \citenamefont {Li},\ and\
  \citenamefont {Soluyanov}}]{zhuPRX2016}%
  \BibitemOpen
  \bibfield  {author} {\bibinfo {author} {\bibfnamefont {Z.}~\bibnamefont
  {Zhu}}, \bibinfo {author} {\bibfnamefont {G.~W.}\ \bibnamefont {Winkler}},
  \bibinfo {author} {\bibfnamefont {Q.}~\bibnamefont {Wu}}, \bibinfo {author}
  {\bibfnamefont {J.}~\bibnamefont {Li}}, \ and\ \bibinfo {author}
  {\bibfnamefont {A.~A.}\ \bibnamefont {Soluyanov}},\ }\href@noop {} {\bibfield
   {journal} {\bibinfo  {journal} {Physical Review X}\ }\textbf {\bibinfo
  {volume} {6}},\ \bibinfo {pages} {031003} (\bibinfo {year}
  {2016})}\BibitemShut {NoStop}%
\bibitem [{\citenamefont {Chang}\ \emph
  {et~al.}(2017{\natexlab{b}})\citenamefont {Chang}, \citenamefont {Xu},
  \citenamefont {Huang}, \citenamefont {Sanchez}, \citenamefont {Hsu},
  \citenamefont {Bian}, \citenamefont {Yu}, \citenamefont {Belopolski},
  \citenamefont {Alidoust}, \citenamefont {Zheng} \emph
  {et~al.}}]{changSciRep2017}%
  \BibitemOpen
  \bibfield  {author} {\bibinfo {author} {\bibfnamefont {G.}~\bibnamefont
  {Chang}}, \bibinfo {author} {\bibfnamefont {S.-Y.}\ \bibnamefont {Xu}},
  \bibinfo {author} {\bibfnamefont {S.-M.}\ \bibnamefont {Huang}}, \bibinfo
  {author} {\bibfnamefont {D.~S.}\ \bibnamefont {Sanchez}}, \bibinfo {author}
  {\bibfnamefont {C.-H.}\ \bibnamefont {Hsu}}, \bibinfo {author} {\bibfnamefont
  {G.}~\bibnamefont {Bian}}, \bibinfo {author} {\bibfnamefont {Z.-M.}\
  \bibnamefont {Yu}}, \bibinfo {author} {\bibfnamefont {I.}~\bibnamefont
  {Belopolski}}, \bibinfo {author} {\bibfnamefont {N.}~\bibnamefont
  {Alidoust}}, \bibinfo {author} {\bibfnamefont {H.}~\bibnamefont {Zheng}},
  \emph {et~al.},\ }\href@noop {} {\bibfield  {journal} {\bibinfo  {journal}
  {Scientific reports}\ }\textbf {\bibinfo {volume} {7}},\ \bibinfo {pages} {1}
  (\bibinfo {year} {2017}{\natexlab{b}})}\BibitemShut {NoStop}%
\bibitem [{\citenamefont {Ma}\ and\ \citenamefont {Pesin}(2015)}]{ma2015PRB}%
  \BibitemOpen
  \bibfield  {author} {\bibinfo {author} {\bibfnamefont {J.}~\bibnamefont
  {Ma}}\ and\ \bibinfo {author} {\bibfnamefont {D.}~\bibnamefont {Pesin}},\
  }\href@noop {} {\bibfield  {journal} {\bibinfo  {journal} {Physical Review
  B}\ }\textbf {\bibinfo {volume} {92}},\ \bibinfo {pages} {235205} (\bibinfo
  {year} {2015})}\BibitemShut {NoStop}%
\bibitem [{\citenamefont {Zhong}\ \emph {et~al.}(2016)\citenamefont {Zhong},
  \citenamefont {Moore},\ and\ \citenamefont {Souza}}]{zhongPRL2016}%
  \BibitemOpen
  \bibfield  {author} {\bibinfo {author} {\bibfnamefont {S.}~\bibnamefont
  {Zhong}}, \bibinfo {author} {\bibfnamefont {J.~E.}\ \bibnamefont {Moore}}, \
  and\ \bibinfo {author} {\bibfnamefont {I.}~\bibnamefont {Souza}},\
  }\href@noop {} {\bibfield  {journal} {\bibinfo  {journal} {Physical review
  letters}\ }\textbf {\bibinfo {volume} {116}},\ \bibinfo {pages} {077201}
  (\bibinfo {year} {2016})}\BibitemShut {NoStop}%
\bibitem [{\citenamefont {de~Juan}\ \emph {et~al.}(2017)\citenamefont
  {de~Juan}, \citenamefont {Grushin}, \citenamefont {Morimoto},\ and\
  \citenamefont {Moore}}]{dejuanNatComm2017}%
  \BibitemOpen
  \bibfield  {author} {\bibinfo {author} {\bibfnamefont {F.}~\bibnamefont
  {de~Juan}}, \bibinfo {author} {\bibfnamefont {A.~G.}\ \bibnamefont
  {Grushin}}, \bibinfo {author} {\bibfnamefont {T.}~\bibnamefont {Morimoto}}, \
  and\ \bibinfo {author} {\bibfnamefont {J.~E.}\ \bibnamefont {Moore}},\
  }\href@noop {} {\bibfield  {journal} {\bibinfo  {journal} {Nature
  communications}\ }\textbf {\bibinfo {volume} {8}},\ \bibinfo {pages} {15995}
  (\bibinfo {year} {2017})}\BibitemShut {NoStop}%
\bibitem [{\citenamefont {Flicker}\ \emph
  {et~al.}(2018{\natexlab{a}})\citenamefont {Flicker}, \citenamefont {De~Juan},
  \citenamefont {Bradlyn}, \citenamefont {Morimoto}, \citenamefont
  {Vergniory},\ and\ \citenamefont {Grushin}}]{flickerPRB2018}%
  \BibitemOpen
  \bibfield  {author} {\bibinfo {author} {\bibfnamefont {F.}~\bibnamefont
  {Flicker}}, \bibinfo {author} {\bibfnamefont {F.}~\bibnamefont {De~Juan}},
  \bibinfo {author} {\bibfnamefont {B.}~\bibnamefont {Bradlyn}}, \bibinfo
  {author} {\bibfnamefont {T.}~\bibnamefont {Morimoto}}, \bibinfo {author}
  {\bibfnamefont {M.~G.}\ \bibnamefont {Vergniory}}, \ and\ \bibinfo {author}
  {\bibfnamefont {A.~G.}\ \bibnamefont {Grushin}},\ }\href@noop {} {\bibfield
  {journal} {\bibinfo  {journal} {Physical Review B}\ }\textbf {\bibinfo
  {volume} {98}},\ \bibinfo {pages} {155145} (\bibinfo {year}
  {2018}{\natexlab{a}})}\BibitemShut {NoStop}%
\bibitem [{\citenamefont {de~Juan}\ \emph {et~al.}(2020)\citenamefont
  {de~Juan}, \citenamefont {Zhang}, \citenamefont {Morimoto}, \citenamefont
  {Sun}, \citenamefont {Moore},\ and\ \citenamefont
  {Grushin}}]{dejuanarxiv2019}%
  \BibitemOpen
  \bibfield  {author} {\bibinfo {author} {\bibfnamefont {F.}~\bibnamefont
  {de~Juan}}, \bibinfo {author} {\bibfnamefont {Y.}~\bibnamefont {Zhang}},
  \bibinfo {author} {\bibfnamefont {T.}~\bibnamefont {Morimoto}}, \bibinfo
  {author} {\bibfnamefont {Y.}~\bibnamefont {Sun}}, \bibinfo {author}
  {\bibfnamefont {J.~E.}\ \bibnamefont {Moore}}, \ and\ \bibinfo {author}
  {\bibfnamefont {A.~G.}\ \bibnamefont {Grushin}},\ }\href@noop {} {\bibfield
  {journal} {\bibinfo  {journal} {Physical Review Research}\ }\textbf {\bibinfo
  {volume} {2}},\ \bibinfo {pages} {012017} (\bibinfo {year}
  {2020})}\BibitemShut {NoStop}%
\bibitem [{\citenamefont {Chang}\ \emph {et~al.}(2018)\citenamefont {Chang},
  \citenamefont {Wieder}, \citenamefont {Schindler}, \citenamefont {Sanchez},
  \citenamefont {Belopolski}, \citenamefont {Huang}, \citenamefont {Singh},
  \citenamefont {Wu}, \citenamefont {Chang}, \citenamefont {Neupert} \emph
  {et~al.}}]{changNatMat2018}%
  \BibitemOpen
  \bibfield  {author} {\bibinfo {author} {\bibfnamefont {G.}~\bibnamefont
  {Chang}}, \bibinfo {author} {\bibfnamefont {B.~J.}\ \bibnamefont {Wieder}},
  \bibinfo {author} {\bibfnamefont {F.}~\bibnamefont {Schindler}}, \bibinfo
  {author} {\bibfnamefont {D.~S.}\ \bibnamefont {Sanchez}}, \bibinfo {author}
  {\bibfnamefont {I.}~\bibnamefont {Belopolski}}, \bibinfo {author}
  {\bibfnamefont {S.-M.}\ \bibnamefont {Huang}}, \bibinfo {author}
  {\bibfnamefont {B.}~\bibnamefont {Singh}}, \bibinfo {author} {\bibfnamefont
  {D.}~\bibnamefont {Wu}}, \bibinfo {author} {\bibfnamefont {T.-R.}\
  \bibnamefont {Chang}}, \bibinfo {author} {\bibfnamefont {T.}~\bibnamefont
  {Neupert}},  \emph {et~al.},\ }\href@noop {} {\bibfield  {journal} {\bibinfo
  {journal} {Nature materials}\ }\textbf {\bibinfo {volume} {17}},\ \bibinfo
  {pages} {978} (\bibinfo {year} {2018})}\BibitemShut {NoStop}%
\bibitem [{\citenamefont {S{\'a}nchez-Mart{\'\i}nez}\ \emph
  {et~al.}(2019)\citenamefont {S{\'a}nchez-Mart{\'\i}nez}, \citenamefont
  {de~Juan},\ and\ \citenamefont {Grushin}}]{sanchezPRB2019}%
  \BibitemOpen
  \bibfield  {author} {\bibinfo {author} {\bibfnamefont {M.-{\'A}.}\
  \bibnamefont {S{\'a}nchez-Mart{\'\i}nez}}, \bibinfo {author} {\bibfnamefont
  {F.}~\bibnamefont {de~Juan}}, \ and\ \bibinfo {author} {\bibfnamefont
  {A.~G.}\ \bibnamefont {Grushin}},\ }\href@noop {} {\bibfield  {journal}
  {\bibinfo  {journal} {Physical Review B}\ }\textbf {\bibinfo {volume} {99}},\
  \bibinfo {pages} {155145} (\bibinfo {year} {2019})}\BibitemShut {NoStop}%
\bibitem [{\citenamefont {Wu}\ \emph {et~al.}(2017)\citenamefont {Wu},
  \citenamefont {Patankar}, \citenamefont {Morimoto}, \citenamefont {Nair},
  \citenamefont {Thewalt}, \citenamefont {Little}, \citenamefont {Analytis},
  \citenamefont {Moore},\ and\ \citenamefont {Orenstein}}]{wuNatPhys2017}%
  \BibitemOpen
  \bibfield  {author} {\bibinfo {author} {\bibfnamefont {L.}~\bibnamefont
  {Wu}}, \bibinfo {author} {\bibfnamefont {S.}~\bibnamefont {Patankar}},
  \bibinfo {author} {\bibfnamefont {T.}~\bibnamefont {Morimoto}}, \bibinfo
  {author} {\bibfnamefont {N.~L.}\ \bibnamefont {Nair}}, \bibinfo {author}
  {\bibfnamefont {E.}~\bibnamefont {Thewalt}}, \bibinfo {author} {\bibfnamefont
  {A.}~\bibnamefont {Little}}, \bibinfo {author} {\bibfnamefont {J.~G.}\
  \bibnamefont {Analytis}}, \bibinfo {author} {\bibfnamefont {J.~E.}\
  \bibnamefont {Moore}}, \ and\ \bibinfo {author} {\bibfnamefont
  {J.}~\bibnamefont {Orenstein}},\ }\href@noop {} {\bibfield  {journal}
  {\bibinfo  {journal} {Nature Physics}\ }\textbf {\bibinfo {volume} {13}},\
  \bibinfo {pages} {350} (\bibinfo {year} {2017})}\BibitemShut {NoStop}%
\bibitem [{\citenamefont {Patankar}\ \emph {et~al.}(2018)\citenamefont
  {Patankar}, \citenamefont {Wu}, \citenamefont {Lu}, \citenamefont {Rai},
  \citenamefont {Tran}, \citenamefont {Morimoto}, \citenamefont {Parker},
  \citenamefont {Grushin}, \citenamefont {Nair}, \citenamefont {Analytis} \emph
  {et~al.}}]{patankarPRB2018}%
  \BibitemOpen
  \bibfield  {author} {\bibinfo {author} {\bibfnamefont {S.}~\bibnamefont
  {Patankar}}, \bibinfo {author} {\bibfnamefont {L.}~\bibnamefont {Wu}},
  \bibinfo {author} {\bibfnamefont {B.}~\bibnamefont {Lu}}, \bibinfo {author}
  {\bibfnamefont {M.}~\bibnamefont {Rai}}, \bibinfo {author} {\bibfnamefont
  {J.~D.}\ \bibnamefont {Tran}}, \bibinfo {author} {\bibfnamefont
  {T.}~\bibnamefont {Morimoto}}, \bibinfo {author} {\bibfnamefont {D.~E.}\
  \bibnamefont {Parker}}, \bibinfo {author} {\bibfnamefont {A.~G.}\
  \bibnamefont {Grushin}}, \bibinfo {author} {\bibfnamefont {N.}~\bibnamefont
  {Nair}}, \bibinfo {author} {\bibfnamefont {J.}~\bibnamefont {Analytis}},
  \emph {et~al.},\ }\href@noop {} {\bibfield  {journal} {\bibinfo  {journal}
  {Physical Review B}\ }\textbf {\bibinfo {volume} {98}},\ \bibinfo {pages}
  {165113} (\bibinfo {year} {2018})}\BibitemShut {NoStop}%
\bibitem [{\citenamefont {Ma}\ \emph {et~al.}(2017)\citenamefont {Ma},
  \citenamefont {Xu}, \citenamefont {Chan}, \citenamefont {Zhang},
  \citenamefont {Chang}, \citenamefont {Lin}, \citenamefont {Xie},
  \citenamefont {Palacios}, \citenamefont {Lin}, \citenamefont {Jia} \emph
  {et~al.}}]{maNatPhys2017}%
  \BibitemOpen
  \bibfield  {author} {\bibinfo {author} {\bibfnamefont {Q.}~\bibnamefont
  {Ma}}, \bibinfo {author} {\bibfnamefont {S.-Y.}\ \bibnamefont {Xu}}, \bibinfo
  {author} {\bibfnamefont {C.-K.}\ \bibnamefont {Chan}}, \bibinfo {author}
  {\bibfnamefont {C.-L.}\ \bibnamefont {Zhang}}, \bibinfo {author}
  {\bibfnamefont {G.}~\bibnamefont {Chang}}, \bibinfo {author} {\bibfnamefont
  {Y.}~\bibnamefont {Lin}}, \bibinfo {author} {\bibfnamefont {W.}~\bibnamefont
  {Xie}}, \bibinfo {author} {\bibfnamefont {T.}~\bibnamefont {Palacios}},
  \bibinfo {author} {\bibfnamefont {H.}~\bibnamefont {Lin}}, \bibinfo {author}
  {\bibfnamefont {S.}~\bibnamefont {Jia}},  \emph {et~al.},\ }\href@noop {}
  {\bibfield  {journal} {\bibinfo  {journal} {Nature Physics}\ }\textbf
  {\bibinfo {volume} {13}},\ \bibinfo {pages} {842} (\bibinfo {year}
  {2017})}\BibitemShut {NoStop}%
\bibitem [{\citenamefont {Osterhoudt}\ \emph {et~al.}(2019)\citenamefont
  {Osterhoudt}, \citenamefont {Diebel}, \citenamefont {Gray}, \citenamefont
  {Yang}, \citenamefont {Stanco}, \citenamefont {Huang}, \citenamefont {Shen},
  \citenamefont {Ni}, \citenamefont {Moll}, \citenamefont {Ran} \emph
  {et~al.}}]{osterhoudtNatMat2019}%
  \BibitemOpen
  \bibfield  {author} {\bibinfo {author} {\bibfnamefont {G.~B.}\ \bibnamefont
  {Osterhoudt}}, \bibinfo {author} {\bibfnamefont {L.~K.}\ \bibnamefont
  {Diebel}}, \bibinfo {author} {\bibfnamefont {M.~J.}\ \bibnamefont {Gray}},
  \bibinfo {author} {\bibfnamefont {X.}~\bibnamefont {Yang}}, \bibinfo {author}
  {\bibfnamefont {J.}~\bibnamefont {Stanco}}, \bibinfo {author} {\bibfnamefont
  {X.}~\bibnamefont {Huang}}, \bibinfo {author} {\bibfnamefont
  {B.}~\bibnamefont {Shen}}, \bibinfo {author} {\bibfnamefont {N.}~\bibnamefont
  {Ni}}, \bibinfo {author} {\bibfnamefont {P.~J.}\ \bibnamefont {Moll}},
  \bibinfo {author} {\bibfnamefont {Y.}~\bibnamefont {Ran}},  \emph {et~al.},\
  }\href@noop {} {\bibfield  {journal} {\bibinfo  {journal} {Nature materials}\
  }\textbf {\bibinfo {volume} {18}},\ \bibinfo {pages} {471} (\bibinfo {year}
  {2019})}\BibitemShut {NoStop}%
\bibitem [{\citenamefont {Ji}\ \emph {et~al.}(2019)\citenamefont {Ji},
  \citenamefont {Liu}, \citenamefont {Addison}, \citenamefont {Liu},
  \citenamefont {Yu}, \citenamefont {Gao}, \citenamefont {Liu}, \citenamefont
  {Rappe}, \citenamefont {Kane}, \citenamefont {Mele} \emph
  {et~al.}}]{jiNatMat2019}%
  \BibitemOpen
  \bibfield  {author} {\bibinfo {author} {\bibfnamefont {Z.}~\bibnamefont
  {Ji}}, \bibinfo {author} {\bibfnamefont {G.}~\bibnamefont {Liu}}, \bibinfo
  {author} {\bibfnamefont {Z.}~\bibnamefont {Addison}}, \bibinfo {author}
  {\bibfnamefont {W.}~\bibnamefont {Liu}}, \bibinfo {author} {\bibfnamefont
  {P.}~\bibnamefont {Yu}}, \bibinfo {author} {\bibfnamefont {H.}~\bibnamefont
  {Gao}}, \bibinfo {author} {\bibfnamefont {Z.}~\bibnamefont {Liu}}, \bibinfo
  {author} {\bibfnamefont {A.~M.}\ \bibnamefont {Rappe}}, \bibinfo {author}
  {\bibfnamefont {C.~L.}\ \bibnamefont {Kane}}, \bibinfo {author}
  {\bibfnamefont {E.~J.}\ \bibnamefont {Mele}},  \emph {et~al.},\ }\href@noop
  {} {\bibfield  {journal} {\bibinfo  {journal} {Nature materials}\ ,\ \bibinfo
  {pages} {1}} (\bibinfo {year} {2019})}\BibitemShut {NoStop}%
\bibitem [{\citenamefont {Ma}\ \emph {et~al.}(2019)\citenamefont {Ma},
  \citenamefont {Gu}, \citenamefont {Liu}, \citenamefont {Lai}, \citenamefont
  {Yu}, \citenamefont {Zhuo}, \citenamefont {Liu}, \citenamefont {Chen},
  \citenamefont {Feng},\ and\ \citenamefont {Sun}}]{maNatMat2019}%
  \BibitemOpen
  \bibfield  {author} {\bibinfo {author} {\bibfnamefont {J.}~\bibnamefont
  {Ma}}, \bibinfo {author} {\bibfnamefont {Q.}~\bibnamefont {Gu}}, \bibinfo
  {author} {\bibfnamefont {Y.}~\bibnamefont {Liu}}, \bibinfo {author}
  {\bibfnamefont {J.}~\bibnamefont {Lai}}, \bibinfo {author} {\bibfnamefont
  {P.}~\bibnamefont {Yu}}, \bibinfo {author} {\bibfnamefont {X.}~\bibnamefont
  {Zhuo}}, \bibinfo {author} {\bibfnamefont {Z.}~\bibnamefont {Liu}}, \bibinfo
  {author} {\bibfnamefont {J.-H.}\ \bibnamefont {Chen}}, \bibinfo {author}
  {\bibfnamefont {J.}~\bibnamefont {Feng}}, \ and\ \bibinfo {author}
  {\bibfnamefont {D.}~\bibnamefont {Sun}},\ }\href@noop {} {\bibfield
  {journal} {\bibinfo  {journal} {Nat. Mater}\ }\textbf {\bibinfo {volume}
  {18}},\ \bibinfo {pages} {476} (\bibinfo {year} {2019})}\BibitemShut
  {NoStop}%
\bibitem [{\citenamefont {Sirica}\ \emph {et~al.}(2019)\citenamefont {Sirica},
  \citenamefont {Tobey}, \citenamefont {Zhao}, \citenamefont {Chen},
  \citenamefont {Xu}, \citenamefont {Yang}, \citenamefont {Shen}, \citenamefont
  {Yarotski}, \citenamefont {Bowlan}, \citenamefont {Trugman} \emph
  {et~al.}}]{siricaPRL2019}%
  \BibitemOpen
  \bibfield  {author} {\bibinfo {author} {\bibfnamefont {N.}~\bibnamefont
  {Sirica}}, \bibinfo {author} {\bibfnamefont {R.}~\bibnamefont {Tobey}},
  \bibinfo {author} {\bibfnamefont {L.}~\bibnamefont {Zhao}}, \bibinfo {author}
  {\bibfnamefont {G.}~\bibnamefont {Chen}}, \bibinfo {author} {\bibfnamefont
  {B.}~\bibnamefont {Xu}}, \bibinfo {author} {\bibfnamefont {R.}~\bibnamefont
  {Yang}}, \bibinfo {author} {\bibfnamefont {B.}~\bibnamefont {Shen}}, \bibinfo
  {author} {\bibfnamefont {D.}~\bibnamefont {Yarotski}}, \bibinfo {author}
  {\bibfnamefont {P.}~\bibnamefont {Bowlan}}, \bibinfo {author} {\bibfnamefont
  {S.}~\bibnamefont {Trugman}},  \emph {et~al.},\ }\href@noop {} {\bibfield
  {journal} {\bibinfo  {journal} {Physical review letters}\ }\textbf {\bibinfo
  {volume} {122}},\ \bibinfo {pages} {197401} (\bibinfo {year}
  {2019})}\BibitemShut {NoStop}%
\bibitem [{\citenamefont {Rees}\ \emph {et~al.}(2020)\citenamefont {Rees},
  \citenamefont {Manna}, \citenamefont {Lu}, \citenamefont {Morimoto},
  \citenamefont {Borrmann}, \citenamefont {Felser}, \citenamefont {Moore},
  \citenamefont {Torchinsky},\ and\ \citenamefont {Orenstein}}]{reesarxiv2019}%
  \BibitemOpen
  \bibfield  {author} {\bibinfo {author} {\bibfnamefont {D.}~\bibnamefont
  {Rees}}, \bibinfo {author} {\bibfnamefont {K.}~\bibnamefont {Manna}},
  \bibinfo {author} {\bibfnamefont {B.}~\bibnamefont {Lu}}, \bibinfo {author}
  {\bibfnamefont {T.}~\bibnamefont {Morimoto}}, \bibinfo {author}
  {\bibfnamefont {H.}~\bibnamefont {Borrmann}}, \bibinfo {author}
  {\bibfnamefont {C.}~\bibnamefont {Felser}}, \bibinfo {author} {\bibfnamefont
  {J.}~\bibnamefont {Moore}}, \bibinfo {author} {\bibfnamefont {D.~H.}\
  \bibnamefont {Torchinsky}}, \ and\ \bibinfo {author} {\bibfnamefont
  {J.}~\bibnamefont {Orenstein}},\ }\href@noop {} {\bibfield  {journal}
  {\bibinfo  {journal} {Science advances}\ }\textbf {\bibinfo {volume} {6}},\
  \bibinfo {pages} {eaba0509} (\bibinfo {year} {2020})}\BibitemShut {NoStop}%
\bibitem [{\citenamefont {Gao}\ \emph {et~al.}(2020)\citenamefont {Gao},
  \citenamefont {Kaushik}, \citenamefont {Philip}, \citenamefont {Li},
  \citenamefont {Qin}, \citenamefont {Liu}, \citenamefont {Zhang},
  \citenamefont {Su}, \citenamefont {Chen}, \citenamefont {Weng} \emph
  {et~al.}}]{gaoNatComm2020}%
  \BibitemOpen
  \bibfield  {author} {\bibinfo {author} {\bibfnamefont {Y.}~\bibnamefont
  {Gao}}, \bibinfo {author} {\bibfnamefont {S.}~\bibnamefont {Kaushik}},
  \bibinfo {author} {\bibfnamefont {E.}~\bibnamefont {Philip}}, \bibinfo
  {author} {\bibfnamefont {Z.}~\bibnamefont {Li}}, \bibinfo {author}
  {\bibfnamefont {Y.}~\bibnamefont {Qin}}, \bibinfo {author} {\bibfnamefont
  {Y.}~\bibnamefont {Liu}}, \bibinfo {author} {\bibfnamefont {W.}~\bibnamefont
  {Zhang}}, \bibinfo {author} {\bibfnamefont {Y.}~\bibnamefont {Su}}, \bibinfo
  {author} {\bibfnamefont {X.}~\bibnamefont {Chen}}, \bibinfo {author}
  {\bibfnamefont {H.}~\bibnamefont {Weng}},  \emph {et~al.},\ }\href@noop {}
  {\bibfield  {journal} {\bibinfo  {journal} {Nature communications}\ }\textbf
  {\bibinfo {volume} {11}},\ \bibinfo {pages} {1} (\bibinfo {year}
  {2020})}\BibitemShut {NoStop}%
\bibitem [{\citenamefont {Ni}\ \emph {et~al.}(2020{\natexlab{a}})\citenamefont
  {Ni}, \citenamefont {Wang}, \citenamefont {Zhang}, \citenamefont {Pozo},
  \citenamefont {Xu}, \citenamefont {Han}, \citenamefont {Manna}, \citenamefont
  {Paglione}, \citenamefont {Felser}, \citenamefont {Grushin}, \citenamefont
  {de~Juan}, \citenamefont {Mele},\ and\ \citenamefont {Wu}}]{Ni2020arXivCPGE}%
  \BibitemOpen
  \bibfield  {author} {\bibinfo {author} {\bibfnamefont {Z.}~\bibnamefont
  {Ni}}, \bibinfo {author} {\bibfnamefont {K.}~\bibnamefont {Wang}}, \bibinfo
  {author} {\bibfnamefont {Y.}~\bibnamefont {Zhang}}, \bibinfo {author}
  {\bibfnamefont {O.}~\bibnamefont {Pozo}}, \bibinfo {author} {\bibfnamefont
  {B.}~\bibnamefont {Xu}}, \bibinfo {author} {\bibfnamefont {X.}~\bibnamefont
  {Han}}, \bibinfo {author} {\bibfnamefont {K.}~\bibnamefont {Manna}}, \bibinfo
  {author} {\bibfnamefont {J.}~\bibnamefont {Paglione}}, \bibinfo {author}
  {\bibfnamefont {C.}~\bibnamefont {Felser}}, \bibinfo {author} {\bibfnamefont
  {A.~G.}\ \bibnamefont {Grushin}}, \bibinfo {author} {\bibfnamefont
  {F.}~\bibnamefont {de~Juan}}, \bibinfo {author} {\bibfnamefont {E.~J.}\
  \bibnamefont {Mele}}, \ and\ \bibinfo {author} {\bibfnamefont
  {L.}~\bibnamefont {Wu}},\ }\href@noop {} {\bibfield  {journal} {\bibinfo
  {journal} {arXiv preprint arXiv:2006.09612}\ } (\bibinfo {year}
  {2020}{\natexlab{a}})}\BibitemShut {NoStop}%
\bibitem [{\citenamefont {Ni}\ \emph {et~al.}(2020{\natexlab{b}})\citenamefont
  {Ni}, \citenamefont {Xu}, \citenamefont {Sanchez-Martinez}, \citenamefont
  {Zhang}, \citenamefont {Manna}, \citenamefont {Bernhard}, \citenamefont
  {Venderbos}, \citenamefont {de~Juan}, \citenamefont {Felser}, \citenamefont
  {Grushin},\ and\ \citenamefont {Wu}}]{Ni2020arXiv}%
  \BibitemOpen
  \bibfield  {author} {\bibinfo {author} {\bibfnamefont {Z.}~\bibnamefont
  {Ni}}, \bibinfo {author} {\bibfnamefont {B.}~\bibnamefont {Xu}}, \bibinfo
  {author} {\bibfnamefont {M.~A.}\ \bibnamefont {Sanchez-Martinez}}, \bibinfo
  {author} {\bibfnamefont {Y.}~\bibnamefont {Zhang}}, \bibinfo {author}
  {\bibfnamefont {K.}~\bibnamefont {Manna}}, \bibinfo {author} {\bibfnamefont
  {C.}~\bibnamefont {Bernhard}}, \bibinfo {author} {\bibfnamefont {J.~W.~F.}\
  \bibnamefont {Venderbos}}, \bibinfo {author} {\bibfnamefont {F.}~\bibnamefont
  {de~Juan}}, \bibinfo {author} {\bibfnamefont {C.}~\bibnamefont {Felser}},
  \bibinfo {author} {\bibfnamefont {A.~G.}\ \bibnamefont {Grushin}}, \ and\
  \bibinfo {author} {\bibfnamefont {L.}~\bibnamefont {Wu}},\ }\href@noop {}
  {\bibfield  {journal} {\bibinfo  {journal} {arXiv preprint arXiv:2005.13473}\
  } (\bibinfo {year} {2020}{\natexlab{b}})}\BibitemShut {NoStop}%
\bibitem [{\citenamefont {B{\'a}csi}\ and\ \citenamefont
  {Virosztek}(2013)}]{bacsiPRB2013}%
  \BibitemOpen
  \bibfield  {author} {\bibinfo {author} {\bibfnamefont {{\'A}.}~\bibnamefont
  {B{\'a}csi}}\ and\ \bibinfo {author} {\bibfnamefont {A.}~\bibnamefont
  {Virosztek}},\ }\href@noop {} {\bibfield  {journal} {\bibinfo  {journal}
  {Physical Review B}\ }\textbf {\bibinfo {volume} {87}},\ \bibinfo {pages}
  {125425} (\bibinfo {year} {2013})}\BibitemShut {NoStop}%
\bibitem [{\citenamefont {Kuzmenko}\ \emph {et~al.}(2008)\citenamefont
  {Kuzmenko}, \citenamefont {Van~Heumen}, \citenamefont {Carbone},\ and\
  \citenamefont {Van Der~Marel}}]{kuzmenkoPRL2008}%
  \BibitemOpen
  \bibfield  {author} {\bibinfo {author} {\bibfnamefont {A.}~\bibnamefont
  {Kuzmenko}}, \bibinfo {author} {\bibfnamefont {E.}~\bibnamefont
  {Van~Heumen}}, \bibinfo {author} {\bibfnamefont {F.}~\bibnamefont {Carbone}},
  \ and\ \bibinfo {author} {\bibfnamefont {D.}~\bibnamefont {Van Der~Marel}},\
  }\href@noop {} {\bibfield  {journal} {\bibinfo  {journal} {Physical review
  letters}\ }\textbf {\bibinfo {volume} {100}},\ \bibinfo {pages} {117401}
  (\bibinfo {year} {2008})}\BibitemShut {NoStop}%
\bibitem [{\citenamefont {Nair}\ \emph {et~al.}(2008)\citenamefont {Nair},
  \citenamefont {Blake}, \citenamefont {Grigorenko}, \citenamefont {Novoselov},
  \citenamefont {Booth}, \citenamefont {Stauber}, \citenamefont {Peres},\ and\
  \citenamefont {Geim}}]{nairScience2008}%
  \BibitemOpen
  \bibfield  {author} {\bibinfo {author} {\bibfnamefont {R.~R.}\ \bibnamefont
  {Nair}}, \bibinfo {author} {\bibfnamefont {P.}~\bibnamefont {Blake}},
  \bibinfo {author} {\bibfnamefont {A.~N.}\ \bibnamefont {Grigorenko}},
  \bibinfo {author} {\bibfnamefont {K.~S.}\ \bibnamefont {Novoselov}}, \bibinfo
  {author} {\bibfnamefont {T.~J.}\ \bibnamefont {Booth}}, \bibinfo {author}
  {\bibfnamefont {T.}~\bibnamefont {Stauber}}, \bibinfo {author} {\bibfnamefont
  {N.~M.}\ \bibnamefont {Peres}}, \ and\ \bibinfo {author} {\bibfnamefont
  {A.~K.}\ \bibnamefont {Geim}},\ }\href@noop {} {\bibfield  {journal}
  {\bibinfo  {journal} {Science}\ }\textbf {\bibinfo {volume} {320}},\ \bibinfo
  {pages} {1308} (\bibinfo {year} {2008})}\BibitemShut {NoStop}%
\bibitem [{\citenamefont {Hosur}\ \emph {et~al.}(2012)\citenamefont {Hosur},
  \citenamefont {Parameswaran},\ and\ \citenamefont
  {Vishwanath}}]{hosurPRL2012}%
  \BibitemOpen
  \bibfield  {author} {\bibinfo {author} {\bibfnamefont {P.}~\bibnamefont
  {Hosur}}, \bibinfo {author} {\bibfnamefont {S.}~\bibnamefont {Parameswaran}},
  \ and\ \bibinfo {author} {\bibfnamefont {A.}~\bibnamefont {Vishwanath}},\
  }\href@noop {} {\bibfield  {journal} {\bibinfo  {journal} {Physical review
  letters}\ }\textbf {\bibinfo {volume} {108}},\ \bibinfo {pages} {046602}
  (\bibinfo {year} {2012})}\BibitemShut {NoStop}%
\bibitem [{\citenamefont {Akrap}\ \emph {et~al.}(2016)\citenamefont {Akrap},
  \citenamefont {Hakl}, \citenamefont {Tchoumakov}, \citenamefont {Crassee},
  \citenamefont {Kuba}, \citenamefont {Goerbig}, \citenamefont {Homes},
  \citenamefont {Caha}, \citenamefont {Nov{\'a}k}, \citenamefont {Teppe} \emph
  {et~al.}}]{akrapPRL2016}%
  \BibitemOpen
  \bibfield  {author} {\bibinfo {author} {\bibfnamefont {A.}~\bibnamefont
  {Akrap}}, \bibinfo {author} {\bibfnamefont {M.}~\bibnamefont {Hakl}},
  \bibinfo {author} {\bibfnamefont {S.}~\bibnamefont {Tchoumakov}}, \bibinfo
  {author} {\bibfnamefont {I.}~\bibnamefont {Crassee}}, \bibinfo {author}
  {\bibfnamefont {J.}~\bibnamefont {Kuba}}, \bibinfo {author} {\bibfnamefont
  {M.}~\bibnamefont {Goerbig}}, \bibinfo {author} {\bibfnamefont
  {C.}~\bibnamefont {Homes}}, \bibinfo {author} {\bibfnamefont
  {O.}~\bibnamefont {Caha}}, \bibinfo {author} {\bibfnamefont {J.}~\bibnamefont
  {Nov{\'a}k}}, \bibinfo {author} {\bibfnamefont {F.}~\bibnamefont {Teppe}},
  \emph {et~al.},\ }\href@noop {} {\bibfield  {journal} {\bibinfo  {journal}
  {Physical review letters}\ }\textbf {\bibinfo {volume} {117}},\ \bibinfo
  {pages} {136401} (\bibinfo {year} {2016})}\BibitemShut {NoStop}%
\bibitem [{\citenamefont {Neubauer}\ \emph {et~al.}(2016)\citenamefont
  {Neubauer}, \citenamefont {Carbotte}, \citenamefont {Nateprov}, \citenamefont
  {L{\"o}hle}, \citenamefont {Dressel},\ and\ \citenamefont
  {Pronin}}]{neubauerPRB2016}%
  \BibitemOpen
  \bibfield  {author} {\bibinfo {author} {\bibfnamefont {D.}~\bibnamefont
  {Neubauer}}, \bibinfo {author} {\bibfnamefont {J.}~\bibnamefont {Carbotte}},
  \bibinfo {author} {\bibfnamefont {A.}~\bibnamefont {Nateprov}}, \bibinfo
  {author} {\bibfnamefont {A.}~\bibnamefont {L{\"o}hle}}, \bibinfo {author}
  {\bibfnamefont {M.}~\bibnamefont {Dressel}}, \ and\ \bibinfo {author}
  {\bibfnamefont {A.}~\bibnamefont {Pronin}},\ }\href@noop {} {\bibfield
  {journal} {\bibinfo  {journal} {Physical Review B}\ }\textbf {\bibinfo
  {volume} {93}},\ \bibinfo {pages} {121202} (\bibinfo {year}
  {2016})}\BibitemShut {NoStop}%
\bibitem [{\citenamefont {Chen}\ \emph {et~al.}(2015)\citenamefont {Chen},
  \citenamefont {Zhang}, \citenamefont {Schneeloch}, \citenamefont {Zhang},
  \citenamefont {Li}, \citenamefont {Gu},\ and\ \citenamefont
  {Wang}}]{chenPRB2015}%
  \BibitemOpen
  \bibfield  {author} {\bibinfo {author} {\bibfnamefont {R.}~\bibnamefont
  {Chen}}, \bibinfo {author} {\bibfnamefont {S.}~\bibnamefont {Zhang}},
  \bibinfo {author} {\bibfnamefont {J.}~\bibnamefont {Schneeloch}}, \bibinfo
  {author} {\bibfnamefont {C.}~\bibnamefont {Zhang}}, \bibinfo {author}
  {\bibfnamefont {Q.}~\bibnamefont {Li}}, \bibinfo {author} {\bibfnamefont
  {G.}~\bibnamefont {Gu}}, \ and\ \bibinfo {author} {\bibfnamefont
  {N.}~\bibnamefont {Wang}},\ }\href@noop {} {\bibfield  {journal} {\bibinfo
  {journal} {Physical Review B}\ }\textbf {\bibinfo {volume} {92}},\ \bibinfo
  {pages} {075107} (\bibinfo {year} {2015})}\BibitemShut {NoStop}%
\bibitem [{\citenamefont {Xu}\ \emph {et~al.}(2018)\citenamefont {Xu},
  \citenamefont {Zhao}, \citenamefont {Marsik}, \citenamefont {Sheveleva},
  \citenamefont {Lyzwa}, \citenamefont {Dai}, \citenamefont {Chen},
  \citenamefont {Qiu},\ and\ \citenamefont {Bernhard}}]{xuPRL2018}%
  \BibitemOpen
  \bibfield  {author} {\bibinfo {author} {\bibfnamefont {B.}~\bibnamefont
  {Xu}}, \bibinfo {author} {\bibfnamefont {L.}~\bibnamefont {Zhao}}, \bibinfo
  {author} {\bibfnamefont {P.}~\bibnamefont {Marsik}}, \bibinfo {author}
  {\bibfnamefont {E.}~\bibnamefont {Sheveleva}}, \bibinfo {author}
  {\bibfnamefont {F.}~\bibnamefont {Lyzwa}}, \bibinfo {author} {\bibfnamefont
  {Y.}~\bibnamefont {Dai}}, \bibinfo {author} {\bibfnamefont {G.}~\bibnamefont
  {Chen}}, \bibinfo {author} {\bibfnamefont {X.}~\bibnamefont {Qiu}}, \ and\
  \bibinfo {author} {\bibfnamefont {C.}~\bibnamefont {Bernhard}},\ }\href@noop
  {} {\bibfield  {journal} {\bibinfo  {journal} {Physical review letters}\
  }\textbf {\bibinfo {volume} {121}},\ \bibinfo {pages} {187401} (\bibinfo
  {year} {2018})}\BibitemShut {NoStop}%
\bibitem [{\citenamefont {Martino}\ \emph {et~al.}(2019)\citenamefont
  {Martino}, \citenamefont {Crassee}, \citenamefont {Eguchi}, \citenamefont
  {Santos-Cottin}, \citenamefont {Zhong}, \citenamefont {Gu}, \citenamefont
  {Berger}, \citenamefont {Rukelj}, \citenamefont {Orlita}, \citenamefont
  {Homes} \emph {et~al.}}]{martinoPRL2019}%
  \BibitemOpen
  \bibfield  {author} {\bibinfo {author} {\bibfnamefont {E.}~\bibnamefont
  {Martino}}, \bibinfo {author} {\bibfnamefont {I.}~\bibnamefont {Crassee}},
  \bibinfo {author} {\bibfnamefont {G.}~\bibnamefont {Eguchi}}, \bibinfo
  {author} {\bibfnamefont {D.}~\bibnamefont {Santos-Cottin}}, \bibinfo {author}
  {\bibfnamefont {R.}~\bibnamefont {Zhong}}, \bibinfo {author} {\bibfnamefont
  {G.}~\bibnamefont {Gu}}, \bibinfo {author} {\bibfnamefont {H.}~\bibnamefont
  {Berger}}, \bibinfo {author} {\bibfnamefont {Z.}~\bibnamefont {Rukelj}},
  \bibinfo {author} {\bibfnamefont {M.}~\bibnamefont {Orlita}}, \bibinfo
  {author} {\bibfnamefont {C.~C.}\ \bibnamefont {Homes}},  \emph {et~al.},\
  }\href@noop {} {\bibfield  {journal} {\bibinfo  {journal} {Physical review
  letters}\ }\textbf {\bibinfo {volume} {122}},\ \bibinfo {pages} {217402}
  (\bibinfo {year} {2019})}\BibitemShut {NoStop}%
\bibitem [{\citenamefont {Xu}\ \emph {et~al.}(2016{\natexlab{b}})\citenamefont
  {Xu}, \citenamefont {Dai}, \citenamefont {Zhao}, \citenamefont {Wang},
  \citenamefont {Yang}, \citenamefont {Zhang}, \citenamefont {Liu},
  \citenamefont {Xiao}, \citenamefont {Chen}, \citenamefont {Taylor} \emph
  {et~al.}}]{xuPRB2016}%
  \BibitemOpen
  \bibfield  {author} {\bibinfo {author} {\bibfnamefont {B.}~\bibnamefont
  {Xu}}, \bibinfo {author} {\bibfnamefont {Y.}~\bibnamefont {Dai}}, \bibinfo
  {author} {\bibfnamefont {L.}~\bibnamefont {Zhao}}, \bibinfo {author}
  {\bibfnamefont {K.}~\bibnamefont {Wang}}, \bibinfo {author} {\bibfnamefont
  {R.}~\bibnamefont {Yang}}, \bibinfo {author} {\bibfnamefont {W.}~\bibnamefont
  {Zhang}}, \bibinfo {author} {\bibfnamefont {J.}~\bibnamefont {Liu}}, \bibinfo
  {author} {\bibfnamefont {H.}~\bibnamefont {Xiao}}, \bibinfo {author}
  {\bibfnamefont {G.}~\bibnamefont {Chen}}, \bibinfo {author} {\bibfnamefont
  {A.}~\bibnamefont {Taylor}},  \emph {et~al.},\ }\href@noop {} {\bibfield
  {journal} {\bibinfo  {journal} {Physical Review B}\ }\textbf {\bibinfo
  {volume} {93}},\ \bibinfo {pages} {121110} (\bibinfo {year}
  {2016}{\natexlab{b}})}\BibitemShut {NoStop}%
\bibitem [{\citenamefont {Kimura}\ \emph {et~al.}(2017)\citenamefont {Kimura},
  \citenamefont {Yokoyama}, \citenamefont {Watanabe}, \citenamefont
  {Sichelschmidt}, \citenamefont {S{\"u}{\ss}}, \citenamefont {Schmidt},\ and\
  \citenamefont {Felser}}]{kimuraPRB2017}%
  \BibitemOpen
  \bibfield  {author} {\bibinfo {author} {\bibfnamefont {S.-i.}\ \bibnamefont
  {Kimura}}, \bibinfo {author} {\bibfnamefont {H.}~\bibnamefont {Yokoyama}},
  \bibinfo {author} {\bibfnamefont {H.}~\bibnamefont {Watanabe}}, \bibinfo
  {author} {\bibfnamefont {J.}~\bibnamefont {Sichelschmidt}}, \bibinfo {author}
  {\bibfnamefont {V.}~\bibnamefont {S{\"u}{\ss}}}, \bibinfo {author}
  {\bibfnamefont {M.}~\bibnamefont {Schmidt}}, \ and\ \bibinfo {author}
  {\bibfnamefont {C.}~\bibnamefont {Felser}},\ }\href@noop {} {\bibfield
  {journal} {\bibinfo  {journal} {Physical Review B}\ }\textbf {\bibinfo
  {volume} {96}},\ \bibinfo {pages} {075119} (\bibinfo {year}
  {2017})}\BibitemShut {NoStop}%
\bibitem [{\citenamefont {Maulana}\ \emph {et~al.}(2020)\citenamefont
  {Maulana}, \citenamefont {Manna}, \citenamefont {Uykur}, \citenamefont
  {Felser}, \citenamefont {Dressel},\ and\ \citenamefont
  {Pronin}}]{maulanaPRR2020}%
  \BibitemOpen
  \bibfield  {author} {\bibinfo {author} {\bibfnamefont {L.}~\bibnamefont
  {Maulana}}, \bibinfo {author} {\bibfnamefont {K.}~\bibnamefont {Manna}},
  \bibinfo {author} {\bibfnamefont {E.}~\bibnamefont {Uykur}}, \bibinfo
  {author} {\bibfnamefont {C.}~\bibnamefont {Felser}}, \bibinfo {author}
  {\bibfnamefont {M.}~\bibnamefont {Dressel}}, \ and\ \bibinfo {author}
  {\bibfnamefont {A.}~\bibnamefont {Pronin}},\ }\href@noop {} {\bibfield
  {journal} {\bibinfo  {journal} {Physical Review Research}\ }\textbf {\bibinfo
  {volume} {2}},\ \bibinfo {pages} {023018} (\bibinfo {year}
  {2020})}\BibitemShut {NoStop}%
\bibitem [{\citenamefont {Sanchez}\ \emph
  {et~al.}(2019{\natexlab{b}})\citenamefont {Sanchez}, \citenamefont
  {Belopolski}, \citenamefont {Cochran}, \citenamefont {Xu}, \citenamefont
  {Yin}, \citenamefont {Chang}, \citenamefont {Xie}, \citenamefont {Manna},
  \citenamefont {S{\"u}{\ss}}, \citenamefont {Huang} \emph
  {et~al.}}]{sanchez2019topological}%
  \BibitemOpen
  \bibfield  {author} {\bibinfo {author} {\bibfnamefont {D.~S.}\ \bibnamefont
  {Sanchez}}, \bibinfo {author} {\bibfnamefont {I.}~\bibnamefont {Belopolski}},
  \bibinfo {author} {\bibfnamefont {T.~A.}\ \bibnamefont {Cochran}}, \bibinfo
  {author} {\bibfnamefont {X.}~\bibnamefont {Xu}}, \bibinfo {author}
  {\bibfnamefont {J.-X.}\ \bibnamefont {Yin}}, \bibinfo {author} {\bibfnamefont
  {G.}~\bibnamefont {Chang}}, \bibinfo {author} {\bibfnamefont
  {W.}~\bibnamefont {Xie}}, \bibinfo {author} {\bibfnamefont {K.}~\bibnamefont
  {Manna}}, \bibinfo {author} {\bibfnamefont {V.}~\bibnamefont {S{\"u}{\ss}}},
  \bibinfo {author} {\bibfnamefont {C.-Y.}\ \bibnamefont {Huang}},  \emph
  {et~al.},\ }\href@noop {} {\bibfield  {journal} {\bibinfo  {journal}
  {Nature}\ }\textbf {\bibinfo {volume} {567}},\ \bibinfo {pages} {500}
  (\bibinfo {year} {2019}{\natexlab{b}})}\BibitemShut {NoStop}%
\bibitem [{\citenamefont {Habe}(2019)}]{habePRB2019}%
  \BibitemOpen
  \bibfield  {author} {\bibinfo {author} {\bibfnamefont {T.}~\bibnamefont
  {Habe}},\ }\href@noop {} {\bibfield  {journal} {\bibinfo  {journal} {Physical
  Review B}\ }\textbf {\bibinfo {volume} {100}},\ \bibinfo {pages} {245131}
  (\bibinfo {year} {2019})}\BibitemShut {NoStop}%
\bibitem [{\citenamefont {{van der Marel}}\ \emph {et~al.}(1998)\citenamefont
  {{van der Marel}}, \citenamefont {Damascelli}, \citenamefont {Schulte},\ and\
  \citenamefont {Menovsky}}]{VANDERMAREL1998}%
  \BibitemOpen
  \bibfield  {author} {\bibinfo {author} {\bibfnamefont {D.}~\bibnamefont {{van
  der Marel}}}, \bibinfo {author} {\bibfnamefont {A.}~\bibnamefont
  {Damascelli}}, \bibinfo {author} {\bibfnamefont {K.}~\bibnamefont {Schulte}},
  \ and\ \bibinfo {author} {\bibfnamefont {A.}~\bibnamefont {Menovsky}},\
  }\href {\doibase https://doi.org/10.1016/S0921-4526(97)00476-6} {\bibfield
  {journal} {\bibinfo  {journal} {Physica B: Condensed Matter}\ }\textbf
  {\bibinfo {volume} {244}},\ \bibinfo {pages} {138 } (\bibinfo {year}
  {1998})},\ \bibinfo {note} {proceedings of the International Conference on
  Low Energy Electrodynamics in Solids}\BibitemShut {NoStop}%
\bibitem [{\citenamefont {Mena}\ \emph {et~al.}(2006)\citenamefont {Mena},
  \citenamefont {DiTusa}, \citenamefont {van~der Marel}, \citenamefont
  {Aeppli}, \citenamefont {Young}, \citenamefont {Damascelli},\ and\
  \citenamefont {Mydosh}}]{MenaPRB2006}%
  \BibitemOpen
  \bibfield  {author} {\bibinfo {author} {\bibfnamefont {F.~P.}\ \bibnamefont
  {Mena}}, \bibinfo {author} {\bibfnamefont {J.~F.}\ \bibnamefont {DiTusa}},
  \bibinfo {author} {\bibfnamefont {D.}~\bibnamefont {van~der Marel}}, \bibinfo
  {author} {\bibfnamefont {G.}~\bibnamefont {Aeppli}}, \bibinfo {author}
  {\bibfnamefont {D.~P.}\ \bibnamefont {Young}}, \bibinfo {author}
  {\bibfnamefont {A.}~\bibnamefont {Damascelli}}, \ and\ \bibinfo {author}
  {\bibfnamefont {J.~A.}\ \bibnamefont {Mydosh}},\ }\href {\doibase
  10.1103/PhysRevB.73.085205} {\bibfield  {journal} {\bibinfo  {journal} {Phys.
  Rev. B}\ }\textbf {\bibinfo {volume} {73}},\ \bibinfo {pages} {085205}
  (\bibinfo {year} {2006})}\BibitemShut {NoStop}%
\bibitem [{\citenamefont {Petrova}\ \emph {et~al.}(2010)\citenamefont
  {Petrova}, \citenamefont {Krasnorussky}, \citenamefont {Shikov},
  \citenamefont {Yuhasz}, \citenamefont {Lograsso}, \citenamefont {Lashley},\
  and\ \citenamefont {Stishov}}]{petrovaPRB2010}%
  \BibitemOpen
  \bibfield  {author} {\bibinfo {author} {\bibfnamefont {A.~E.}\ \bibnamefont
  {Petrova}}, \bibinfo {author} {\bibfnamefont {V.~N.}\ \bibnamefont
  {Krasnorussky}}, \bibinfo {author} {\bibfnamefont {A.~A.}\ \bibnamefont
  {Shikov}}, \bibinfo {author} {\bibfnamefont {W.~M.}\ \bibnamefont {Yuhasz}},
  \bibinfo {author} {\bibfnamefont {T.~A.}\ \bibnamefont {Lograsso}}, \bibinfo
  {author} {\bibfnamefont {J.~C.}\ \bibnamefont {Lashley}}, \ and\ \bibinfo
  {author} {\bibfnamefont {S.~M.}\ \bibnamefont {Stishov}},\ }\href@noop {}
  {\bibfield  {journal} {\bibinfo  {journal} {Physical Review B}\ }\textbf
  {\bibinfo {volume} {82}},\ \bibinfo {pages} {155124} (\bibinfo {year}
  {2010})}\BibitemShut {NoStop}%
\bibitem [{\citenamefont {Xu}\ \emph {et~al.}(2019)\citenamefont {Xu},
  \citenamefont {Wang}, \citenamefont {Cochran}, \citenamefont {Sanchez},
  \citenamefont {Chang}, \citenamefont {Belopolski}, \citenamefont {Wang},
  \citenamefont {Liu}, \citenamefont {Tien}, \citenamefont {Gui} \emph
  {et~al.}}]{xuXPRB2019}%
  \BibitemOpen
  \bibfield  {author} {\bibinfo {author} {\bibfnamefont {X.}~\bibnamefont
  {Xu}}, \bibinfo {author} {\bibfnamefont {X.}~\bibnamefont {Wang}}, \bibinfo
  {author} {\bibfnamefont {T.~A.}\ \bibnamefont {Cochran}}, \bibinfo {author}
  {\bibfnamefont {D.~S.}\ \bibnamefont {Sanchez}}, \bibinfo {author}
  {\bibfnamefont {G.}~\bibnamefont {Chang}}, \bibinfo {author} {\bibfnamefont
  {I.}~\bibnamefont {Belopolski}}, \bibinfo {author} {\bibfnamefont
  {G.}~\bibnamefont {Wang}}, \bibinfo {author} {\bibfnamefont {Y.}~\bibnamefont
  {Liu}}, \bibinfo {author} {\bibfnamefont {H.-J.}\ \bibnamefont {Tien}},
  \bibinfo {author} {\bibfnamefont {X.}~\bibnamefont {Gui}},  \emph {et~al.},\
  }\href@noop {} {\bibfield  {journal} {\bibinfo  {journal} {Physical Review
  B}\ }\textbf {\bibinfo {volume} {100}},\ \bibinfo {pages} {045104} (\bibinfo
  {year} {2019})}\BibitemShut {NoStop}%
\bibitem [{\citenamefont {Pshenay-Severin}\ \emph {et~al.}(2018)\citenamefont
  {Pshenay-Severin}, \citenamefont {Ivanov}, \citenamefont {Burkov},\ and\
  \citenamefont {Burkov}}]{SeverinJPCM2018}%
  \BibitemOpen
  \bibfield  {author} {\bibinfo {author} {\bibfnamefont {D.}~\bibnamefont
  {Pshenay-Severin}}, \bibinfo {author} {\bibfnamefont {Y.~V.}\ \bibnamefont
  {Ivanov}}, \bibinfo {author} {\bibfnamefont {A.}~\bibnamefont {Burkov}}, \
  and\ \bibinfo {author} {\bibfnamefont {A.}~\bibnamefont {Burkov}},\
  }\href@noop {} {\bibfield  {journal} {\bibinfo  {journal} {Journal of
  Physics: Condensed Matter}\ }\textbf {\bibinfo {volume} {30}},\ \bibinfo
  {pages} {135501} (\bibinfo {year} {2018})}\BibitemShut {NoStop}%
\bibitem [{\citenamefont {Mostofi}\ \emph {et~al.}(2014)\citenamefont
  {Mostofi}, \citenamefont {Yates}, \citenamefont {Pizzi}, \citenamefont {Lee},
  \citenamefont {Souza}, \citenamefont {Vanderbilt},\ and\ \citenamefont
  {Marzari}}]{Mostofi2014}%
  \BibitemOpen
  \bibfield  {author} {\bibinfo {author} {\bibfnamefont {A.~A.}\ \bibnamefont
  {Mostofi}}, \bibinfo {author} {\bibfnamefont {J.~R.}\ \bibnamefont {Yates}},
  \bibinfo {author} {\bibfnamefont {G.}~\bibnamefont {Pizzi}}, \bibinfo
  {author} {\bibfnamefont {Y.-S.}\ \bibnamefont {Lee}}, \bibinfo {author}
  {\bibfnamefont {I.}~\bibnamefont {Souza}}, \bibinfo {author} {\bibfnamefont
  {D.}~\bibnamefont {Vanderbilt}}, \ and\ \bibinfo {author} {\bibfnamefont
  {N.}~\bibnamefont {Marzari}},\ }\href@noop {} {\bibfield  {journal} {\bibinfo
   {journal} {Computer Physics Communications}\ }\textbf {\bibinfo {volume}
  {185}},\ \bibinfo {pages} {2309} (\bibinfo {year} {2014})}\BibitemShut
  {NoStop}%
\bibitem [{\citenamefont {Ashby}\ and\ \citenamefont
  {Carbotte}(2014)}]{Ashby2014}%
  \BibitemOpen
  \bibfield  {author} {\bibinfo {author} {\bibfnamefont {P.~E.~C.}\
  \bibnamefont {Ashby}}\ and\ \bibinfo {author} {\bibfnamefont {J.~P.}\
  \bibnamefont {Carbotte}},\ }\href@noop {} {\bibfield  {journal} {\bibinfo
  {journal} {Phys. Rev. B}\ }\textbf {\bibinfo {volume} {89}},\ \bibinfo
  {pages} {245121} (\bibinfo {year} {2014})}\BibitemShut {NoStop}%
\bibitem [{\citenamefont {Wang}\ \emph {et~al.}(2020)\citenamefont {Wang},
  \citenamefont {Xu}, \citenamefont {Rischau}, \citenamefont {Bachar},
  \citenamefont {Michon}, \citenamefont {Teyssier}, \citenamefont {Qiu},
  \citenamefont {Ohtsuki}, \citenamefont {Cheng}, \citenamefont {Armitage},
  \citenamefont {Nakatsuji},\ and\ \citenamefont {van~der Marel}}]{Wang2020NP}%
  \BibitemOpen
  \bibfield  {author} {\bibinfo {author} {\bibfnamefont {K.}~\bibnamefont
  {Wang}}, \bibinfo {author} {\bibfnamefont {B.}~\bibnamefont {Xu}}, \bibinfo
  {author} {\bibfnamefont {C.~W.}\ \bibnamefont {Rischau}}, \bibinfo {author}
  {\bibfnamefont {N.}~\bibnamefont {Bachar}}, \bibinfo {author} {\bibfnamefont
  {B.}~\bibnamefont {Michon}}, \bibinfo {author} {\bibfnamefont
  {J.}~\bibnamefont {Teyssier}}, \bibinfo {author} {\bibfnamefont
  {Y.}~\bibnamefont {Qiu}}, \bibinfo {author} {\bibfnamefont {T.}~\bibnamefont
  {Ohtsuki}}, \bibinfo {author} {\bibfnamefont {B.}~\bibnamefont {Cheng}},
  \bibinfo {author} {\bibfnamefont {N.~P.}\ \bibnamefont {Armitage}}, \bibinfo
  {author} {\bibfnamefont {S.}~\bibnamefont {Nakatsuji}}, \ and\ \bibinfo
  {author} {\bibfnamefont {D.}~\bibnamefont {van~der Marel}},\ }\href {\doibase
  10.1038/s41567-020-0955-0} {\bibfield  {journal} {\bibinfo  {journal} {Nat.
  Phys.}\ } (\bibinfo {year} {2020}),\ 10.1038/s41567-020-0955-0}\BibitemShut
  {NoStop}%
\bibitem [{\citenamefont {Chang}\ \emph {et~al.}(2020)\citenamefont {Chang},
  \citenamefont {Yin}, \citenamefont {Neupert}, \citenamefont {Sanchez},
  \citenamefont {Belopolski}, \citenamefont {Zhang}, \citenamefont {Cochran},
  \citenamefont {Ch{\'e}ng}, \citenamefont {Hsu}, \citenamefont {Huang} \emph
  {et~al.}}]{changPRL2020}%
  \BibitemOpen
  \bibfield  {author} {\bibinfo {author} {\bibfnamefont {G.}~\bibnamefont
  {Chang}}, \bibinfo {author} {\bibfnamefont {J.-X.}\ \bibnamefont {Yin}},
  \bibinfo {author} {\bibfnamefont {T.}~\bibnamefont {Neupert}}, \bibinfo
  {author} {\bibfnamefont {D.~S.}\ \bibnamefont {Sanchez}}, \bibinfo {author}
  {\bibfnamefont {I.}~\bibnamefont {Belopolski}}, \bibinfo {author}
  {\bibfnamefont {S.~S.}\ \bibnamefont {Zhang}}, \bibinfo {author}
  {\bibfnamefont {T.~A.}\ \bibnamefont {Cochran}}, \bibinfo {author}
  {\bibfnamefont {Z.}~\bibnamefont {Ch{\'e}ng}}, \bibinfo {author}
  {\bibfnamefont {M.-C.}\ \bibnamefont {Hsu}}, \bibinfo {author} {\bibfnamefont
  {S.-M.}\ \bibnamefont {Huang}},  \emph {et~al.},\ }\href@noop {} {\bibfield
  {journal} {\bibinfo  {journal} {Physical Review Letters}\ }\textbf {\bibinfo
  {volume} {124}},\ \bibinfo {pages} {166404} (\bibinfo {year}
  {2020})}\BibitemShut {NoStop}%
\bibitem [{\citenamefont {Homes}\ \emph {et~al.}(1993)\citenamefont {Homes},
  \citenamefont {Reedyk}, \citenamefont {Cradles},\ and\ \citenamefont
  {Timusk}}]{Homes1993}%
  \BibitemOpen
  \bibfield  {author} {\bibinfo {author} {\bibfnamefont {C.~C.}\ \bibnamefont
  {Homes}}, \bibinfo {author} {\bibfnamefont {M.}~\bibnamefont {Reedyk}},
  \bibinfo {author} {\bibfnamefont {D.~A.}\ \bibnamefont {Cradles}}, \ and\
  \bibinfo {author} {\bibfnamefont {T.}~\bibnamefont {Timusk}},\ }\href@noop {}
  {\bibfield  {journal} {\bibinfo  {journal} {Appl. Opt.}\ }\textbf {\bibinfo
  {volume} {32}},\ \bibinfo {pages} {2976} (\bibinfo {year}
  {1993})}\BibitemShut {NoStop}%
\bibitem [{\citenamefont {Ramer}\ and\ \citenamefont
  {Rappe}(1999)}]{RamerPRB99}%
  \BibitemOpen
  \bibfield  {author} {\bibinfo {author} {\bibfnamefont {N.~J.}\ \bibnamefont
  {Ramer}}\ and\ \bibinfo {author} {\bibfnamefont {A.~M.}\ \bibnamefont
  {Rappe}},\ }\href@noop {} {\bibfield  {journal} {\bibinfo  {journal} {Phys.
  Rev. B}\ }\textbf {\bibinfo {volume} {59}},\ \bibinfo {pages} {12471}
  (\bibinfo {year} {1999})}\BibitemShut {NoStop}%
\bibitem [{\citenamefont {Rappe}\ \emph {et~al.}(1990)\citenamefont {Rappe},
  \citenamefont {Rabe}, \citenamefont {Kaxiras},\ and\ \citenamefont
  {Joannopoulos}}]{RappePRB1990}%
  \BibitemOpen
  \bibfield  {author} {\bibinfo {author} {\bibfnamefont {A.~M.}\ \bibnamefont
  {Rappe}}, \bibinfo {author} {\bibfnamefont {K.~M.}\ \bibnamefont {Rabe}},
  \bibinfo {author} {\bibfnamefont {E.}~\bibnamefont {Kaxiras}}, \ and\
  \bibinfo {author} {\bibfnamefont {J.~D.}\ \bibnamefont {Joannopoulos}},\
  }\href@noop {} {\bibfield  {journal} {\bibinfo  {journal} {Phys. Rev. B}\
  }\textbf {\bibinfo {volume} {41}},\ \bibinfo {pages} {1227} (\bibinfo {year}
  {1990})}\BibitemShut {NoStop}%
\bibitem [{\citenamefont {Giannozzi}\ \emph {et~al.}(2009)\citenamefont
  {Giannozzi}, \citenamefont {Baroni}, \citenamefont {Bonini}, \citenamefont
  {Calandra}, \citenamefont {Car}, \citenamefont {Cavazzoni}, \citenamefont
  {Ceresoli}, \citenamefont {Chiarotti}, \citenamefont {Cococcioni},
  \citenamefont {Dabo} \emph {et~al.}}]{GiannozziJPCM2009}%
  \BibitemOpen
  \bibfield  {author} {\bibinfo {author} {\bibfnamefont {P.}~\bibnamefont
  {Giannozzi}}, \bibinfo {author} {\bibfnamefont {S.}~\bibnamefont {Baroni}},
  \bibinfo {author} {\bibfnamefont {N.}~\bibnamefont {Bonini}}, \bibinfo
  {author} {\bibfnamefont {M.}~\bibnamefont {Calandra}}, \bibinfo {author}
  {\bibfnamefont {R.}~\bibnamefont {Car}}, \bibinfo {author} {\bibfnamefont
  {C.}~\bibnamefont {Cavazzoni}}, \bibinfo {author} {\bibfnamefont
  {D.}~\bibnamefont {Ceresoli}}, \bibinfo {author} {\bibfnamefont {G.~L.}\
  \bibnamefont {Chiarotti}}, \bibinfo {author} {\bibfnamefont {M.}~\bibnamefont
  {Cococcioni}}, \bibinfo {author} {\bibfnamefont {I.}~\bibnamefont {Dabo}},
  \emph {et~al.},\ }\href@noop {} {\bibfield  {journal} {\bibinfo  {journal}
  {Journal of Physics: Condensed Matter}\ }\textbf {\bibinfo {volume} {21}}
  (\bibinfo {year} {2009})}\BibitemShut {NoStop}%
\bibitem [{Opi()}]{Opium}%
  \BibitemOpen
  \href@noop {} {}\bibinfo {howpublished}
  {http://opium.sourceforge.net}\BibitemShut {NoStop}%
\bibitem [{\citenamefont {Perdew}\ \emph {et~al.}(1996)\citenamefont {Perdew},
  \citenamefont {Burke},\ and\ \citenamefont {Ernzerhof}}]{PerdewPRL1996}%
  \BibitemOpen
  \bibfield  {author} {\bibinfo {author} {\bibfnamefont {J.~P.}\ \bibnamefont
  {Perdew}}, \bibinfo {author} {\bibfnamefont {K.}~\bibnamefont {Burke}}, \
  and\ \bibinfo {author} {\bibfnamefont {M.}~\bibnamefont {Ernzerhof}},\
  }\href@noop {} {\bibfield  {journal} {\bibinfo  {journal} {Phys. Rev. Lett.}\
  }\textbf {\bibinfo {volume} {77}},\ \bibinfo {pages} {3865} (\bibinfo {year}
  {1996})}\BibitemShut {NoStop}%
\bibitem [{\citenamefont {Dressel}\ and\ \citenamefont
  {Gr\"uner}(2002)}]{Dressel2002}%
  \BibitemOpen
  \bibfield  {author} {\bibinfo {author} {\bibfnamefont {M.}~\bibnamefont
  {Dressel}}\ and\ \bibinfo {author} {\bibfnamefont {G.}~\bibnamefont
  {Gr\"uner}},\ }\href@noop {} {\emph {\bibinfo {title} {{Electrodynamics of
  Solids}}}}\ (\bibinfo  {publisher} {Cambridge University press},\ \bibinfo
  {year} {2002})\BibitemShut {NoStop}%
\bibitem [{\citenamefont {Chang}\ \emph
  {et~al.}(2017{\natexlab{c}})\citenamefont {Chang}, \citenamefont {Xu},
  \citenamefont {Wieder}, \citenamefont {Sanchez}, \citenamefont {Huang},
  \citenamefont {Belopolski}, \citenamefont {Chang}, \citenamefont {Zhang},
  \citenamefont {Bansil}, \citenamefont {Lin},\ and\ \citenamefont
  {Hasan}}]{chang_unconventional_2017}%
  \BibitemOpen
  \bibfield  {author} {\bibinfo {author} {\bibfnamefont {G.}~\bibnamefont
  {Chang}}, \bibinfo {author} {\bibfnamefont {S.-Y.}\ \bibnamefont {Xu}},
  \bibinfo {author} {\bibfnamefont {B.~J.}\ \bibnamefont {Wieder}}, \bibinfo
  {author} {\bibfnamefont {D.~S.}\ \bibnamefont {Sanchez}}, \bibinfo {author}
  {\bibfnamefont {S.-M.}\ \bibnamefont {Huang}}, \bibinfo {author}
  {\bibfnamefont {I.}~\bibnamefont {Belopolski}}, \bibinfo {author}
  {\bibfnamefont {T.-R.}\ \bibnamefont {Chang}}, \bibinfo {author}
  {\bibfnamefont {S.}~\bibnamefont {Zhang}}, \bibinfo {author} {\bibfnamefont
  {A.}~\bibnamefont {Bansil}}, \bibinfo {author} {\bibfnamefont
  {H.}~\bibnamefont {Lin}}, \ and\ \bibinfo {author} {\bibfnamefont {M.~Z.}\
  \bibnamefont {Hasan}},\ }\href@noop {} {\bibfield  {journal} {\bibinfo
  {journal} {Physical Review Letters}\ }\textbf {\bibinfo {volume} {119}}
  (\bibinfo {year} {2017}{\natexlab{c}})}\BibitemShut {NoStop}%
\bibitem [{\citenamefont {Flicker}\ \emph
  {et~al.}(2018{\natexlab{b}})\citenamefont {Flicker}, \citenamefont {de~Juan},
  \citenamefont {Bradlyn}, \citenamefont {Morimoto}, \citenamefont
  {Vergniory},\ and\ \citenamefont {Grushin}}]{flicker_chiral_2018}%
  \BibitemOpen
  \bibfield  {author} {\bibinfo {author} {\bibfnamefont {F.}~\bibnamefont
  {Flicker}}, \bibinfo {author} {\bibfnamefont {F.}~\bibnamefont {de~Juan}},
  \bibinfo {author} {\bibfnamefont {B.}~\bibnamefont {Bradlyn}}, \bibinfo
  {author} {\bibfnamefont {T.}~\bibnamefont {Morimoto}}, \bibinfo {author}
  {\bibfnamefont {M.~G.}\ \bibnamefont {Vergniory}}, \ and\ \bibinfo {author}
  {\bibfnamefont {A.~G.}\ \bibnamefont {Grushin}},\ }\href@noop {} {\bibfield
  {journal} {\bibinfo  {journal} {Physical Review B}\ }\textbf {\bibinfo
  {volume} {98}} (\bibinfo {year} {2018}{\natexlab{b}})}\BibitemShut {NoStop}%
\bibitem [{\citenamefont {Manes}(2012)}]{manes_existence_2012}%
  \BibitemOpen
  \bibfield  {author} {\bibinfo {author} {\bibfnamefont {J.~L.}\ \bibnamefont
  {Manes}},\ }\href@noop {} {\bibfield  {journal} {\bibinfo  {journal}
  {Physical Review B}\ }\textbf {\bibinfo {volume} {85}} (\bibinfo {year}
  {2012})}\BibitemShut {NoStop}%
\end{thebibliography}
\end{document}